\documentclass[12pt]{article}
\usepackage{url}
\input{epsf}

\font\bm=cmmib10 at 10pt
\font\bms=cmmib10 at 7pt \textfont9=\bm \scriptfont9=\bms
\usepackage{amsfonts}
\usepackage{multirow}
\mathchardef\balpha= "790B
\mathchardef\bbeta= "790C
\mathchardef\bTheta= "7902
\mathchardef\bzeta= "7910
\mathchardef\bOmega= "790A
\mathchardef\bGamma= "7900
\mathchardef\bDelta= "7901
\mathchardef\bPhi= "7908
\mathchardef\bphi= "791E
\mathchardef\bomega= "7921
\mathchardef\bxi= "7918
\mathchardef\bet= "7911
\mathchardef\brho= "791A
\mathchardef\btau= "791C
\mathchardef\bmu= "7916
\mathchardef\bvarpi= "7924

\def \lvec{(\kern-.26em(}
\def\pmb#1{\setbox0=\hbox{#1}%
\def \lvec{(\kern-.26em(}
\def\mb {\hbox{ $\mu$b}}
\def\perthou{^o\!\!/_{\!\!oo}}

\kern-.025em\copy0\kern-\wd0
\kern.05em\copy0\kern-\wd0
\kern-.025em\raise.0433em\box0 }
\mathchardef\btheta= "7912

\def\XXint#1#2#3{{\setbox0=\hbox{$#1{#2#3}{\int}$}
     \vcenter{\hbox{$#2#3$}}\kern-.5\wd0}}

\usepackage{amsmath}
\usepackage{authblk}
\usepackage{url}
\providecommand{\keywords}[1]{\textbf{\textit{Keywords:  }} #1}
\usepackage{graphicx}% Include figure files
\usepackage{dcolumn}% Align table columns on decimal point
\begin{document}

\title{Acidification of Water by CO$_2$}
\author[1]{ W. A. van Wijngaarden}
\author[2]{P. Ridd}
\author[3]{ M. Cornell}
\author[4] {W. Happer}
\affil[1]{Department of Physics and Astronomy, York University, Canada}
\affil[2]{Physicist, Adjunct Fellow, Institute of Public Affairs, Malanda, Queensland, Australia}
\affil[3]{Industrial Chemist, Lake Jackson, Texas, USA}
\affil[4]{Department of Physics, Princeton University, USA}
\renewcommand\Affilfont{\itshape\small}
\date{\today}
\maketitle
\begin{abstract}
Fundamental inorganic chemistry shows that
increasing concentrations of atmospheric CO$_2$ will have no harmful effect on organisms that live in the natural waters of the Earth\,\cite{CH}, and may well benefit them. Alkalinity and dissolved CO$_2$ give high buffering capacity to most natural waters and minimize the change of pH from external influences. For example,  doubling the atmospheric concentration of CO$_2$ from 430 ppm to 860 ppm would reduce the pH of representative sea water at a temperature of 25 C from pH = 8.18 to pH = 7.93. This change is comparable to diurnal pH changes in biologically productive surface waters, due to photosynthetic fixation of dissolved inorganic carbon during the day and respiration at night. The change is also less than the variations of pH with latitude, longitude and depth in the oceans. This paper includes a quantitative review of the carbonate chemistry of seawater and freshwater, the buffering capacity, the Revelle factor, the transport of calcium carbonate in ground water, the formation of flowstone, and the classic use of limewater to detect gaseous CO$_2$. The paper concludes with a brief review of those parts of chemical thermodynamics that are involved in ocean acidification.
\end{abstract}
\keywords{ocean acidification, pH, alkalinity, buffering, carbonate chemistry, groundwater}
\newpage
\tableofcontents
\newpage
\section{Introduction}
Natural surface waters  and rainwater contain dissolved CO$_2$, in equilibrium with atmospheric CO$_2$.  As  atmospheric CO$_2$ concentrations increase, the increase of the dissolved weak acid, CO$_2$, will lower the pH of the waters.
Fig. \ref{CH} shows how the pH of representative seawater and rainwater depend on the partial pressure $P$ of CO$_2$\,\cite{CH}.

The lowering of the pH of natural surface waters by more atmospheric CO$_2$ is often called {\it acidification}, but this term can be misleading. For example, representative seawater at a temperature $T = 25$ C is quite basic, with a pH = 8.18 
at current partial pressures, $P_c = 430$ $\mu$b of CO$_2$.  Doubling the partial pressure of CO$_2$ to $P_{\rm d} = 860$
$\mu$b would decrease the pH of the ocean to pH = 7.93, a decrease of $\Delta{\rm pH} = - 0.25$. This would leave the oceans almost as basic as today.  At  the same temperature, $T = 25$ C, doubling of CO$_2$ would reduce the pH of  rainwater,  which is slightly acidic because of dissolved CO$_2$ and with no compensating alkalinity, from pH = 5.59  to pH = 5.44, an even smaller decrease of  $\Delta{\rm pH} = - 0.15$. Although contemporary seawater has absorbed about 120 times more CO$_2$ per unit volume than rainwater in equilibrium with the same partial pressure $P$ of CO$_2$, the relatively large alkalinity [A] = 2.4 mM of seawater keeps it basic. 

In subsequent sections we will show straightforward ways to calculate the numbers cited here.  Lord Kelvin once said\,\cite{Kelvin}:
\begin{quotation}
When you can measure what you are speaking about, and express it in numbers, you know something about it, when you cannot express it in numbers, your knowledge is of a meager and unsatisfactory kind; it may be the beginning of knowledge, but you have scarcely, in your thoughts advanced to the stage of science.
\end{quotation}
As Lord Kelvin urged, much of the discussion below is in terms of numbers, equations and graphs. We have simplified discussions of the complicated chemistry of the oceans and the biochemistry of ocean life to keep the paper understandable to readers who recall some high school mathematics, chemistry and biology. More details can be found in the literature cited.

The oceans  will remain non-acidic for any credible scenarios of human emissions of CO$_2$ in the future.
This  paper gives a quantitative discussion of why increasing concentrations of atmospheric CO$_2$ will not cause harmful acidification of natural waters on Earth,  and will have little effect on the biochemistry of aquatic organisms. 
\section{Surface Seawater and Freshwater}

\begin{figure}
\begin{centering}
%\postscriptscale{GGNTa.eps}{1.2}
\includegraphics[height=90mm,width=.8\columnwidth]{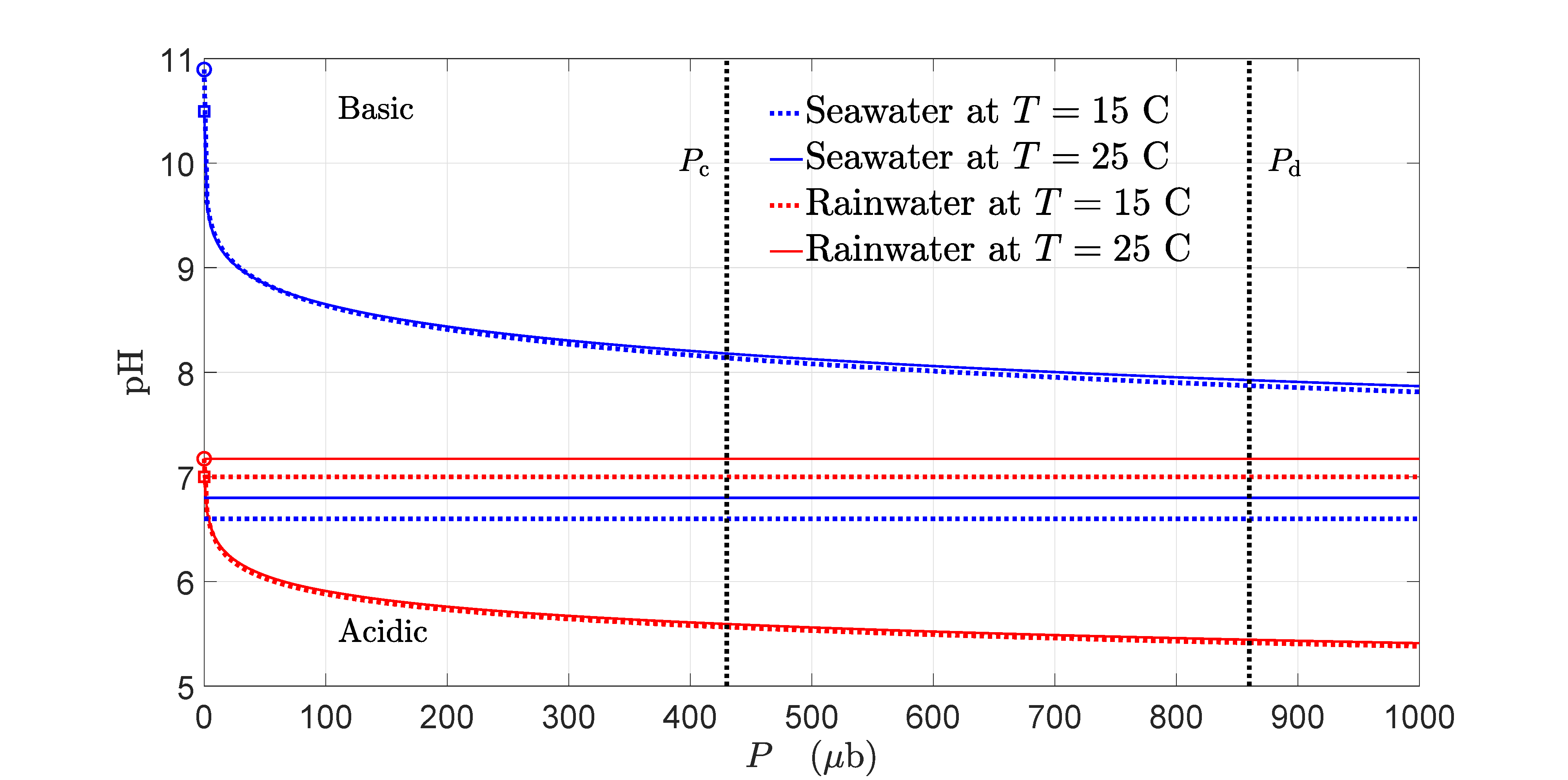}
\caption {The curved lines are pH values,  pH($P$), of seawater (blue) and rainwater (red) in equilibrium with atmospheric partial pressure $P$ of CO$_2$. In the year 2025,  the pressure was $P_{\rm c}= 430$ $\mu$b.  The doubled value is $P_{\rm d}= 860$ $\mu$b.  The representative seawater has a salinity $S = 35$\textperthousand, an alkalinity [A] = 2.4 mM and a boron molality [B] = 0.43 mM.  Rainwater contains only dissolved CO$_2$. pH values are given for temperatures, $T= 15$ C and $T=25$ C. The horizontal lines are neutral pH values, pH$_n$, for the four combinations of salinity and temperature. Waters with pH $<$ pH$_{\rm n}$ are acidic and waters with pH $>$ pH$_{\rm n}$ are basic.  In Section {\bf \ref{cb}} we show how to calculate these curves. Some numerical details are given in Table \ref{pH}.}
\label{CH}
\end{centering}
\end{figure}

\begin{figure}
\begin{centering}
%\postscriptscale{GGNTa.eps}{1.2}
\includegraphics[height=100mm,width=.9\columnwidth]{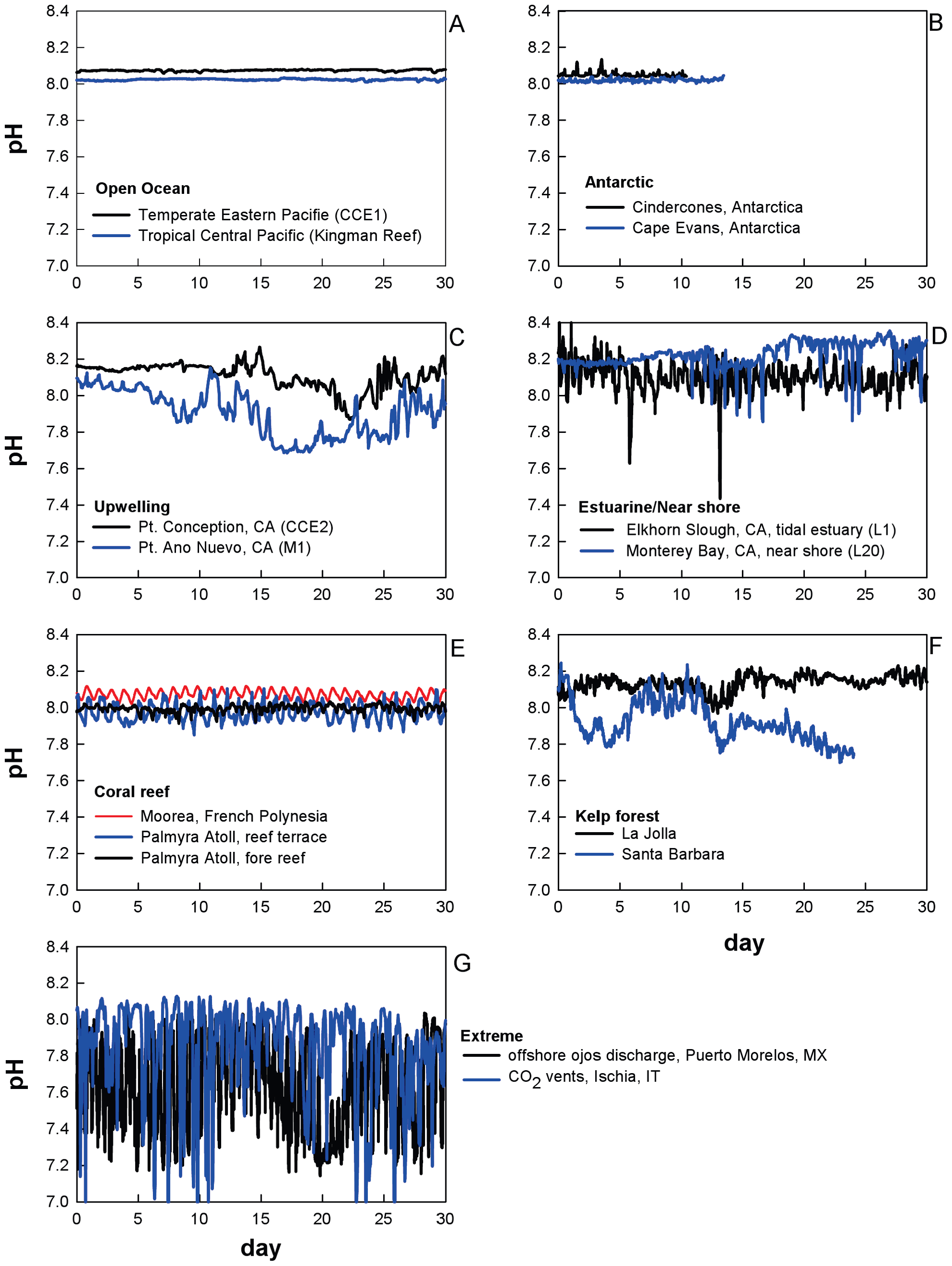}
%\postscriptscale{Hofmann.eps}{0.5}
\caption {Time dependence of the pH of ocean surface waters from reference\,\cite{Hofmann}. The very small pH decrease, 0.25 units from doubling atmospheric concentrations of CO$_2$, that is shown in Fig. \ref{CH}, is smaller than the day-to-night  fluctuations of pH in biologically active parts of the ocean, shown here. Doubling the atmospheric concentration of CO$_2$ at current rates of increase, would require nearly two centuries.}
\label{Hofmann}
\end{centering}
\end{figure}

\begin{table}
\begin{center}
\begin{tabular}{|l|c|c|c|c|}
\hline
&$S = 0$\textperthousand &$S = 0$\textperthousand &$S = 35$\textperthousand &$S = 35$\textperthousand\\
&$T=15$ C& $T=25$ C&$T=15$ C& $T=25$ C\\
\hline\hline
pH$_{\rm n}$&7.17&7.00&6.80&6.60\\
\hline
pH($0$)&7.17&7.00&10.9&10.5\\
\hline
pH($P_{\rm c}$)&5.56&5.59&8.14&8.18\\
\hline
pH($P_{\rm d}$)&5.41&5.44&7.87&7.93\\
\hline
\end{tabular}
\end{center}
\caption{  Numerical  pH values, pH($P$),  from Fig. \ref{CH}. The first row shows neutral pH values,  pH$_{\rm n}$, for water with a given combination of salinity and temperature, when the hydrogen-ion and hydroxyl-ion molalities are equal, 
$[{\rm H}^+] = [{\rm OH}^-]= e^{-{\rm pH}_{\rm n}} $ M.  The second line gives pH values, pH(0) for CO$_2$-free waters. The third line gives the pH values, pH($P_{\rm c}$), of water in equilibrium with the contemporary partial pressure of CO$_2$, $P_{\rm c}= 430$ $\mu$b. The fourth line gives the pH values, pH($P_{\rm d}$), of water in equilibrium with the double that value, $P_{\rm d}= 860$ $\mu$b.
\label{pH}}
\end{table}
An overview of how the pH values of natural waters depend on the atmospheric partial pressure $P$ of CO$_2$, with which the waters are in equilibrium, is shown in Fig. \ref{CH}.   Some numerical values from Fig. \ref{CH} are summarized in Table \ref{pH}.
The decrease,  $\Delta \hbox{pH} = -0.25$, of the pH of seawater at a temperature $T= 25$ C, from doubling the CO$_2$ partial pressure $P$,  that is shown in Fig. \ref{CH} and summarized in Table \ref{pH}, would take more than a century at the current rates of CO$_2$ increase.  As shown in Fig. \ref{Hofmann} from Hofmann {\it et al.}\,\cite{Hofmann} or Table 2.2 of Jokiel {\it et al.}\,\cite{Jokiel}, this century-long pH decrease is comparable to, or smaller, than typical day-night fluctuations of pH in biologically active parts of the ocean,  like coral reefs.  Photosynthesis converts dissolved inorganic CO$_2$ into organic carbon in the tissues of living organisms or CaCO$_3$(s) biominerals. This mainly decreases the dissolved inorganic carbon  (DIC) but carbonate biomineralization also removes Ca$^{2+}$ ions from solution and slightly decreases the alkalinity. During the day, photosynthesis removes CO$_2$  faster than it can be replaced by eddy diffusion from the atmosphere above or from deeper water below. With less of the dissolved weak acid CO$_2$, less of the natural alkalinity of the seawater is cancelled and the pH can increase by several fractions of a pH unit, or more. During the night,  aerobic respiration by living organisms converts organic compounds back to dissolved inorganic CO$_2$, and this  approximately cancels the daytime pH increase.

\begin{figure}
\begin{centering}
%\postscriptscale{GGNTa.eps}{1.2}
\includegraphics[height=90mm,width=.9\columnwidth]{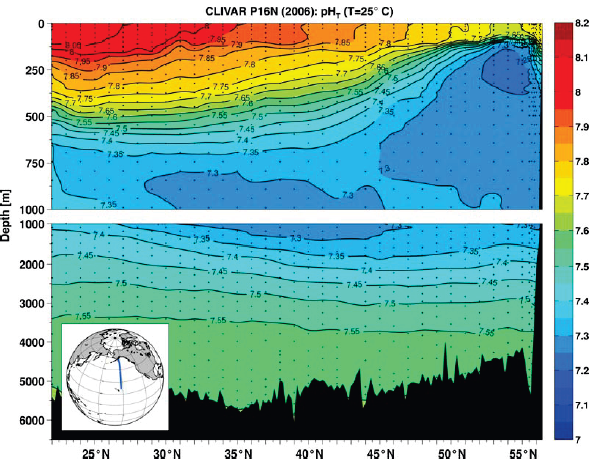}
%\postscriptscale{pH-depth.eps}{0.8}
\caption {The pH of the North Pacific Ocean\, \cite{Byrne} along the longitude 152$^{\circ}$ W in 2006.  The pH changes with latitude and depth are much larger than the pH change $\Delta\hbox{pH} = - 0.25$ shown in Fig. \ref{CH}  for doubling the atmospheric concentration of CO$_2$.}
\label{pH-depth}
\end{centering}
\end{figure}

The pH of the North Pacific Ocean along the longitude 152$^{\circ}$\,W between Hawaii and Alaska\,\cite{Byrne}, measured in 2006, is shown in Fig. \ref{pH-depth}.  The decrease of the pH with moderate depths below the surface is mostly due to the increase of dissolved inorganic carbon (DIC) as a result of the {\it biological pump}\,\cite{biopump}, that is, the aerobic decomposition of sinking organic debris created by photosynthesis at the surface, and by the dissolution of mineralized skeletons at depths below the lysocline\,\cite{Lys}. The decrease of surface pH with increasingly northerly latitudes is due to the decrease in alkalinity, [A], which is correlated with the decrease in surface salinity. Much of the very deep water of the North Pacific comes from cold, salty water that was formed off of Antarctica and other parts of the globe\,\cite{THC}. The bottom water is little affected by properties of the water above.

\begin{table}
\begin{center}
\begin{tabular}{|c|c|c|}
\hline
 Water & [A] (mM)&pH\\[0.5ex]
 \hline
\hline
Lake Magadi\cite{Golan}&4,200&11.1 \\
\hline
North Great Salt Lake\cite{Golan}&16&7.7\\
\hline
Dead Sea\cite{Golan}&4.8&6.3\\
\hline
Limestone Lake &2.8&8.2 \\
\hline
Ocean\cite{SurfaceA} &2.4&8.2\\
\hline
Lake Michigan\cite{Phillips} &2.2&8.2\\
\hline
Mississippi River\cite{Mississippi} &2.0 &8.2\\
\hline 
Lake Ontario\cite{Phillips} &1.8&8.1\\
\hline
Lake Superior\cite{Phillips} &0.8&7.8\\
\hline
Rainwater&0&5.6\\
\hline
Kawa Ijen Lake\cite{Delmelle} &$-800$& 0.1\\
\hline
\end{tabular}
\end{center}
\caption{Representative surface alkalinities, [A], and pH values of natural waters. See the text for more details.
\label{table1}}
\end{table}

\begin{figure}[t]
\begin{centering}
\includegraphics[height=90mm,width=.8\columnwidth]{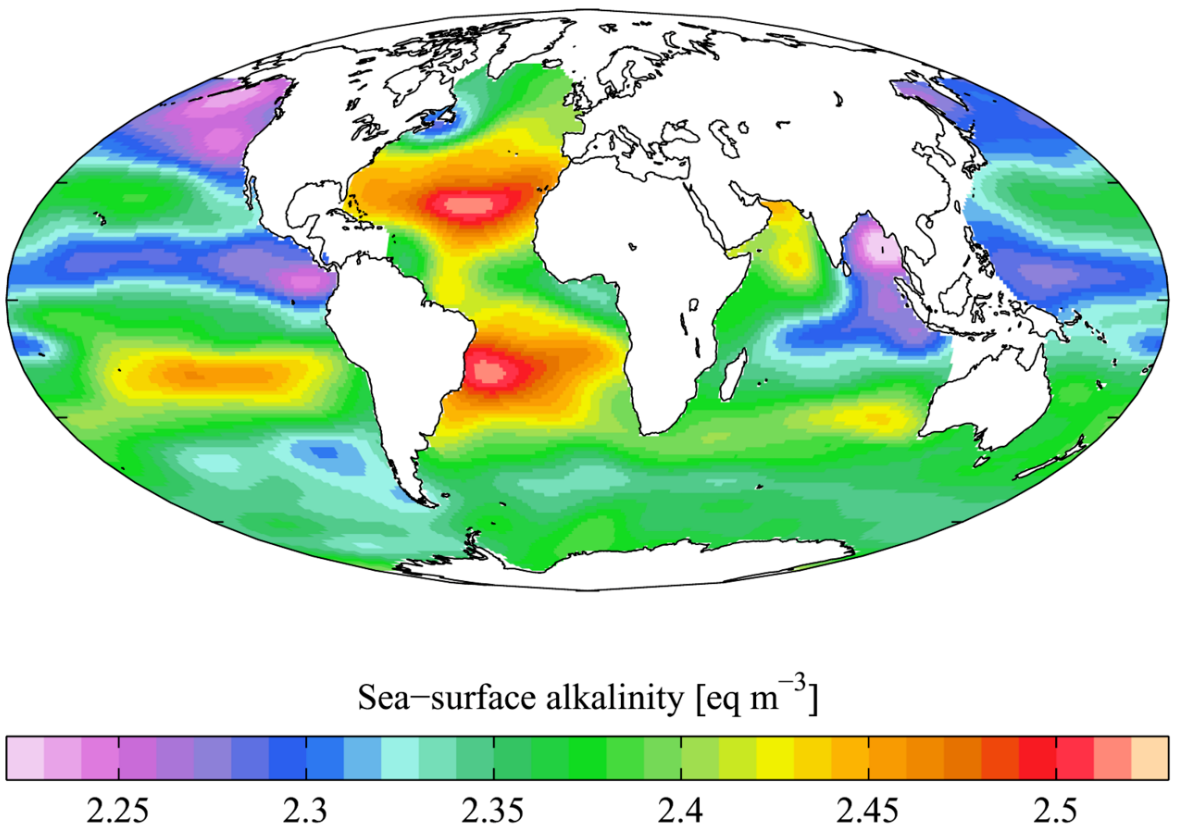}
%\postscriptscale{SurfaceA.eps}{0.8}
\caption {Surface akalinity, [A], of the world oceans\cite{SurfaceA}. The alkalinity is very nearly proportional to the salinity. The alkalinity is higher in the saline regions centered at latitudes  30$^{\circ}$ N and 30$^{\circ}$ S, under the subsiding dry air of the Hadley circulation. The alkalinity is lower near the equator, under the rainy tropical interconvergence zone, and in the extreme northern Pacific ocean. The numerical value of a given alkalinity in the units of this figure, eq m$^{-3}$ (equivalents per cubic meter), is  nearly the same as its numerical value in the units used in this paper (mM =  $10^{-3}$ moles kg$^{-1}$).}
\label{SurfaceA}
\end{centering}
\end{figure}

For natural waters the total alkalinity, [A] = [TALK], which we will discuss in more quantitative detail in Section {\bf \ref{al}}, is the difference between the equivalent molalities of strong bases and strong acids. We will quote alkalinities in the units commonly used for chemical oceanography, moles per kilogram, or molality, a unit we denote with the symbol M.  In terms of the molecules of interest, unit molality is

\begin{equation}
1 \hbox{ M} = 1\hbox{ mol kg}^{-1} =  N_{\rm A}\hbox{ molecules  kg}^{-1}.
\label{int2}
\end{equation}
Here Avogadro's number is 
\begin{equation}
N_{\rm A}=6.022\,\times\, 10^{23}.
\label{int4}
\end{equation}
The closely-related unit of moles per liter is called the molarity.

The values of the  alkalinities [A] of most drinking waters  and oceans\,\cite{Phillips} are a few millimoles per kilogram (mM). But the alkalinities of soda lakes\,\cite{Golan} with high concentrations of sodium ions, Na$^+$,  can be hundreds and even thousands of mM. Lakes in volcanic regions, like Indonesia's crater lake, Kawa Ijen \,\cite{Delmelle}, with high concentrations of sulfate ions, SO$_4^{2-}$, can have  negative alkalinities (acidities) of nearly a thousand mM.  Some representative alkalinities and pH values of natural water bodies are shown in Table \ref{table1}. 
The alkalinities of the Great Lakes  are from Table 1 of reference \,\cite{Phillips}. The corresponding pH values come from solving (\ref{al38})  at a CO$_2$ partial pressure, $P = 430$ $\mu$b, and a temperature of $T= 15$ C. 
``Limestone lake" is a saturated solution of calcite in freshwater at temperature of 15 C,  as discussed in Section {\bf\ref{LL}}.  The value pH = 5.6 for ``rainwater," comes from the solution of (\ref{al38})  at a CO$_2$ partial pressure, $P = 430$ $\mu$b, a temperature of $T= 25$ C and an alkalinity [A] = 0.    Kenya's Magadi Lake, Utah's Great Salt Lake, the Dead Sea, and other endorheic  water bodies\,\cite{Golan} have  large internal variations of salinity, pH and alkalinity. 

A global map of the average alkalinity of ocean surface water\,\cite{SurfaceA} is shown in Fig. \ref{SurfaceA}. A representative ocean alkalinity we will use in this paper is
\begin{equation}
[\rm A] = 2.4 \hbox{ mM}.
\label{int4a}
\end{equation}

The oceans are famously salty, and their salinity, $S$, can be specified by grams of salt contained by a kilogram of solution. The units of salinity have traditionally been taken to be parts per thousand, denoted by the symbol \textperthousand.  A representative ocean salinity we will use in this paper is
\begin{equation}
S = 35\hbox{ gm kg$^{-1}$}= 35\, {^o\!\!/_{\!\!oo}}.
\label{int5a}
\end{equation}
If we were to model the ocean as a solution of table-salt, with a gram molecular weight  $M$(NaCl) = 58.44 gm mol$^{-1}$, the NaCl molality, $[\hbox{NaCl}]$, corresponding to 
the salinity (\ref{int5a}) would be
\begin{equation}
[\hbox{NaCl}] =\frac{S}{M(\hbox{NaCl})}= \frac{35 \hbox{ gm kg$^{-1}$}}{58.44 \hbox{ gm mol$^{-1}$}} = 599 \hbox{ mM}.
\label{int5b}
\end{equation}
So we could think of the alkalinity (\ref{int4a}) of the ocean as due to adding slightly more equivalents, (599 +2.4) mM of the strong base, NaOH, to 599 mM of the strong acid,  HCl, leaving the ocean alkaline.  Much of that residual alkalinity is neutralized by the the weak acid CO$_2$ that is dissolved in the ocean.

The CO$_2$ partial pressure, $P$, that is in equilbrium with CO$_2$ dissolved in surface water can be taken to be that of air just above the water surface. In the year 2025, which we denote with the subscript c for {\it contemporary},  the  sea-level partial pressure was approximately
\begin{equation}
P_{\rm c}= 430  \, \mu{\rm b}.
\label{int6}
\end{equation}
The unit of pressure, $\mu$b = microbar, is related to other common pressure units by
\begin{equation}
1 \mu{\rm b} = 0.1 \hbox{ Pa}= 10^{-6} \hbox{ bar}= 1 \hbox{ erg cm}^{-3}.
\label{int8}
\end{equation}
Deep waters of oceans, lakes, rivers or groundwater, can be ``carbonated" and have higher partial pressures of CO$_2$ than the atmospheric value (\ref{int6}).  For example, values of $P$ in the groundwaters of limestone areas can exceed surface partial pressures by factors of 100 or more because of the CO$_2$ produced by respiration in plant roots, by aerobic decomposition of organic detritus near the surface, and because of the dissolution of limestone or dolomite by this acidic, biologically carbonated water\,\cite{Macpherson}. The waters of the upper  Mississippi River\cite{Mississippi} are carbonated to an average CO$_2$ partial pressure of about $P = 1600$ $\mu$b. For similar reasons, the partial pressure of CO$_2$ in the deep oceans can be much higher than the atmospheric value (\ref{int8}) because of the biological pump\,\cite{biopump} and because below the lysocline\,\cite{Lys}, the solubility product has increased enough from the increasing hydrostatic pressure to force the dissolution of sinking CaCO$_3$(s) crystals, produced by biomineralization near the surface.  Bottom waters of a few deep volcanic lakes, like infamous Lake Nyos\cite{Nyos}, can have $P \ge 10$ bar, greater than that in a champagne bottle\,\cite{champagne}.

\subsection{Chemical reactions\label{dc}}
Many reversible chemical reactions take place in natural waters. Here we review the most important ones for the carbonate chemistry of seawater and freshwater. 
\subsubsection{Self-ionization of water}
Recall\,\cite{Kw} that a small fraction of the H$_2$O molecules of liquid water break up into positive hydrogen ions ${\rm H}^+$ and negative hydroxyl ions, ${\rm OH}^-$, as described by the chemical equation,
\begin{equation}
{\rm H}_2{\rm O}  \rightleftharpoons {\rm H}^++{\rm OH}^-\quad\hbox{or}\quad {\rm H}_2{\rm O} +{\rm H}_2{\rm O} \rightleftharpoons \hbox{H$_3$O$^{+}$}+{\rm OH}^-.
\label{in2}
\end{equation}
Although the expression on the right of (\ref{in2}) is a more physically correct representation of self ionization, where H$^+$ ions are almost always attached to water molecules to form a hydronium ion, H$_3$O$^{+}$, we will adhere to common practice and use the expression on the left to simplify the notation for chemical equations.

In thermal equilibrium, the concentration of the ions of (\ref{in2}) are related by the law of mass action
\begin{equation}
 [{\rm H}^+][{\rm OH}^-]=K_{\rm w},
\label{in4}
\end{equation}
The concentration  of ${\rm H}^+$ ions in solution is often written in terms of the dimensionless pH number
\begin{equation}
 [{\rm H}^+]=10^{-{\rm pH}} \hbox{ M},
\label{in10}
\end{equation}
or
\begin{equation}
{\rm pH} = -\log \left( [{\rm H}^+]\hbox{ M}^{-1}\right),
\label{in12}
\end{equation}
where $\log(x)$ is the base-10 or common logarithm of $x$.
In physical chemistry 
 it is common to use {\it decimal cologarithms} like (\ref{in12})  to quantify molalities, and by analogy, to quantify equilibrium constants, for example
\begin{equation}
pK_{\rm w} = -\log \left( K_{\rm w}\hbox{ M}^{-2}\right)\quad\hbox{or}\quad \hbox{pOH} =  -\log \left( [\hbox{OH$^{-}$}]\hbox{ M}^{-1}\right).
\label{in14}
\end{equation}
The relative molalities of hydrogen ions $[{\rm H}^+]$ and hydroxyl ions $[{\rm OH}^-]$ determine whether an aqueous solution is acidic, neutral or basic:
\begin{eqnarray}
\hbox{H$^{+}$}&>&\hbox{OH$^{-}$}\quad\hbox{or}\quad \hbox{pH}<\hbox{pH}_{\rm n}\quad\leftrightarrow\quad\hbox{acidic},\label{in14a}\\
\hbox{H$^{+}$}&=&\hbox{OH$^{-}$}\quad\hbox{or}\quad \hbox{pH}=\hbox{pH}_{\rm n}\quad\leftrightarrow\quad\hbox{neutral},\label{in14b}\\
\hbox{H$^{+}$}&<&\hbox{OH$^{-}$}\quad\hbox{or}\quad \hbox{pH}>\hbox{pH}_{\rm n}\quad\leftrightarrow\quad\hbox{basic}.\label{in14c}
\end{eqnarray}
Here neutral pH is defined by
\begin{equation}
\hbox{pH}_{\rm n} = -\log\left(\sqrt{K_{\rm w}\hbox{ M}^{-2}}\right)
\label{in14d}
\end{equation}
The equilibrium ``constant,"  $K_{\rm w}$ depends on many solution parameters, most notably on temperature $T$, total pressure $p$ and salinity $S$. 
 Representative values of the $K_{\rm w}$ can be found in Table \ref{K012}.
\subsubsection{pH and activity}
For seawater, and even more for the the liquids inside of living cells, the simple definition (\ref{in12}) for pH must be sharpened to
\begin{equation}
{\rm pH} = -\log a,
\label{ac2}
\end{equation}
where the {\it activity} of hydrogen ions ${\rm H}^+$ is
\begin{equation}
a=\gamma [{\rm H}^+]\hbox{ M}^{-1},
\label{ac2}
\end{equation}
Here the  dimensionless number $\gamma$ is the {\it activity coefficient}.  For freshwater it is often a good approximation to take $\gamma = 1$, which would make the two definitions (\ref{in12}) and (\ref{ac2}) of pH identical. But for sea water and for the water in living cells the activity coefficient can differ markedly from unity. A useful comment\,\cite{Nordstrom} on pH is 
\begin{quote}
The pH scale for aqueous solutions and natural waters is often given as 0 - 14 without any explanation. It is an arbitrary and convenient range because it places the value for neutrality of pure water at 25 $^{\circ}$C (pH = 7.0) squarely in the middle. Values of pH less than 0.0 and greater than 14.0 not only are possible but have been prepared frequently in chemical laboratories.
\end{quote}
We will have more to say about activities in later sections.
\subsubsection{Dissolution of gaseous carbon dioxide in water}
Assume that CO$_2$ molecules  in the atmosphere  have come to thermal equilibrium with CO$_2$(aq) molecules dissolved in the surface water. 
A small fraction of dissolved, linear CO$_2$ molecules, with a concentration $[\hbox{CO$_2$(aq)}]$, will attach to water molecules to form approximately  triangular {\it carbonic acid} molecules, H$_2$CO$_3$,
\begin{equation}
{\rm H}_2{\rm O}+\hbox{CO$_2$(aq)}  \rightleftharpoons \hbox{H$_2$CO$_3$} .
\label{in14}
\end{equation}
The concentrations of the two uncharged forms of CO$_2$ are related  by 
\begin{eqnarray}
[\hbox{H$_2$CO$_3$}] &=&K_h[\hbox{CO$_2$(aq)}].
\label{in18a}
\end{eqnarray}
We will not make direct use of the carbonic acid fraction, $K_h$, but according to reference \,\cite{Wiki}, its value is very small,
\begin{equation}
K_h=1.70\,\times\, 10^{-3}.
\label{in20}
\end{equation}

The non-catalyzed rate of the reaction (\ref{in14}) in carbonated water is exceptionally slow compared to protonation or deprotonation rates of other carbonate species.  The rate of conversion of aqueous CO$_2$(aq) molecules to  carbonic acid molecules, H$_2$CO$_3$, depends on pH, temperature, salinity and other factors, but without catalysts, it is on the order of one conversion per second per CO$_2$ molecule\,\cite{Schulz}. In contrast, the rate of conversion of  a bicarbonate ion ${\rm H}_2{\rm CO}_3^-$  to a carbonate ion, 
${\rm CO}_3^{2-}$, and a hydrogen ion, ${\rm H}^+$, in the reaction (\ref{in40}) discussed below, is on the  order of one conversion per picosecond  per  ${\rm HCO}_3^-$ ion, some 12 orders of magnitude faster.   Living organisms have evolved hundreds of {\it carbonic anhydrase} enzymes to speed up the hydration-dehydration rates of (\ref{in14}) by orders of magnitude so this reaction does not bottleneck other biochemistry\,\cite{DiMario}. 

From  (\ref{in18a}) we see that the total concentration of the two uncharged species of dissolved CO$_2$ is 
\begin{eqnarray}
\hbox{[CO$_2^*$]} &=&[\hbox{CO$_2$(aq)}]+[\hbox{H$_2$CO$_3$}]\nonumber\\
&=&\frac{1+K_h}{K_h}[\hbox{H$_2$CO$_3$}].
\label{in22}
\end{eqnarray}
According to Henry's law, the equilibrium concentration of uncharged species of CO$_2$ in surface water is proportional to the partial pressure, $P$, of atmospheric CO$_2$, 
\begin{equation}
\hbox{[CO$_2^*$]}=K_0P.
\label{in24}
\end{equation}
Representative values of the equilibrium constant $K_0$ are listed in Table \ref{K012}.

An uncharged carbonic acid molecule H$_2$CO$_3$ can dissociate into a positive hydrogen ion  and a negative  bicarbonate ion
\begin{equation}
\hbox{H$_2$CO$_3$}  \rightleftharpoons {\rm H}^++{\rm HCO}_3^-.
\label{in32}
\end{equation}
The equilibrium concentrations are related  by the law of mass action,
\begin{eqnarray}
[{\rm H}^+][{\rm HCO}_3^-]&=&K_{a1}[\hbox{H$_2$CO$_3$}],
\label{in34}
\end{eqnarray}
or 
\begin{eqnarray}
[{\rm H}^+][{\rm HCO}_3^-]&=&K_1\hbox{[CO$_2^*$]},
\label{in34a}
\end{eqnarray}
where
\begin{equation}
K_1 = \frac{K_{a1}K_h}{1+K_h}.
\label{in34b}
\end{equation}
Representative values of the equilibrium constant $K_1$ are listed in Table \ref{K012}.
According to (\ref{in34a}) the equilibrium  concentration of bicarbonate ions is
\begin{eqnarray}
[{\rm HCO}_3^-]&=&\frac{K_1[\hbox{CO$_2^*$]}}{[{\rm H}^+]}\nonumber\\
&=&\frac{K_1K_0P}{[{\rm H}^+]}.
\label{in38}
\end{eqnarray}

A dissolved bicarbonate ion, HCO$_3^-$ can dissociate into a positive hydrogen ion and a doubly negative carbonate ion
\begin{equation}
{\rm HCO}_3^{-}  \rightleftharpoons {\rm H}^++{\rm CO}_3^{2-}.
\label{in40}
\end{equation}
The law of mass action for the reaction (\ref{in40}) is
\begin{equation}
 [{\rm H}^+][{\rm CO}_3^{2-}]=K_{2}[{\rm HCO}_3^-].
\label{in42}
\end{equation}
Representative values of the equilibrium constant $K_2$ are listed in Table \ref{K012}.
Eqs.  (\ref{in42}) and (\ref{in38}) imply that the concentration of carbonate ions is
\begin{eqnarray}
[{\rm CO}_3^{2-}]&=&\frac{K_2 K_1 [\hbox{CO$_2^*$}]}{[{\rm H}^+]^2}\nonumber\\
&=&\frac{K_2 K_1 K_0P}{[{\rm H}^+]^2}.
\label{in46}
\end{eqnarray}
\subsubsection{Dissociation of boric acid\label{ba}}
After carbon dioxide, CO$_2$, boric acid, B(OH)$_3$, is normally the second most important weak acid in seawater\,\cite{Dickson}. B(OH)$_3$ has a much smaller effect on pH than CO$_2$ at contemporary partial pressures of CO$_2$.
In aqueous solution, the approximately flat,  triangular   B(OH)$_3$ molecule  can attach a hydroxyl ion, OH$^{-}$ from a water molecule to liberate a proton ${\rm H}^+$ and to form a borate (tetrahydroxyborate) anion  B(OH)$_4^-$, which has an approximately tetrahedral structure similar to the ammonia cation NH$_4^+$. The reaction can be described by the chemical equation
\begin{equation}
\hbox{B(OH)$_3$} +\hbox{ H$_2$O}\rightleftharpoons {\rm H}^++\hbox{B(OH)$_4^-$}. 
\label{ba2}
\end{equation}
The law of mass action for the reaction (\ref{ba2})   can be written as
\begin{equation}
\frac{[{\rm H}^+][\hbox{B(OH)$_4^-$]}}{[\hbox{B(OH)$_3$}]}=K_{\rm B}.
\label{ba4}
\end{equation}
Representative values of the equilibrium constant $K_{\rm B}$ are listed in Table \ref{K012}.
We will denote the {\it boron molality} by
\begin{equation}
[\hbox{B}]=[\hbox{B(OH)$_3$}]+[\hbox{B(OH)$_4^{-}$}].
\label{ba6}
\end{equation}
For the year 2025, a representative value\,\cite{Dickson}  of [B] in seawater  is
\begin{equation}
[\hbox{B]}=0.43\hbox{ mM}.
\label{ba8}
\end{equation}
\subsubsection{Dissolution of solid calcium carbonate  or calcite,  CaCO$_3$(s)  } 
Solid crystals of calcium carbonate, which we denote by CaCO$_3$(s), can dissolve or grow  in solution as described by the chemical equation
\begin{equation}
\hbox {CaCO$_3$(s)} \rightleftharpoons\hbox{Ca$^{2+}$}\, +\, \hbox{CO$_3^{2-}$}.
\label{ss2}
\end{equation}
For a subsaturated solution, the forward rate of (\ref{ss2}) will be larger than the reverse rate and crystals will tend to dissolve. For a supersaturated solution the reverse rate will be greater than the forward rate and the crystals will tend to grow.  For a saturated solution, containing calcium carbonate crystals, CaCO$_3$(s), in equilibrium with dissolved calcium ions, Ca$^{2+}$, and carbonate ions, CO$_3^{2-}$, the forward and reverse rates of the reaction  (\ref{ss2}) are equal, the crystals will have no natural tendency to dissolve or grow, and the ion concentrations will obey  the law of mass action
\begin{equation}
[\hbox{Ca}^{2+}][{\rm CO}_3^{2-}]=K_{\rm sp},
\label{ss4}
\end{equation}
Here $K_{\rm sp}$ is the {\it concentration solubility product}. It has units of M$^2$, and it depends on temperature and pressure.
Representative values of the equilibrium constant $K_{\rm sp}$ are listed in Table \ref{K012}. For basic theoretical discussions, a more convenient quantity is the dimensionless  {\it thermodynamic solubility product} $\mathcal{K}_{\rm sp}$, which we will say a little more about in Section {\bf \ref{g}}. For practical calculations of this paper, we will always use concentration solubility products like (\ref{ss4}) that have been determined from experimental measurements. More discussion of 
concentration and thermodynamic solubility products can be found in references\,\cite{Ktc,SolEq}

For Earth's oceans and surface waters, calcium carbonate normally forms as trigonal calcite or orthorhombic aragonite crystals. Biomineralization may produce either calcite or aragonite or sometimes both in the same organism. The biomineralization of aragonite in seawater is facilitated by high ratios of the magnesium to calcium molalities [Mg$^{2+}$]/[Ca$^{2+}$]$> 2$. Today's  oceans\,\cite{seawater}, with [Mg$^{2+}$]/[Ca$^{2+}$]$= 5.2$, are said to be {\it aragonite seas}\,\cite{aragonite_sea}. There have been earlier geological periods, like the Devonian or Cretaceous, where the magnesium to calcium ratios of seawater were exceptionally low,  [Mg$^{2+}$]/[Ca$^{2+}$]$< 2$, and when Earth's oceans were said to be {\it calcite seas}\,\cite{calcite_sea}.  Rapid seafloor spreading may lead to calcite seas, since Mg$^{2+}$ ions are exchanged for  Ca$^{2+}$ during circulation of water  through hydrothermal vents near seafloor spreading ridges,\,\cite{aragonite_sea, calcite_sea, Broecker}. 

Following Mucci\,\cite{Mucci}, in Fig. \ref{MP} we will assume that the solubility constant of aragonite is 50\% larger than that of calcite,
\begin{equation}
K_{\rm sp}(\hbox{aragonite}) = 1.5\, \times\, K_{\rm sp}(\hbox{calcite}).  \label{ss11cc}
\end{equation}
The simplifying assumption (\ref{ss11cc}) is accurate enough for our purposes.

\subsubsection{Dissolution of solid calcium hydroxide or portlandite, Ca(OH)$_2$(s)} 
Solid crystals of calcium hydroxide, which we denote by Ca(OH)$_2$(s), can dissolve or grow  in solution as described by the chemical equation
\begin{equation}
\hbox {Ca(OH)$_2$(s)}\rightleftharpoons \hbox{Ca$^{2+}$}\, +\, 2\,{\rm OH}^- .
\label{SL2}
\end{equation}
For a saturated solution, containing calcium hydroxide crystals, Ca(OH)$_2$(s), in equilibrium with dissolved calcium ions, Ca$^{2+}$, and hydoxide ions, OH$^{-}$, the forward and reverse rates of the reaction  (\ref{SL2}) are equal; the crystals will neither dissolve nor grow, and the ion concentrations will obey  the law of mass action
\begin{equation}
\hbox{[Ca$^{2+}$][OH$^{-}$]$^2$}= K_{\rm sl}.
\label{SL4}
\end{equation}
Naturally occurring crystals of calcium hydroxide are called {\it portlandite}. Slaked lime also consists of crystals of calcium hydroxide.
\subsubsection{Dissolution of solid magnesium hydroxide or brucite, Mg(OH)$_2$(s)} 

Solid crystals of magnesium hydroxide, which we denote by Mg(OH)$_2$(s), can dissolve or grow  in solution as described by the chemical equation
\begin{equation}
\hbox {Mg(OH)$_2$(s)}\rightleftharpoons \hbox{Mg$^{2+}$}\, +\, 2\,{\rm OH}^- .
\label{MM2}
\end{equation}
{\it Milk of magnesia} is a saturated solution containing magnesium hydroxide crystals, Mg(OH)$_2$(s), in equilibrium with dissolved magnesium ions, Mg$^{2+}$, and hydoxide ions, OH$^{-}$. For saturated solutions, the forward and reverse rates of the reaction  (\ref{ss2}) are equal; the crystals will neither dissolve nor grow, and the ion concentrations will obey  the law of mass action
\begin{equation}
\hbox{[Mg$^{2+}$][OH$^{-}$]$^2$}= K_{\rm sb}.
\label{MM4}
\end{equation}
Naturally occurring crystals of magnesium hydroxide are called {\it brucite}.  Values of the equilibrium constant $K_{\rm sb}$ quoted by Wattenberg and Timmerman\,\cite{WT}  are listed in Table \ref{K012}.
\begin{table}
\begin{center}
\begin{tabular}{|l|l|l|l|l|l|}
\hline
Constant&\qquad Unit&$S = 0$\textperthousand&$S = 0$\textperthousand&$S = 35$\textperthousand&$S = 35$\textperthousand\\
&&$T=15$ C& $T=25$ C&$T=15$ C& $T=25$ C\\
\hline\hline
$K_{\rm w}$&$10^{-14}$ M$^2$&0.450\,\cite{Kw}&1.00\,\cite{Kw}&2.51\,\cite{Culberson}&6.31\,\cite{Culberson}\\
\hline
$K_0$& $10^{-2}$\,\, M bar$^{-1}$&4.56\,\cite{Weiss}&3.40\,\cite{Weiss}&3.85\,\cite{Weiss}&2.84\,\cite{Weiss} \\
\hline
$K_1$&$10^{-7}$ \,\,M&3.80\,\cite{Harned-D}&4.45\,\cite{Harned-D}&8.73\,\cite{Mehrbach}&9.99\,\cite{Mehrbach} \\
\hline
$K_2$&  $10^{-11}$ M&3.71\,\cite{Harned-S}&4.69\,\cite{Harned-S}&55.8\,\cite{Mehrbach}&76.8\,\cite{Mehrbach}\\
\hline
$K_{\rm B}$& $10^{-10}$\,\,M$^2$&4.70\,\cite{Dickson}&5.80\,\cite{Dickson}&19.2\,\cite{Dickson}&25.3\,\cite{Dickson}\\
\hline
$K_{\rm sp}$& $10^{-9}$\,\,\,\,M$^2$&3.71\,\cite{Mucci}& 3.31\,\cite{Mucci}&431\,\cite{Mucci}&427\,\cite{Mucci}\\
\hline
$K_{\rm sl}$& $10^{-5}$\,\,\,\,M$^3$&\,5.53\cite{Solubility_table}& 4.59\,\cite{Solubility_table}&&\\
\hline
$K_{\rm sb}$& $10^{-11}$\,\,M$^3$&  & 1\,\cite{WT}&  &5\,\cite{WT}\\
\hline
\end{tabular}
\end{center}
\caption{ Estimated equilibrium constants for representative freshwater with salinity $S = 0${\rm \textperthousand} and seawater with salinity $S = 35${\rm \textperthousand}, at temperatures $T=15$ C and $T=25$ C,  and a total pressure $p = 1$ bar. The equilibrium constants include  $K_{\rm w}$ of (\ref{in4}) for self-ionization of water;  $K_0$ of (\ref{in24}) for interconversion CO$_2$(g) from the air to CO$_2^*$, uncharged dissolved  CO$_2$;  $K_1$ of (\ref{in34a}) for interconversion of CO$_2^*$ and bicarbonate HCO$_3^-$;  $K_2$ of (\ref{in42}) for interconversion of HCO$_3^-$ and carbonate CO$_3^{2-}$;  $K_{\rm B}$ of (\ref{ba2}) for dissociation of boric acid;  $K_{\rm sp}$ of (\ref{ss4}), the solubility product for calcite, CaCO$_3$(s);   $K_{\rm sl}$ of (\ref{SL4}), the solubility product for portlandite, Ca(OH)$_2$(s); and  $K_{\rm sb}$ of (\ref{MM4}), the solubility product for brucite, Mg(OH)$_2$(s).
\label{K012}}
\end{table}
\subsubsection{Representative equilibrium constants}
Representative values of the equilibrium ``constants" mentioned above are listed in Table \ref{K012}.   These are {\it concentration} equilibrium contants, not {\it thermodynamic} equilibrium constants\,\cite{Ktc, SolEq}.  We will normally quote equilibrium constants for a sea-level total pressure, $p = 1$ bar. For representative oceans we will assume the salinity (\ref{int5a}) of $S = 35$\textperthousand\ and for representative  freshwater will assume $S=0$\textperthousand. As briefly discussed in Section {\bf\ref{K}},
chemical thermodynamics helps one to quantitatively understand how  equilibrium constants depend on temperature and pressure. A qualitative discussion of  how salinity affects equilibrium constants can be found in reference\,\cite{ionpair}.  The solubility constant $K_0$ is slightly smaller for seawater than freshwater but all the other equilibrium constants are much larger for seawater than freshwater.

Using the values of $K_{\rm w}$ from Table \ref{K012} at a temperature $T = 25 $ C with  (\ref{in14d}) implies that neutral pH  values for freshwater and seawater are  
\begin{equation}
 \hbox{pH}_{\rm n}=-\log\sqrt{K_{\rm w}\hbox{ M}^{-2}} =\left \{\begin{array}{rl}7&\mbox{for  $S = 0$\textperthousand\ or freshwater} ,\\
6.6&\mbox{for $S = 35$\textperthousand\ or seawater}. \end{array}\right .
\label{in48}
\end{equation}
\begin{figure}[t]
\begin{centering}
\includegraphics[height=90mm,width=1\columnwidth]{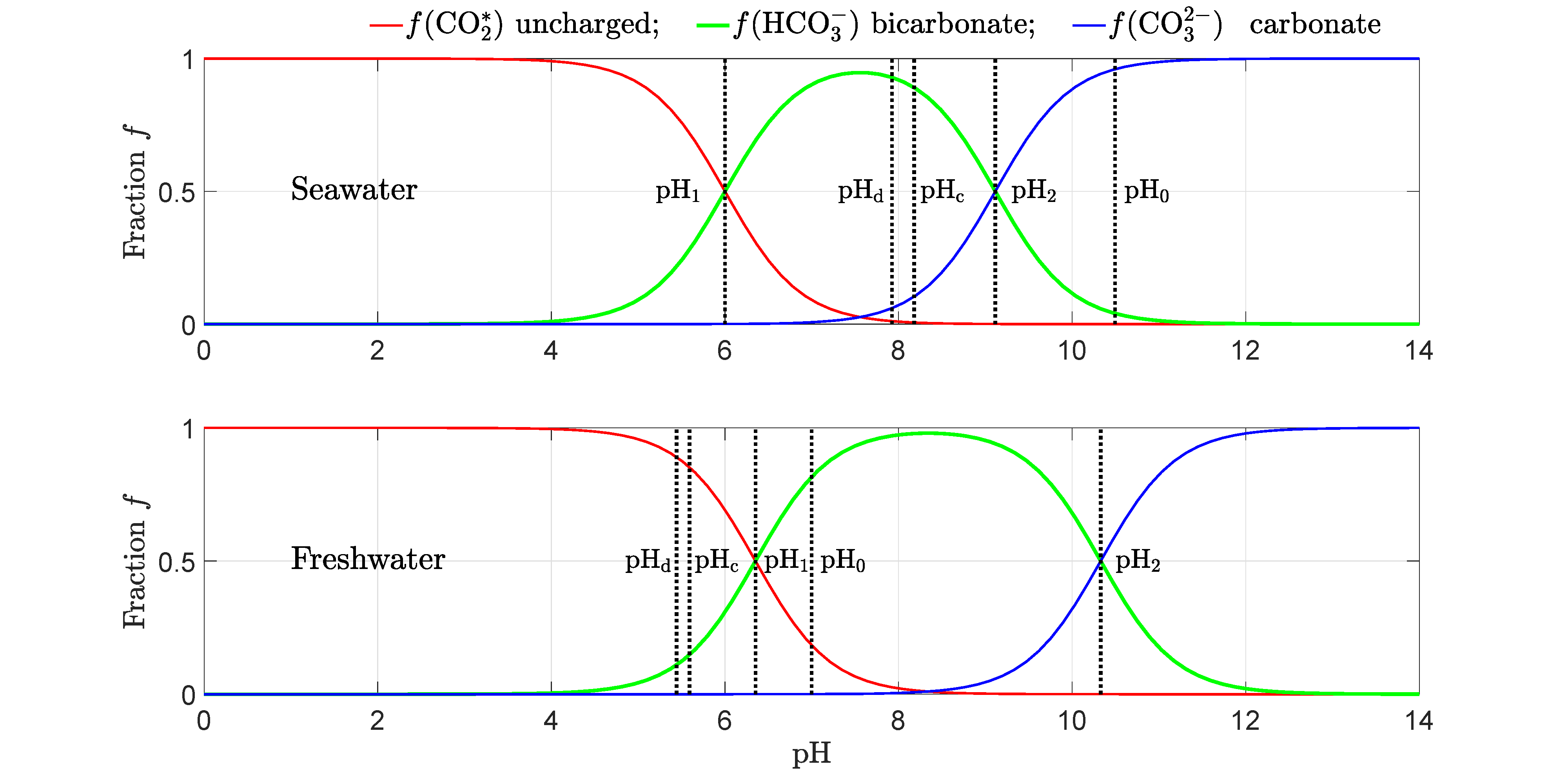}
\caption {The fractions $f (\hbox{CO$_2^*$})$, $f (\hbox{HCO$_3^-$})$  and $f (\hbox{CO$_3^{2-}$})$  of uncharged, bicarbonate and  carbonate forms of dissolved CO$_2$,   given by (\ref{if4}),  (\ref{if6}) and  (\ref{if8}), versus the pH for seawater (top), with salinity $S = 35$\textperthousand, and freshwater  (bottom) with $S = 0$\textperthousand. The temperature is $T=25$ C and the total pressure is $p=1$ bar.  At pH$_1$ of (\ref{if10}), the molalities of uncharged species and bicarbonate ions are equal.  
At pH$_2$ of (\ref{if12}) the molalities of bicarbonate and carbonate ions are equal. 
For seawater,  one can use  (\ref{al34}) with alkalinity [A] = 2.4 mM and boron molality [B] = 0.43 mM and the contemporary CO$_2$ partial pressure $P_{\rm c} = 430\ \mu$b to find the  hydrogen ion  molality $[{\rm H}^+]_{\rm c}$  and the corresponding pH$_{\rm c}$.
For freshwater, one can use  (\ref{al38}) with [A] = 0. In like manner, one can evaluate  pH$_{\rm d}$ for double the contemporary partial pressure, $P_{\rm d} = 860\ \mu$b,
or pH$_0$ for water with no dissolved CO$_2$ at all and $P = 0$.  Table \ref{pHi} summarizes numerical values of the parameters.}
\label{Fraction}
\end{centering}
\end{figure}
\subsubsection{Fractions of dissolved CO$_2$ species \label{if}} 
We denote the total molality of all dissolved species of CO$_2$, the {\it dissolved inorganic carbon} or DIC\,\cite{DIC} by
\begin{eqnarray}
[\hbox{DIC}]&=&[\hbox{CO$_2^*$}]+[{\rm HCO}_3^-]+[{\rm CO}_3^{2-}]\nonumber\\
&=&\bigg([{\rm H}^+]^2+ K_1[{\rm H}^+]+K_2 K_1\bigg) \frac{K_0P}{[{\rm H}^+]^2}
\label{if2}
\end{eqnarray}
To write the second line of (\ref{if2}) we used (\ref{in24}),  (\ref{in38}) and (\ref{in46}). The terms in parentheses give the relative molalities of [CO$_2^*$],
[HCO$_3^-$] and [CO$_3^{2-}$].
The two uncharged species of (\ref{in22}) make up a fraction
\begin{equation}
f(\hbox{CO$_2^*$})=\frac{[\hbox{CO$_2^*$}]}{[\hbox{DIC}]}=\frac{[{\rm H}^+]^2}{[{\rm H}^+]^2+K_1[{\rm H}^+]+K_2 K_1}.
\label{if4}
\end{equation}
of the dissolved inorganic carbon.
The fraction of bicarbonate ions is
\begin{equation}
f({\rm HCO}_3^-)=\frac{[{\rm HCO}_3^-]}{\hbox{[DIC]}}=\frac{K_1[{\rm H}^+]}{[{\rm H}^+]^2+K_1[{\rm H}^+]+K_2 K_1 }.
\label{if6}
\end{equation}
Finally, the fraction of carbonate ions is
\begin{equation}
f({\rm CO}_3^{2-})=\frac{[{\rm CO}_3^{2-}]}{[\hbox{DIC]}}=\frac{K_2K_1}{[{\rm H}^+]^2+K_1[{\rm H}^+]+K_2 K_1 }.
\label{if8}
\end{equation}
The ion fractions (\ref{if4}), (\ref{if6}) and (\ref{if8}) are shown as the {\it Bjerrum plots}\,\cite{Bjerrum} of  Fig. \ref{Fraction}. 

From inspection of (\ref{if4}) and (\ref{if6}) we see that the fractions of uncharged and bicarbonate forms of dissolved CO$_2$ are equal to each other when
\begin{equation}
[{\rm H}^+]= K_1\quad\hbox{or}\quad \hbox{pH}_1=-\log \left (K_1{\rm M}^{-1}\right).
\label{if10}
\end{equation}
From  inspection of (\ref{if6}) and  (\ref{if8}) we see that the fractions of bicarbonate and carbonate species of dissolved CO$_2$ are equal to each other   when
\begin{equation}
[{\rm H}^+]= K_2\quad\hbox{or}\quad \hbox{pH}_2 = -\log \left (K_2{\rm M}^{-1}\right).
\label{if12}
\end{equation}
\begin{table}
\begin{center}
\begin{tabular}{|c|l|l|}
\hline
&$S = 0$\textperthousand&$S = 35$\textperthousand\\
\hline\hline
pH$_0$&7.00&10.5\\
\hline
pH$_1$&6.35&6.00\\
\hline
pH$_2$&10.33&9.11 \\
\hline
pH$_c$&5.59&8.18 \\
\hline
pH$_d$&5.44&7.93\\
\hline
$P_1\, (\mu{\rm b})$&12.4&84,400\\
\hline
$P_2\, (\mu{\rm b})$&&17.9\\
\hline
[\hbox{DIC}]$_{\rm c}$  (mM)&0.0172&2.07\\
\hline
[\hbox{DIC}]$_{\rm d}$  (mM)&0.0329&2.21\\
\hline
\end{tabular}
\end{center}
\caption{Numerical pH values of Fig. \ref{Fraction}.  The CO$_2$ partial pressures $P_1$ and $P_2$ needed to produce the pH values pH$_1$ and pH$_2$ are calculated with
 (\ref{al40}). The value pH$_2$ = 10.33 for freshwater cannot be produced by physically acceptable positive molalities of the weak acid CO$_2$.  Eq. (\ref{if2}) was used to calculate the molalities of dissolved inorganic carbon, [DIC]$_{\rm c}$ for the contemporary  CO$_2$ partial pressures, $P_{\rm c} = 430\ \mu$b, and [DIC]$_{\rm d}$ for double the contemporary  value, $P_{\rm d} = 830\ \mu$b. A temperature $T = 25$ C was assumed for both freshwater and seawater.
\label{pHi}}
\end{table}
\subsection {Alkalinity\label{al}}
In addition to  dissolved species like  ${\rm H}^+$, ${\rm OH}^-$, ${\rm HCO}_3^-$ discussed in Section {\bf \ref{dc}}, for which the concentrations [${\rm H}^+$], [${\rm OH}^-$], [${\rm HCO}_3^-$] depend on pH, water bodies can contain  cations, like Na$^+$, Mg$^{2+}$,  Ca$^{2+}$, K$^+$, Sr$^{2+}$, {\it etc.} for most of which the concentrations are independent of normal pH variations in natural waters.  The cations originate from the dissolution of the strong bases NaOH, Mg(OH)$_2$, Ca(OH)$_2$, KOH, Sr(OH)$_2$,  {\it etc.}
 The most abundant, pH-independent cations of representative seawater \cite{seawater} are shown in Table \ref{table2}. Seawater also contains pH-independent anions like Cl$^{-}$,   SO$_4^{2-} $, Br$^{-}$, {\it etc.} These originate from the dissolution of  strong acids like HCl, H$_2$SO$_4$,  HBr, {\it etc.}.  The most abundant, pH-independent anions  are shown in Table \ref{table3}.
As one can see from Tables \ref{table2} and {\ref{table3}, the sum of the mass fractions of pH-independent ions is 
\begin{eqnarray}
\sum f = 34.3 \hbox{ g kg}^{-1}\approx S
\label{al0}
\end{eqnarray}
or very nearly equal to the salinity (\ref{int5a}), $S= 35$\textperthousand, of representative seawater.

The difference between the charge-weighted molalities of pH-independent cations and anions is the {\it total alkalinity} of the water or [TALK]. We will denote this alkalinity with the symbol  A], and we write it as
\begin{eqnarray}
\hbox{[A]}&=&\hbox{ [TALK]}\nonumber\\
&=&\hbox{[Na$^+$]}+2\hbox{[Mg$^{2+}$]}+2\hbox{[Ca$^{2+}$]}+\hbox{[K$^+$]}+\cdots\nonumber\\
&&-\hbox{[Cl$^-$]}-2\hbox{[SO$_4^{2-}$]}-\hbox{[Br$^-$]}-\cdots
\label{al6}
\end{eqnarray}
From Table \ref{table2} and Table \ref{table3} one can verify that the alkalinity (\ref{al6}) has the value [A] = 2.4 mM of (\ref{int4a}) and is consistent with Fig. \ref{SurfaceA}.

The total alkalinity [A] of (\ref{al6}) can be measured by titration, which we will discuss in Section {\bf \ref{t}}.  When we refer to the ``alkalinity" of water bodies below, we will normally mean the total alkalinity [A] of (\ref{al6}).  Representative surface alkalinities of the oceans were shown in Fig. \ref{SurfaceA}. The dissolution of biomineralized calcium carbonate crystals from surface debris that has sunk below the lysocline\,\cite{Lys} slightly increases the alkalinity of the deep oceans by increasing the calcium-ion molality [Ca$^{2+}$].

\begin{table}
\begin{center}
\begin{tabular}{|l|c|c|r|c|r|}
\hline
I$^{q+}$&$q$&$m$&$f\qquad$&[I$^{q+}$]$=f/m$&$q$[I$^{q+}$]\\
\hline
& &(g mole$^{-1}$)&(g kg$^{-1}$)&(mol kg$^{-1}$)&(mF kg$^{-1}$)\\
\hline\hline
Na$^+$&1&22.99&10.556&0.4592&459.2\\
\hline
Mg$^{2+}$&2&24.31&1.272&0.0523&104.7\\
\hline
Ca$^{2+}$&2&40.08&0.400&0.0100&20.0\\
\hline
K$^{+}$&1&39.10&0.380&0.0097&9.7\\
\hline
Sr$^{2+}$&2&87.62&0.013&0.0001&0.3\\
\hline\hline
$\Sigma$&&&12.621&&593.8\\
\hline
\end{tabular}
\end{center}
\caption{The most abundant pH-independent cations I$^{q+}$ dissolved in seawater, from reference\,\cite{seawater}. The second column gives the positive charge per ion, $q$. The third column is the mass of one mole of cations. The fourth column gives the mass fractions $f$ in representative seawater. The fifth column gives the molality. The sixth column gives the positive charge density in  millifaradays per kilogram. One faraday (F) is the charge of one mole of protons, H$^+$.
\label{table2}}
\end{table}
\begin{table}
\begin{center}
\begin{tabular}{|l|c|c|r|c|r|}
\hline
I$^{q-}$&$q$&$m$&$f\qquad$&[I$^{q-}$]$=f/m$&$q$[I$^{q-}$]\\
\hline
& &(g mol$^{-1}$)&(g kg$^{-1}$)&(mol kg$^{-1}$)&(mF kg$^{-1}$)\\
\hline\hline
Cl$^-$&1&35.45&18.980&0.5354&535.4\\
\hline
SO$_4^{2-}$&2&96.06&2.649&0.0276&55.2\\
\hline
Br$^{-}$&1&79.90&0.065&0.0008&0.8\\
%\hline
%F$^{-}$&1&19.0&0.001&0.1\\
\hline\hline
$\Sigma$&&&21.695&&591.4\\
\hline
\end{tabular}
\end{center}
\caption{The most abundant pH-independent anions I$^{q-}$ dissolved in seawater, from reference\,\cite{seawater}. The second column gives the negative charge per ion, $q$. The third column is the mass of one mole of anions. The fourth column gives the mass fractions $f$ in representative seawater. The fifth column gives the molality. The sixth column gives the magnitude of the negative charge density.
\label{table3}}
\end{table}
To maintain charge neutrality of the water, the charge-weighted concentration of pH-dependent ions, H$^+$, NH$_4^+,\,  $OH$^{-}$, HCO$_3^-$, CO$_3^{2-}$, B(OH)$_4^-$, {\it etc.}, must be equal and opposite to the total alkalinity of (\ref{al6}) that is,
\begin{eqnarray}
\hbox{[A]}&=&\hbox{ [TALK]}\nonumber\\
&=&\hbox{[OH$^-$]}+\hbox{[HCO$_3^-$]}+2\hbox{[CO$_3^{2-}$]}+\hbox{[B(OH)$_4^{-}$]}+\cdots\nonumber\\
&&-\hbox{[H$^+$]}-[\hbox{NH}_4^+] -\cdots\nonumber\\
&=&\hbox{[WALK]} + \hbox{[CALK]} + \hbox{[BALK]} + \cdots
\label{al8}
\end{eqnarray}
Weak bases  like ammonia, NH$_3$, seldom have much effect on the pH of open ocean waters, but for estuaries and  terrestrial waters  ammonia, can significantly increase the pH. 
Dissolved ammonia can greatly increase the productivity of natural waters by providing living organisms with nitrogen for the synthesis of amino acids and other important nitrogen-containing molecules. But for special circumstances like freshwater fish farms, ammonia concentrations can become high enough to be toxic\,\cite{Edwards}.

Special names can be given  to the terms on the right side of Eq.  (\ref{al8}), as indicated by the third line. For example, the {\it water alkalinity} can be defined to be
\begin{eqnarray}
\hbox{[WALK]}&=&[{\rm OH}^-] - [{\rm H}^+]\nonumber\\
&=&\frac{K_{\rm w}}{[{\rm H}^+]}-[{\rm H}^+].\label{al10}
\end{eqnarray}
The {\it carbonate alkalinity} can be defined to be
\begin{eqnarray}
\hbox{[CALK]}&=&\hbox{[HCO$_3^-$]}+2\hbox{[CO$_3^{2-}$]}\nonumber\\
&=&\frac{K_1K_0P}{ [{\rm H}^+]}+\frac{2K_2 K_{1}K_0P}{ [{\rm H}^+]^2}.\label{al12}
\end{eqnarray}
The {\it borate alkalinity} can be defined to be
\begin{eqnarray}
\hbox{[BALK]}&=& \hbox{[B(OH)$_4^{-}$]}\nonumber\\
&=&\frac{[{\rm B}]K_{\rm B}}{[{\rm H}^+]+K_{\rm B}} .\label{al14}
\end{eqnarray}
Eq. (\ref{in4}) was used to write the second line of (\ref{al10}). Eqs. (\ref{in38}) and (\ref{in46}) were used to write the second line of (\ref{al12}). To get the second line of (\ref{al14}) we used (\ref{ba4}) and (\ref{ba6}).

Unlike the total alkalinity [A] of (\ref{al6}), which does not dependent on pH, the partial alkalinities [WALK], [CALK] and [BALK] on the right of (\ref{al8}) do depend on pH.   For acidic water [WALK] is negative, and for basic water [WALK] is positive.  Both [CALK] and [BALK] are nonnegative for any pH. Fig. \ref{talk2} shows how the carbonate alkalinity CALK of (\ref{al12}), the water alkalinity WALK of (\ref{al10}), and the borate alkality BALK of (\ref{al14})  depend on the partial pressure $P$ of CO$_2$.  The total alkalinity of (\ref{al6}) is [A] = 2.4 mM and the total dissolved boron (\ref{ba8}) is [B] = 0.43 mM. 

Assuming that all weak acids or bases except CO$_2$ and B(OH)$_3$ make  negligible contributions to the pH, we can write  (\ref{al8}) as
\begin{equation}
\hbox{[A]}=\frac{K_{\rm w}}{\hbox{[H$^+$]}}-\hbox{[H$^+$]}+\frac{K_1K_0P}{ [{\rm H}^+]}+\frac{2K_2 K_{1}K_0P}{ [{\rm H}^+]^2}+\frac{[{\rm B}]K_{\rm B}}{[{\rm H}^+]+K_{\rm B}}.
\label{al32}
\end{equation}
Eq. (\ref{al32}) has four explicit unknowns:  (1) the alkalinity [A], (2) the hydrogen-ion molality [H$^+$], (3) the CO$_2$ partial pressure $P$, and (4) the boron molality [B].  The equilibrium constants $K_{\rm w}$, $K_0$,  $K_1$, $K_2$ and $K_{\rm B}$ are known, but depend on salinity $S$ and temperature $T$, as shown in Table \ref{K012}. We can use (\ref{al32}) to find any one of the four unknowns in terms of the other three.  Eq. (\ref{al32})  gives an explicit formula for [A] in terms of [H$^+$], $P$ and [B].  To find [H$^+$] if [A], [B] and $P$ are known,  we multiply  both sides of (\ref{al32}) by the factor $[{\rm H}^+]^2\left([{\rm H}^+]+K_{\rm B}\right)$, and arrange the resulting terms as a quartic {\it seawater polynomial} equation in [${\rm H}^+$],
\begin{eqnarray}
0&=&[{\rm H}^+]^4+\left(\hbox{[A]}+K_{\rm B}\right)[{\rm H}^+]^3-\left(K_{\rm w}+K_{\rm B}\{[{\rm B}]-[{\rm A}]\} +K_{1}K_0P \right)[{\rm H}^+]^2\nonumber\\
&&- \left(K_wK_{\rm B}+\{K_{\rm B}+2K_{2}\}K_{1}K_0  P\right)[{\rm H}^+]-2 K_{\rm B}K_2K_1K_0P.
\label{al34}
\end{eqnarray}
The coefficients of the quartic equation  (\ref{al34}), which was given by Eq. (28) of reference\,\cite{CH}, are determined by the variables, [A], [B], $P$ and by  the equilibrium constants, $K_w$, $K_0$, $K_1$, $K_2$ and $K_{\rm B}$. 
Up  to four different values of  $[{\rm H}^+]$, or ``roots"  will satisfy (\ref{al34}). Only one of the four solutions is a physically acceptable value of $[{\rm H}^+]$, that is, a positive real number. The other three solutions are either negative or have non-zero imaginary parts.

\begin{figure}[t]
\begin{centering}
\includegraphics[height=90mm,width=1\columnwidth]{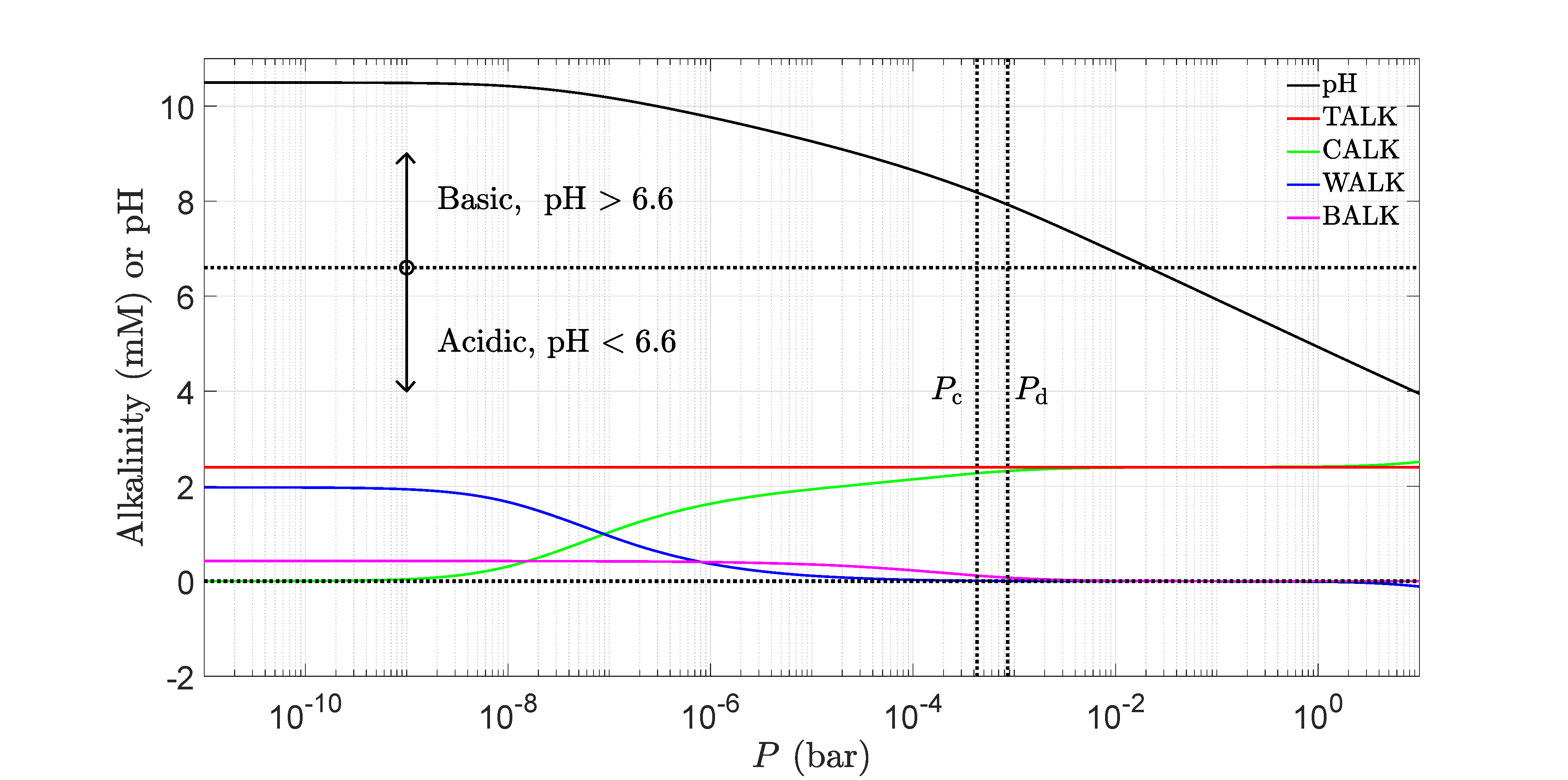}
\caption {Seawater  with a temperature of $T=25$ C, and with the same total alkalinity [A] = 2.4 mM and dissolved boron [B] = 0.43 mM as Fig. \ref{CH}. The carbonate alkalinity CALK $ = [{\rm HCO}_3^-]+2[{\rm CO}_3^{2-}]$ of (\ref{al12}), the borate alkalinity BALK = [B(OH)$_4^-$] of (\ref{al14}), and the water alkalinity WALK $=[{\rm OH}^-]-[{\rm H}^+]$ of (\ref{al10}) vary with $P$, the partial pressure of CO$_2$ in the atmosphere.  Their sum (\ref{al8}),  the total alkalinity [TALK] = [A], is independent of  $P$.  In the year 2025,  the CO$_2$ partial pressure was $P_{\rm c}= 430 \ \mu$b.
The doubled value is $P_{\rm d}= 860 \ \mu$b.}
\label{talk2}
\end{centering}
\end{figure}

Solving (\ref{al32}) for [B] in terms of [A], $[{\rm H}^+]$ and $P$ we find
\begin{eqnarray}
[{\rm B}]=\frac{[{\rm H}^+]+K_{\rm B}}{[{\rm H}^+]^2 K_{\rm B}}\bigg([{\rm H}^+]^3+\hbox{[A]}[{\rm H}^+]^2-(K_w+K_1K_0P)[{\rm H}^+]-2K_2K_1K_0P\bigg).
\label{al36}
\end{eqnarray}
Terrestrial groundwaters usually have negligible boric acid. Then we can set [B] = 0 in (\ref{al36}) to find the cubic {\it groundwater polynomial} equation
\begin{eqnarray}
0=[{\rm H}^+]^3+\hbox{[A]}[{\rm H}^+]^2-(K_w+K_1K_0P)[{\rm H}^+]-2K_2K_1K_0P.
\label{al38}
\end{eqnarray}

If [A], [B] and  [H$^+$] are known, one can solve (\ref{al36}) to find that the equilibrium partial pressure of CO$_2$  must be
\begin{equation}
 P=\frac{[{\rm H}^+]^3+\hbox{[A]}[{\rm H}^+]^2-K_{\rm w}[{\rm H}^+]}{K_{1}K_0 ([{\rm H}^+]+2K_{2})}-\frac{K_{\rm B}[B][{\rm H}^+]^2}{K_{1}K_0 ([{\rm H}^+]+2K_{2})([{\rm H}^+]+K_{\rm B})}.
\label{al40}
\end{equation}

\begin{figure}[t]
\begin{centering}
\includegraphics[height=90mm,width=1\columnwidth]{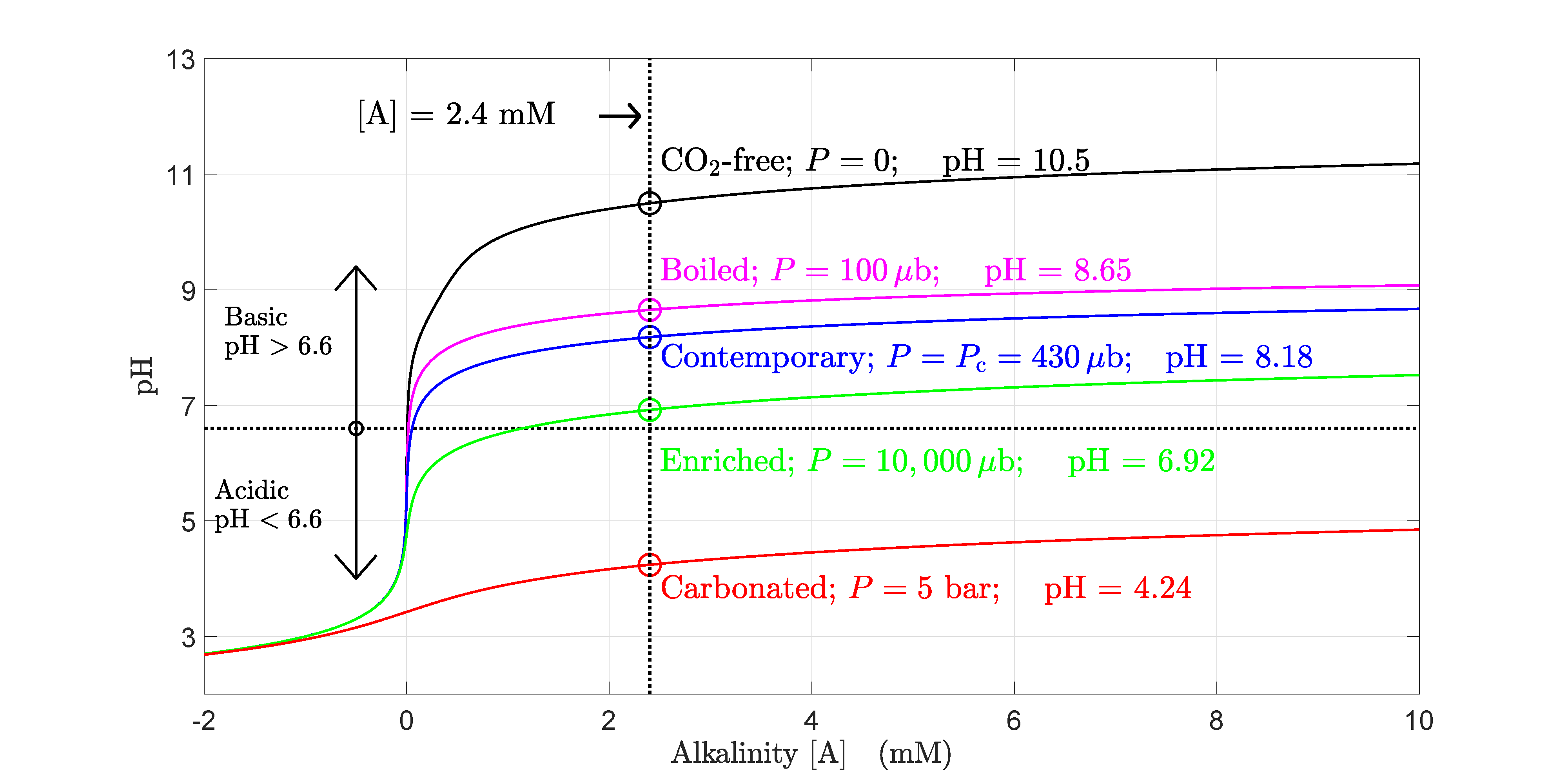}
\caption {The pH of hypothetical seawater with a temperature of $T=25$ C, with a salinity $S = 35$\textperthousand\ and with a boron molality, [B] = 0.43 mM, but with variable total alkalinity [A], and with several different partial pressures $P$ of CO$_2$ in the sealevel atmosphere.  Without dissolved CO$_2$ contemporary seawater with alkalinity [A] = 2.4 mM would be very basic, with pH =10.5.  The dissolved weak acid CO$_2$ reduces the pH to a level that is tolerable for most  oceanic life.}
\label{Carbonated}
\end{centering}
\end{figure}

For water with no CO$_2$ at all, the equilibrium partial pressure, must vanish. For $P=0$, (\ref{al34}) reduces to the cubic polynomial equation
\begin{equation}
0=[{\rm H}^+]^3+\left(\hbox{[A]}+K_{\rm B}\right)[{\rm H}^+]^2-\left(K_{\rm w}+K_{\rm B}\{[{\rm B}]-[{\rm A}]\}\right)[{\rm H}^+]- K_wK_{\rm B}.
\label{al41}
\end{equation}
Solving (\ref{al41}) provides a useful way find the hydrogen-ion molalitiy, $[{\rm H}^+]_0$ for CO$_2$-free water.

Consider hypothetical ocean water at a temperature $T=25$ C, pressure $p= 1$ bar,  salinity $S = 35$\textperthousand, and boron molality [B] = 0.43 mM.   Fig. \ref{Carbonated} gives an overview of how the pH depends on the alkalinity [A] and the partial pressure $P$ of CO$_2$ of this hypothetical water.  Each of the colored curves, which are calculated by solving  (\ref{al34}) for [H$^+$],  shows how the pH of water in equilibrium with a fixed partial pressure $P$ of CO$_2$ depends on the total alkalinity [A].
At the contemporary partial pressure,  $P_{\rm c}= 430\ \mu$b, seawater has pH = 8.18, as shown by the small blue circle in Fig. \ref{Carbonated}.  
The solubility of CO$_2$ decreases by about a factor of 5 between typical laboratory temperatures of 25 C and  100 C\,\cite{eng}.  Therefore, by boiling water samples  one can drive off much of the dissolved CO$_2$. If the water is cooled back down to room temperature before it can reabsorb CO$_2$ from the air, it will be in equilibrium with a much smaller partial pressure of  CO$_2$. In Fig. \ref{Carbonated} we assumed that boiling the water reduces the room-temperature partial pressure to $P= 100\ \mu$b. With about 4.3 times  less of the weak acid, CO$_2$,  the  boiled ocean water is more basic and has a  higher pH = 8.65, although the total alkalinity [A] is the same.  
If one could remove all of the CO$_2$ from seawater without affecting the salinity,  alkalinity or boron molality,  and without inducing the precipitation of any crystals, like CaCO$_3$(s) or Mg(OH)$_2$(s), for which the seawater may be supersaturated,  the pH would increase to pH = 10.5, about the same as milk of magnesia, a suspension of Mg(OH)$_2$ (brucite) crystallites in freshwater, and shown by the small black circle near the top of Fig. \ref{Carbonated}.

Also shown in Fig. \ref{Carbonated} is a hypothetical sample of ocean water that has been allowed to come to equilibrium with air enriched to $P =10,000\ \mu$b of CO$_2$.  This is much more CO$_2$ than could be released by any credible scenario of fossil-fuel combustion. But even for this large value, the ocean water would still be slightly basic, pH = 6.92, as shown by the small green circle for alkalinity [A] = 2.4 mM.  

If one were to artificially carbonate ocean water with a CO$_2$ partial pressure $P= 5 $ bar, representative of the bottling pressure of strongly carbonated seltzer or  champagne\,\cite{champagne},  the pH of the hypothetical seawater would finally become acidic and drop to pH = 4.24, shown by the small red circle near the bottom of the figure.
\subsection{Titration\label{t}}
According to the U.S. Geological Survey\,\cite{USGS}.
\begin{quote}
One common method the U.S. Geological Survey (USGS) uses for measuring alkalinity is to take a water sample and to add acid to it while checking the pH of the water as the acid is added. An initial pH reading of the water is taken and then small amounts of acid are added in increments, the water is stirred, and the pH is taken. This is done many times. In the beginning, the acid added will be neutralized by compounds in the water, such as bicarbonates. As more acid is added, the bicarbonates get ``used up", as it is also being neutralized by the acid. Eventually all the acid-neutralizing compounds are used up. After this point, any acid added to the water will lower the pH in a linear fashion, and the scientist will be able to see this inflection point by viewing a line chart of the amount of acid added to the water and the resulting pH. The point at which the change in pH line becomes linear is used to determine the water's alkalinity.
\end{quote}
Hydrochloric acid, HCl, and sulfuric acid, H$_2$SO$_4$ are commonly used for the {\it titration} described above. For the pH range of most natural waters of the Earth, both are strong acids which dissociate completely into ${\rm H}^+$ ions and the anions Cl$^{-}$ or SO$_4^{2-}$  as indicated by the chemical equations,
\begin{equation}
\hbox{HCl} \rightarrow {\rm H}^++\hbox{Cl$^{-}$},\quad \hbox{or}\quad
\hbox{H$_2$SO$_4$} \rightarrow 2{\rm H}^++\hbox{ SO$_4^{2-}$  }.
\label{b2}
\end{equation}
In thermal equilibrium the concentration of undissociated HCl  or H$_2$SO$_4$ molecules in solution is negligibly small, a fact we have indicated with single right-pointing arrows in (\ref{b2}). Under normal conditions, the reactions are {\it irreversible}, in contrast to {\it reversible} reactions like (\ref{in2}), (\ref{in14}) or (\ref{in40}), where there can be substantial concentrations of reactants on the left sides of the chemical equation and products on the right. We indicate reversibility by including both right-pointing and left-pointing arrows in chemical equations like  (\ref{in32}) or (\ref{in40}).

Adding the strong acids HCl or H$_2$SO$_4$ to water increases the concentration of  pH-independent negative chloride ions, Cl$^{-}$, or doubly negative sulfate ions,  SO$_4^{2-}$.  This decreases the alkalinity of (\ref{al6}),
\begin{equation}
\hbox{[A]} \rightarrow \hbox{[A] - [T]},
\label{b6}
\end{equation}
where the {\it titrant} molality [T] is related to the hypothetical, undissociated molality of the strong acids by
\begin{equation}
[\hbox{T}]=[\hbox{HCl}] \quad\hbox{or}\quad [\hbox{T}]=2[\hbox{H$_2$SO$_4$}] ,\label{b7}\\
\end{equation}
The titration described by (\ref{b6}) changes (\ref{al34}) to
\begin{eqnarray}
0&=&[{\rm H}^+]^4+[{\rm H}^+]^3\left(\hbox{[A] - [T]}+K_{\rm B}\right)-[{\rm H}^+]^2\left(K_{\rm w}+K_{\rm B}\{[{\rm B}] - [{\rm A}]+[{\rm T}]\} +K_{1}K_0P \right)\nonumber\\
&&- [{\rm H}^+]\left(K_wK_{\rm B}+\{K_{\rm B}+2K_{2}\}K_{1}K_0 P\right)-2 K_{\rm B}K_2K_1K_0P.
\label{b10}
\end{eqnarray}
We will call (\ref{b10}), which is a variant of the basic equation (\ref{al34}), the {\it titration equation}. As for (\ref{al34}), there is only one physically acceptable solution of (\ref{b10}), the root with [H]$^+ >0$.

\begin{figure}[t]
\begin{centering}
\includegraphics[height=90mm,width=1\columnwidth]{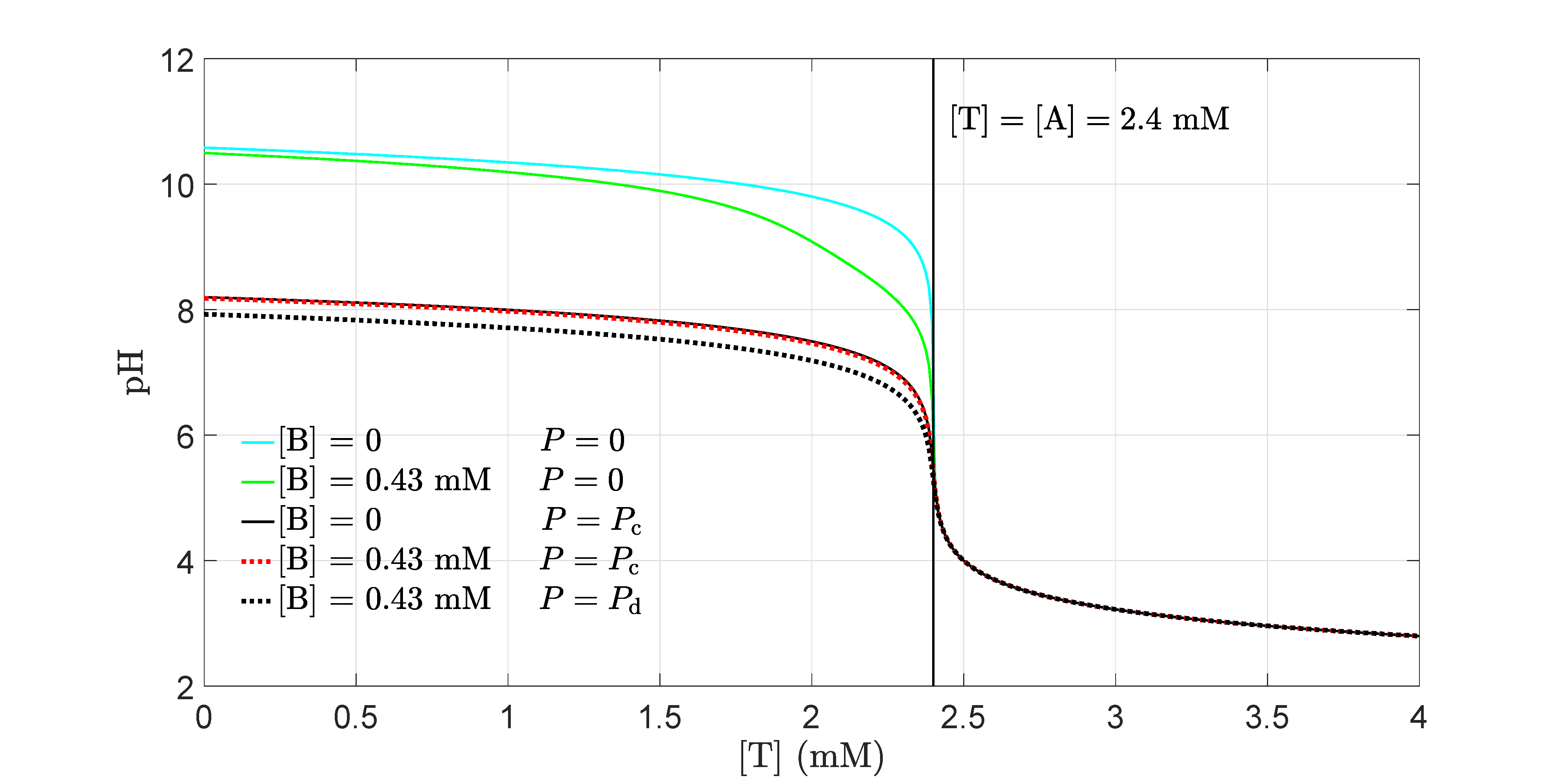}
\caption {The total alkalinity [A]=[TALK] of (\ref{al8}) is often determined by titration with a strong acid. For the example shown here and calculated with (\ref{b10}), hydrochloric acid, HCl, of molality [HCl] = [T], is  added to water with a temperature $T=25$ C and with a total alkalinity  [A] =  2.4  mM, typical of seawater. The molality at which the pH decreases most rapidly with increasing [T] is a good approximation to [A], and is little affected by the  partial pressure $P$ of atmospheric CO$_2$ or by the boron molality [B].  The contemporary partial pressure of CO$_2$ is $P_{\rm c} = 430\ \mu$b.}  
\label{talk3}
\end{centering}
\end{figure}

Modeled titrations of a water sample are shown in Fig. \ref{talk3}.  
As noted in the quote from the USGS above, near the {\it equivalence point}, where [T] = [A],  the pH decreases most rapidly with increasing [T]. The weak acids,  dissolved CO$_2$ and borate, have little effect on the result. 
\subsection{Saturation state}
The tendency of calcium carbonate crystals to dissolve or grow in solution can be described with the {\it saturation state}, $\Omega$, defined by
\begin{equation}
\Omega =\frac{[\hbox{Ca}^{2+}][{\rm CO}_3^{2-}]}{K_{\rm sp}}.
\label{ss8}
\end{equation}
Crystals will tend to grow if $\Omega>1$ and they  will tend to dissolve if $\Omega<1$.  For {\it saturated solutions} where  $\Omega=1$, crystals will have no tendency to either dissolve or grow.  Most surface waters of the oceans are supersaturated, and $\Omega>1$ for both calcite and aragonite. Many ``fresh" terrestrial surface waters are subsaturated, and have $\Omega <1$.  But the rates of precipitation or dissolution can be very slow, and depend on many factors besides the saturation state $\Omega$. For example, Pytkowicz\,\cite{Pytkowicz} has shown Mg$^{2+}$ ions of seawater slow down the precipitation of CaCO$_3$(s) so much that calcite and aragonite formation rates are completely dominated by biomineralization, with negligible contributions from inorganic precipitation. Organic surface coatings of biomineralized CaCO$_3$(s) also suppress dissolution or precipitation.

\begin{figure}[t]
\begin{centering}
\includegraphics[height=90mm,width=1\columnwidth]{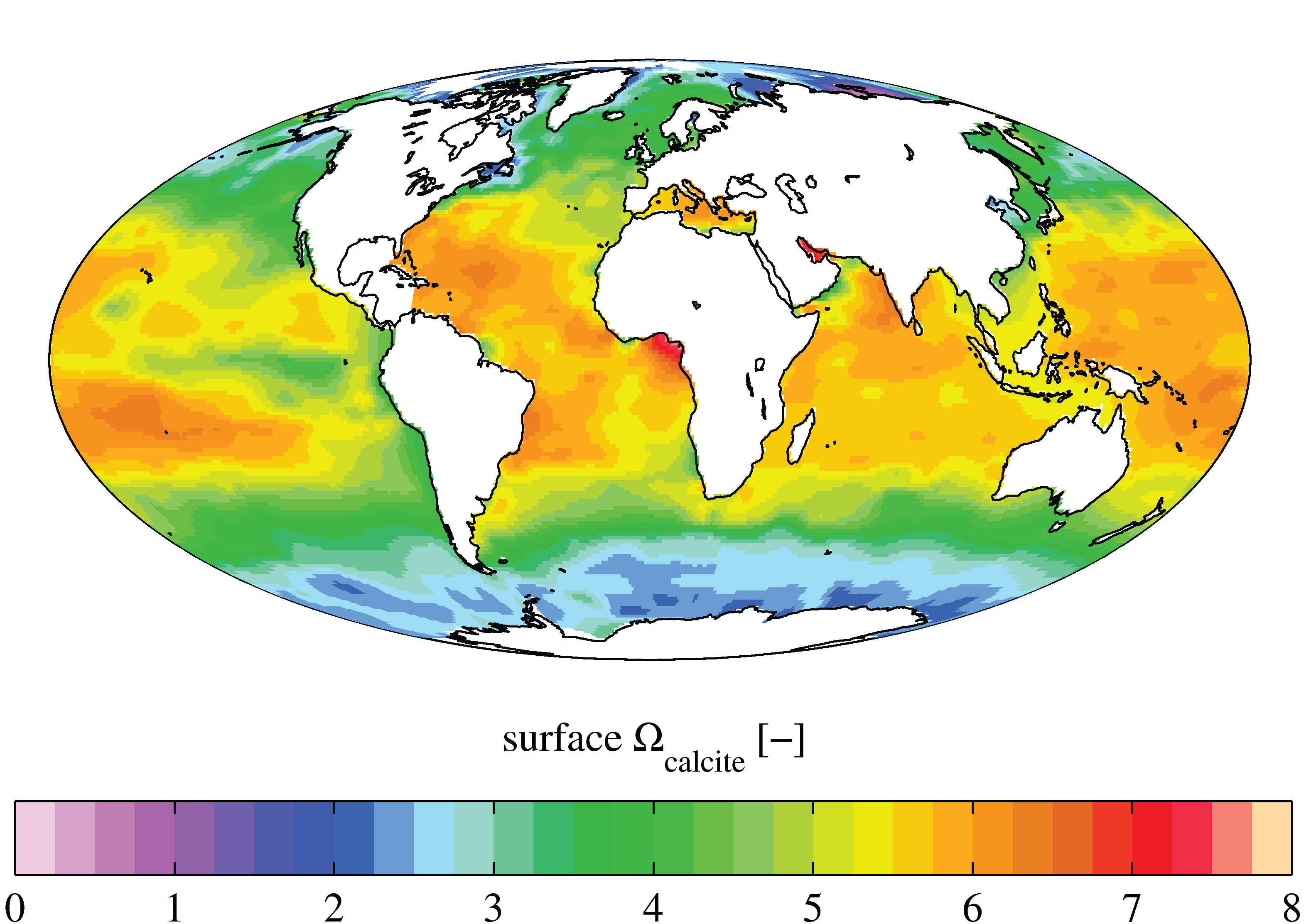}
\caption {Saturation state (\ref{ss8}) of surface ocean water with respect to calcite crystals, CaCO$_3$(s)\,\cite{Lys}.  Note the qualitative similarity to the surface alkalinity of Fig. \ref{SurfaceA}.}
\label{Omega}
\end{centering}
\end{figure}

In representative seawater, the calcium-ion and carbonate-ion molalities are
\begin{eqnarray}
 \hbox{[Ca$^{2+}$]} &=& 10 \hbox{ mM}, \label{ss11a} \quad\hbox{and}\quad
[{\rm CO}_3^{2-}]= 0.214 \hbox{ mM},\label{ss11b}
\end{eqnarray}
In (\ref{ss11b}) the calcium ion molality is from Table\,\ref{table2}. The carbonate-ion  molality comes from solving  (\ref{al34}) for $[{\rm H}^+]$ at the contemporary CO$_2$ partial pressure $P_{\rm c} = 430\ \mu$b, the alkalinity [A] = 2.4 mM, the boron molality [B] = 0.43 mM and for the equilibrium constants of Table \ref{K012}  at a temperature $T = 25$ C  salinity $S = 35$\textperthousand.  Using  (\ref{ss11b}) and the seawater value of $K_{\rm sp}$ for a temperature $T=25$ C and salinity $S = 35$\textperthousand\ from Table \ref{K012} in (\ref{ss8}) we find a representative  saturation state of calcite for the upper layers of the ocean,
\begin{equation}
\Omega = 5.01\label{ss11c}
\end{equation}

\begin{figure}[t]
\begin{centering}
\includegraphics[height=90mm,width=1\columnwidth]{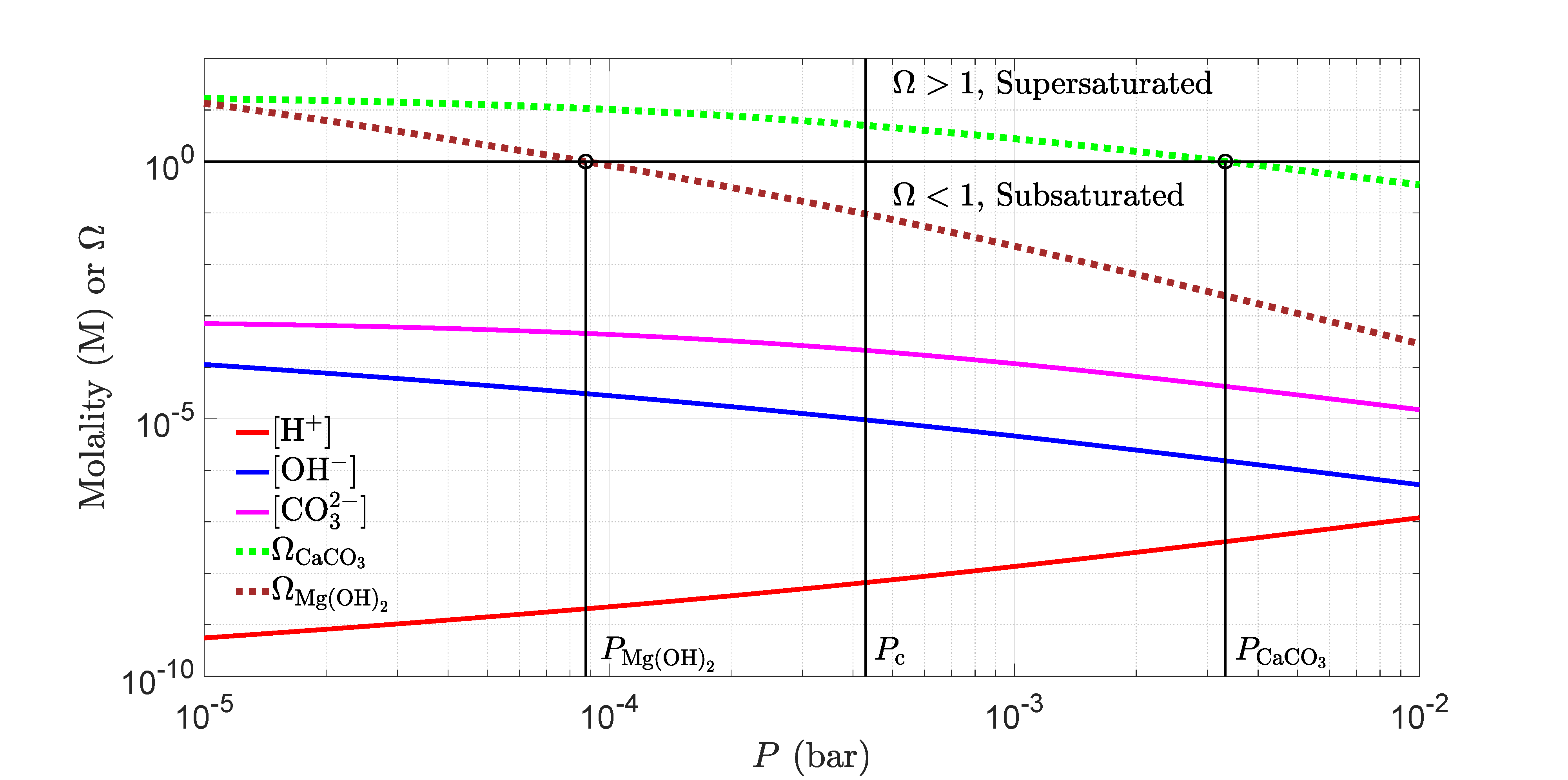}
\caption {Saturation states for calcite, $\Omega =[{\rm Ca}^{2+}][{\rm CO}_3^{2-}]/K_{\rm sp}$ and brucite, $\Omega=[{\rm Mg}][{\rm OH}^{-}]^2/K_{\rm sb}$, versus the partial pressure $P$ of CO$_2$ of air in equilibrium with surface seawater at a temperature $T=25$ C, with alkalinity [A] = 2.4 mM and dissolved inorganic boron [B] = 0.43 mM. According to Table \ref{table2}, the calcium ion molality is [Ca$^{2+}$] = 10.0 mM and the  magnesium ion molality is [Mg$^{2+}$] = 52.3 mM.  Increasing  $P$ decreases the pH. As shown in Fig. \ref{Fraction} decreasing pH decreases the  carbonate molality [CO$_3^{2-}$] and this increases the solubility of CaCO$_3$(s) or calcite.   Similarly, increasing $P$ decreases the hydroxyl molality [OH$^{-}$] and increases the solubility of Mg(OH)$_2$(s) or brucite.  The CO$_2$ partial pressure, $P_{\rm c} = 430\ \mu$b in the year 2025 would have to increase by nearly a factor of 8 to $P_{{\rm CaCO}_3} = 3320\ \mu$b to make the seawater subsaturated with respect to calcite. Seawater is supersaturated with respect to  Mg(OH)$_2$(s)  for CO$_2$ partial pressures less than $P_{{\rm Mg(OH)}_2} = 87.5\ \mu$b.}
\label{MP}
\end{centering}
\end{figure}
An overview of the calcite saturation state of the ocean surface across the globe is shown in Fig. \ref{Omega}. Note the qualitative similarity to the surface alkalinity of Fig. \ref{SurfaceA}.  The solubility constant $K_{\rm sp}$ of (\ref{ss4}) increases with increasing depth and  pressure.  At the lysocline depth, typically around 4 or 5 km\,\cite{Lys}, the calcite solubility product $K_{\rm sp}$ in the denominator of (\ref{ss8}) has increased so much that $\Omega = 1$. Below the lysocline, where the pressure is high enough that $\Omega<1$, inorganic calcite crystals CaCO$_3$(s) dissolve into the surrounding water. But recent research\,\cite{deepshells} has  shown that organisms with biomineralized calcium carbonate crystals are abundant at ocean depths thousands of meters below the lysocline. Evolution has taught most aquatic organisms how to deal with waters that are subsaturated in CaCO$_3$(s). Changes of the concentrations of bicarbonate and carbonate ions due to changing partial pressures of atmospheric CO$_2$ will have negligible effect on them.

Fig. \ref{MP} shows how the saturation state $\Omega$ of (\ref{ss8}) of  representative  seawater at a temperature $T=25$ C depends on the partial pressure $P$ of CO$_2$.  The saturation state of surrounding water means little to aquatic organisms, since their internal saturation state and rate constants for biomineralization are under biological control.  But the representative seawater of Fig. \ref{MP} will remain supersaturated with respect to calcite as long as the CO$_2$ partial pressure remains less than $P< 3320\ \mu$b, nearly 8 times greater than the value $P_{\rm c}=430\ \mu$b, in the year 2025. 
\begin{figure}[t]
\begin{centering}
\includegraphics[height=90mm,width=1\columnwidth]{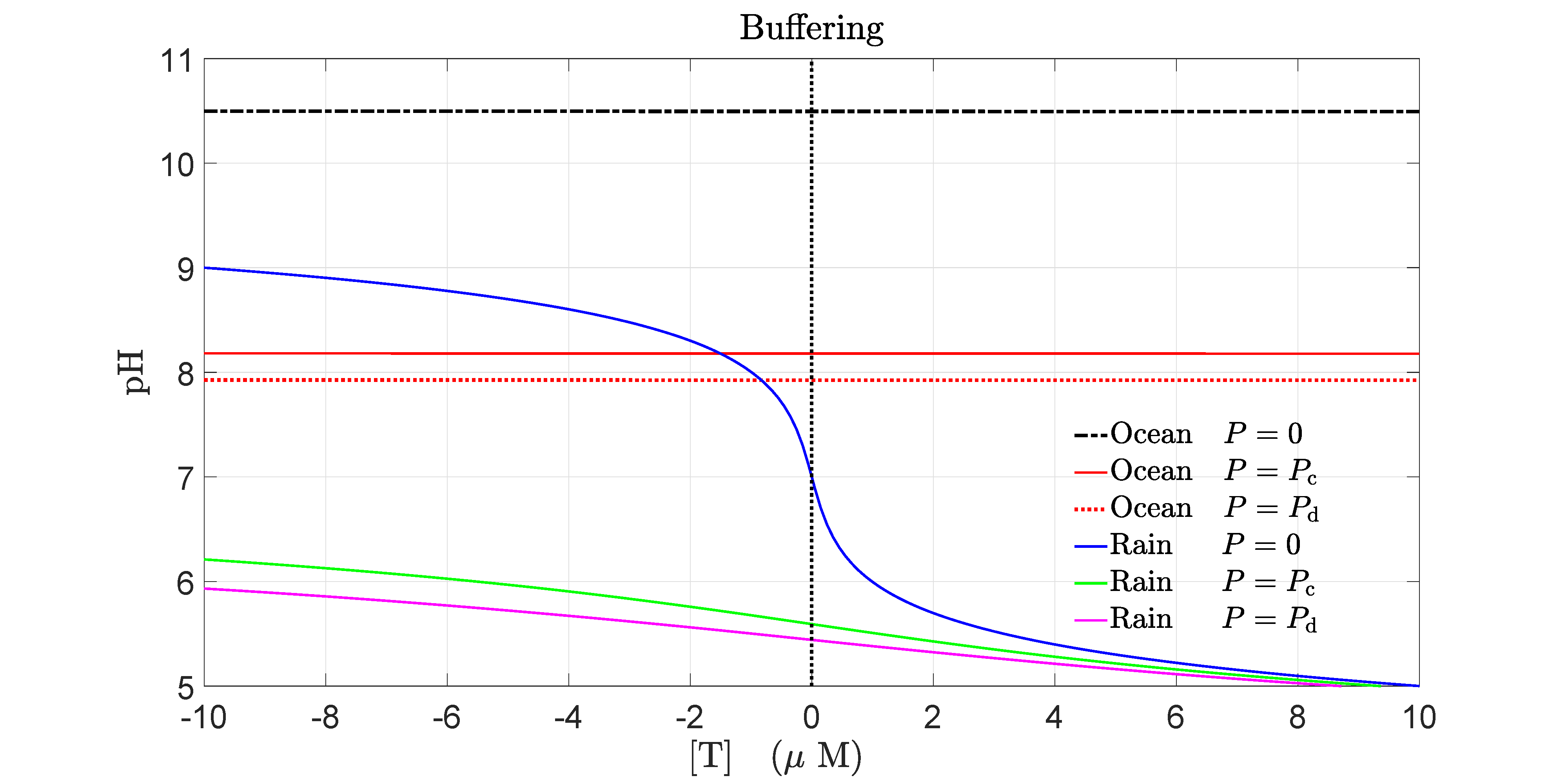}
\caption {Titration of model seawater at a temperature $T= 25$ C, a salinity $S = 35$\textperthousand, a total alkalinity 
[A] = 2.4  mM,  and a boron molality [B] = 0.43 mM. Acidification with small amounts of a strong-acid titrant  like [T] = [HCl], or basification with a small amount of a strong-base titrant like [T] = $-$[ NaOH], have almost no effect on the pH of ocean water in equilibrium with CO$_2$ partial pressures, $P=0$, $P=P_c= 430\ \mu$b, and $P=P_d = 2P_c = 860\ \mu$b.  Also shown are titration curves for rain, pure water at a temperature $T = 15$ C, with vanishing salinity, $S=0$\textperthousand, alkalinity,[A] $=0$, and boron molality [B] $=0$, but with dissolved CO$_2$ in equilibrium with the same three partial pressures of CO$_2$.  Rain is much less resistant to acidification by atmospheric CO$_2$ than ocean water.  The buffering coefficients $\beta$ of (\ref{bc4}) for these titration curves are given in Table \ref{tableb}.}
\label{buffer1}
\end{centering}
\end{figure}
\subsection{Buffering}
The ocean is a {\it buffered solution} that resists changes in pH when acids or bases are added to it. The United States Geological Survey\,\cite{USGSbuffering} defines the buffering capacity qualitatively  as:  
\begin{quote}
The buffering capacity of a water body; a measure of the ability of the water body to neutralize acids and bases and thus maintain a fairly stable pH level.
\end{quote}
The oceans are buffered because they are a mixture of a strong base, with representative alkalinity [A] = 2.4 mM, and a weak acid, mostly dissolved atmospheric CO$_2$.   
The alkalinity comes from a mixture of strong bases like NaOH, KOH, Mg(OH)$_2$, Ca(OH)$_2$ {\it etc.}  and strong acids HCl, H$_2$SO$_4$, {\it etc.} The strong basicity slightly exceeds the strong acidity. How seawater at a temperature $T = 25$ C responds to additions of small amounts of strong acids or bases is shown in Fig. \ref{buffer1}, which was calculated with the titration equation (\ref{b10}).
On the figure one cannot see any effect on the pH from adding 1 $\mu$M of a strong acid like HCl,  or adding 1 $\mu$M of a  strong base like NaOH, to ocean water. The strong buffering is nearly independent of the partial pressure of CO$_2$, as one can see from the curves for  $P=P_{\rm c}= 430\ \mu$b, the current atmospheric value, or double that value,  $P=P_{\rm d} = 860\ \mu$b\,, or for no dissolved  CO$_2$ at all, $P = 0$. 

The curves labeled ``rain" in Fig. \ref{buffer1} refer to water at a temperature, $T = 15$ C, with vanishing alkalinity, [A] = 0, vanishing boron molality, [B] = 0, vanishing  salinity, $S =0$\textperthousand\, and with enough dissolved CO$_2$ to be in equilibrium with the CO$_2$ partrial pressures $P$.  Rain is very sensitive to additions of strong acids or bases. The same 10 $\mu$M additions of acids or bases  that cause negligible changes in the pH of ocean water, causes a relatively large increase of the pH of rainwater.

Adding $\delta$ [T] = 10 $\mu$M of a strong acid to hypothetical rainwater with no dissolved CO$_2$ and $P=0$ decreases the pH  by about $\Delta{\rm pH} = -1.9$. This is equivalent to increasing the hydrogen-ion concentration [${\rm H}^+$] by a factor of $10^{-\Delta{\rm pH}}=79$.  For rainwater in equilibrium with a CO$_2$ partial pressure
 $P_{\rm c}=430\ \mu$b, and double that value, $P_{\rm d}=860\ \mu$b, the additon of [T] = 10 $\mu$M of a strong acid would decrease the pH by $\Delta{\rm pH} = -0.6$ and  $\Delta{\rm pH} = -0.5$, respectively.
Atmospheric CO$_2$ dissolved in rainwater provides some buffering with respect to pure water, but much less than  from the alkalinity and dissolved CO$_2$ of seawater.
\begin{table}
\begin{center}
\begin{tabular}{|l|c|c|c|c|c|r|}
\hline
 Water& $S$(\textperthousand)&[A] (mM)& [B] (mM)&$P$ &pH &$\beta\quad$ \\[0.5ex]
 \hline
\hline
Ocean&35&2.4&0.43&0&10.5&3,940\\
\hline
Ocean  &35&2.4&0.43&$P_{\rm c}$&8.18&5,560\\
\hline
Ocean&35&2.4&0.43&$P_{\rm d}$&7.93&5,280\\
\hline
Rain &0&0&0&0&7 &1\\
\hline
Rain &0&0&0&$P_{\rm c}$&5.59&25.6\\
\hline
Rain &0&0&0&$P_{\rm d}$&5.44&36.1\\
\hline
\end{tabular}
\end{center}
\caption{Buffering coefficients $\beta$ of (\ref{bc4}) for the titration curves of Fig. \ref{buffer1}. A closed-form expression for $\beta$ is given by (\ref{bc12}). For both seawater and rainwater the temperature is $T = 25$ C.  The strong buffering of seawater is due to its large alkalinity, [A] = 2.4 mM.}\label{tableb}
\end{table}
For model seawater, where the hydrogen ion concentration $[{\rm H}^+]$ is the physically acceptable solution of  the seawater polynomial (\ref{al34}),  the pH is a  function of the salinity $S$, the alkalinity, [A],  the boron molality [B] and the partial pressure $P$ of CO$_2$ and
\begin{equation}
\hbox{pH}=\hbox{ pH}(S,[\hbox{A}], [\hbox{B}],P) =-\log[{\rm H}^+].
\label{bc2}
\end{equation}
Adding a small increment $\delta [{\rm T}]$ of a strong acid to the seawater is equivalent to a small alkalinity change, $\delta [\rm A] =-\delta[\rm T]$. A measure of the sensitivity of seawater to acidification by a strong acid is therefore the {\it acidification rate},
\begin{eqnarray}
\partial\,\hbox{pH}/\partial [\rm T]&=&-\partial\,\hbox{pH}/\partial [\rm A]\nonumber\\
&=&\frac{1}{[{\rm H}^+]\ln 10}\left(\frac{\partial[{\rm H}^+]}{\partial [\rm A]}\right).
\label{bc2a}
\end{eqnarray}
We will define the {\it buffering coefficient} $\beta$ as the factor by which the acidification rate (\ref{bc2a}) of the solution of interest is smaller than that of pure, unbuffered water, with $S=0$,  $[\rm A] = 0$, $[\rm B] = 0$ and $P = 0$. We denote the pH of pure water by  pH$_{\rm pw}$. The reference acidification rate is  
\begin{equation}
\partial\, \hbox{pH}_{\rm pw}/\partial\, [\rm T]=-\partial\, \hbox{pH}_{\rm pw}/\partial\, [\rm A].
\label{bc2b}
\end{equation}
Then the buffering coefficient is
\begin{equation}
\beta =\frac{\partial\, \hbox{pH}_{\rm pw}/\partial [\rm A]}{\partial\, \hbox{pH}/\partial [\rm A]}.
\label{bc4}
\end{equation}
Differentiating both sides of (\ref{al32}), solving the resulting equation for $\partial[{\rm H}^+]/\partial[\rm A]$, and substituting the result into
 (\ref{bc2a}) we find that the acidification rate is
\begin{eqnarray}
&&-\partial\,\rm pH/\partial [\hbox{A}]=\nonumber\\
&&\frac{- [{\rm H}^+]/\ln 10}{3 [{\rm H}^+]^2+2[{\rm H}^+][\hbox{A}]-K_w-K_1K_0P-[{\rm H}^+][\rm B]K_{B}([{\rm H}^+]+2K_B)/([{\rm H}^+]+K_B)^{2}}.\quad
\label{bc8}
\end{eqnarray}
For pure, unbuffered water, with [A] = 0, $P = 0$, [B] = 0 and with neutral pH, $[{\rm H}^+]=\sqrt{K_w}$, (\ref{bc8}) becomes
\begin{eqnarray}
-\left(\partial\,\rm pH/\partial [\hbox{A}]\right)^{\circ}&=&\frac{-1}{2\ln 10\sqrt{K_w}}.
\label{bc10}
\end{eqnarray}
Substituting (\ref{bc8}) and (\ref{bc10}) into (\ref{bc4}), we find that the buffering coefficient for model seawater is
\begin{eqnarray}
\beta=
\frac{3 [{\rm H}^+]^2+2[{\rm H}^+][\hbox{A}]-K_w-K_1K_0P-[{\rm H}^+][\rm B]K_{B}([{\rm H}^+]+2K_B)/([{\rm H}^+]+K_B)^{2}}{2 [{\rm H}^+]\sqrt{K_w}}.\qquad
\label{bc12}
\end{eqnarray}

Some representative buffering coefficients $\beta$ at a temperature $T = 25$ C, evaluated with (\ref{bc12}) and the solutions of (\ref{al34}) are shown in Table \ref{tableb}. For ocean water, $\beta >1,000$ and the pH changes more than a thousand times less from small additions of strong acids or bases than does the pH of pure water.  For rainwater dissolved CO$_2$ in equilibrium with the partial pressure $P$ of the air provides some buffering, but much less than the alkalinity of ocean water.

\begin{figure}[t]
\begin{centering}
\includegraphics[height=90mm,width=1\columnwidth]{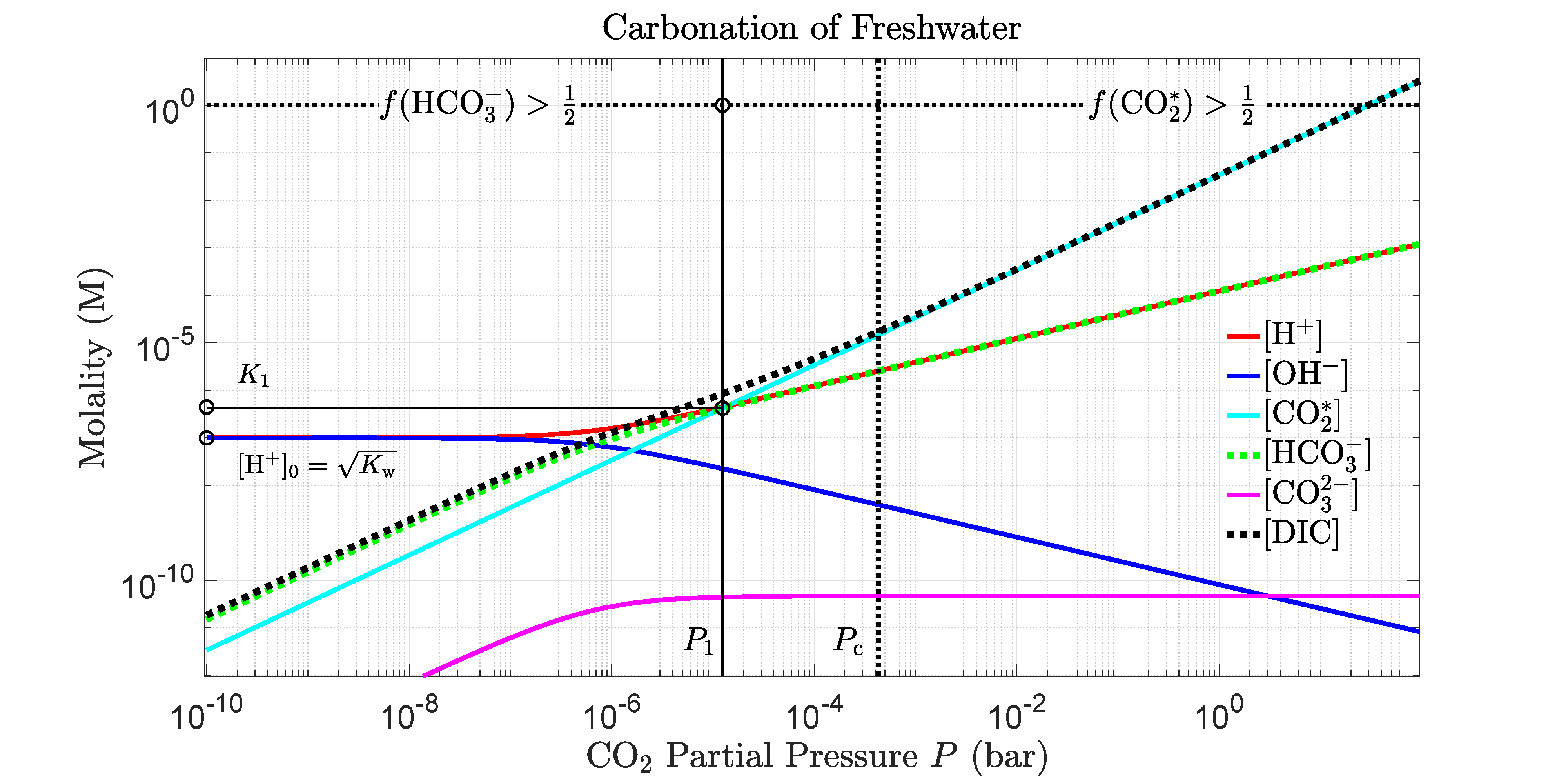}
\caption {The molalities [CO$_2^*]$, $[{\rm HCO}_3^-]$ and $[{\rm CO}_3^{2-}]$ of dissolved CO$_2$, and their sum [DIC] are plotted versus the partial pressure $P$ of atmospheric CO$_2$ which is in equilibrium with freshwater  at a temperature $T=25$ C.  Also shown are the hydrogen-ion and hydroxyl-ion molalities, $[{\rm H}^+]$ and $[{\rm OH}^-]$.
 For $P<P_1= 1.24\,\times\, 10^{-5}$ bar, more than half of the dissolved inorganic carbon, DIC, is bicarbonate ions ${\rm HCO}_3^-$. For $P_1<P$ more than half is uncharged species, CO$_2^*$.  The contemporary partial pressure of CO$_2$ is $P_{\rm c} = 430\ \mu$b.  For $P =  0$, freshwater contains only hydrogen and hydroxyl ions in self-ionization equilibrium with molalities $[{\rm H}^+]_0=[{\rm OH}^-]_0 = \sqrt{K_{\rm w}} = 10^{-{\rm pH}_n}$ where neutral pH of water at a temperature $T=25$ C is given by Table \ref{pHi} as  pH$_n$ = 7.00.}
\label{CPf}
\end{centering}
\end{figure}
\subsection{Carbonation\label{cb}}
To more clearly understand how varying atmospheric partial pressures $P$ of CO$_2$  affect the inorganic chemistry of natural waters, we find it useful to plot {\it carbonation curves}. These  show how the molalities of dissolved substances depend on $P$.  Fig. \ref{CPf} shows a carbonation curve for freshwater at a temperature of 25 C. The curves are found by solving the groundwater polynomial equation (\ref{al38}) for $[{\rm H}^+]$ as a function of the CO$_2$ partial pressure $P$, with vanishing alkalinity,  [A] = 0, vanishing boron molality, [B] = 0,  with vanishing salinity $S=0$\textperthousand,  and with the equilibrium constants of Table \ref{K012} at a temperature $T=25$ C.  The values of $[{\rm H}^+]$ and $P$ are used to find the molalities [CO$_2^*$] of (\ref{in24}), $[{\rm HCO}_3^-]$ of (\ref{in38}), $[{\rm CO}_3^{2-}]$ of (\ref{in46}), and their sum,  [DIC] of (\ref{if2}).  Fig. \ref{CPf}  shows that for the low-pressure limit
\begin{equation}
[{\rm H}^+]_{{\rm f}\,0}\to \sqrt{K_{\rm w}}=10^{-7}\hbox{ M} \quad\hbox{as}\quad P\to 0\quad\hbox{for freshwater at $T=25$ C}.
\label{cb2}
\end{equation}
Using (\ref{cb2}) in (\ref{if2}) we find that the molality of dissolved inorganic carbon approaches the low-pressure limit
\begin{eqnarray}
[\hbox{DIC}]_{{\rm f}\,0}&=&\left(1+\frac{K_1}{\sqrt{K_{\rm w}}}+\frac{K_2K_1}{K_{\rm w}}\right ) K_0 P\nonumber\\
&=&0.185 \hbox{ M bar}^{-1} P\quad \hbox{for freshwater at $T=25$ C and}\quad P\to 0
\label{cb8}
\end{eqnarray}
\begin{figure}[t]
\begin{centering}
\includegraphics[height=90mm,width=1\columnwidth]{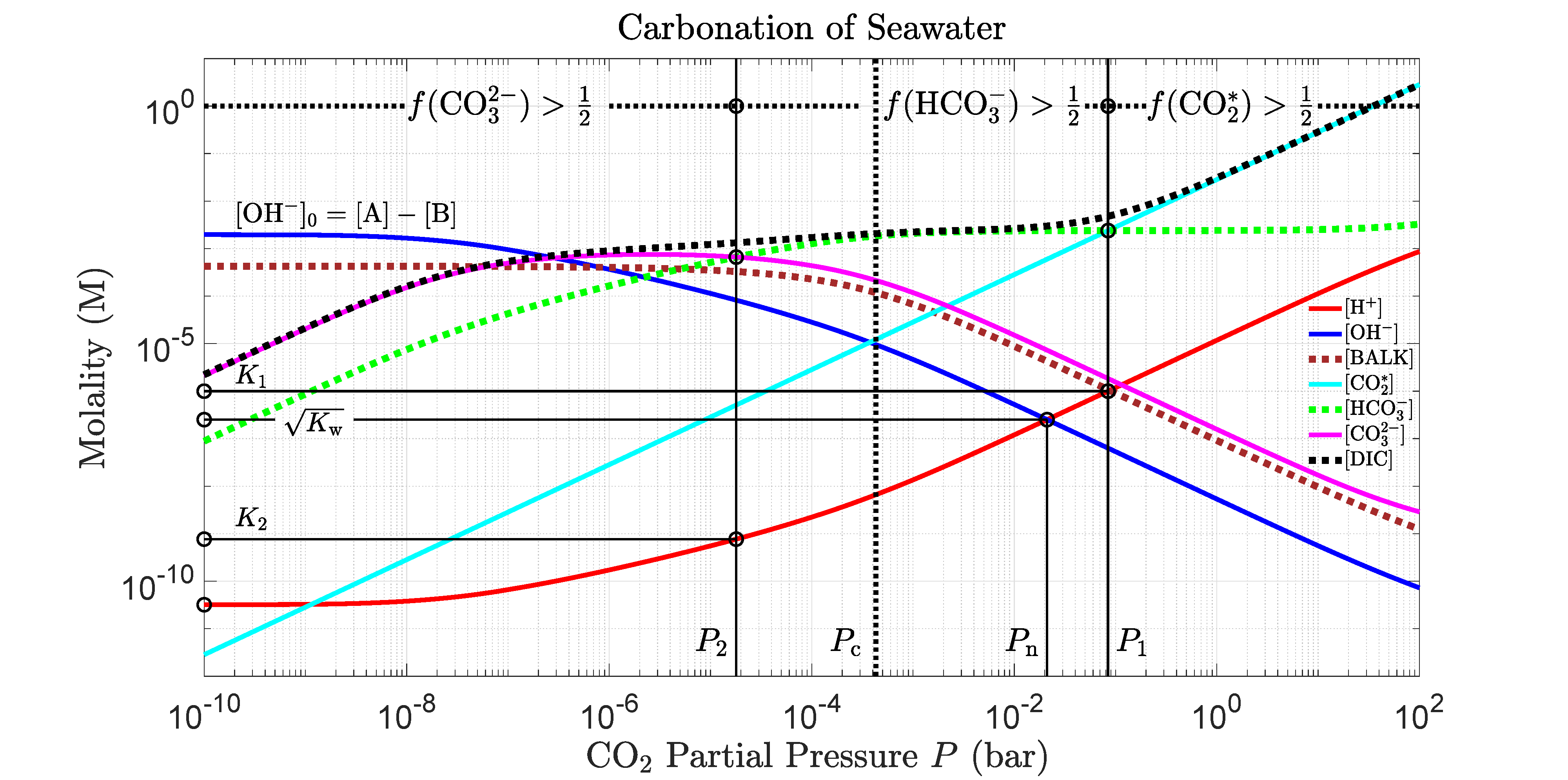}
\caption {The molalities [CO$_2^*]$, $[{\rm HCO}_3^-]$ and $[{\rm CO}_3^{2-}]$ of dissolved CO$_2$, and their sum [DIC] are plotted versus the partial pressure $P$ of atmospheric CO$_2$ which is in equilibrium with seawater  at a temperature $T=25$ C.  Also shown are the hydrogen-ion and hydroxyl-ion molalities, $[{\rm H}^+]$ and $[{\rm OH}^-]$, and the borate molality [B(OH)$_3^{-}$]=[BALK]. The total alkalinity of the water is [A] = 2.4 mM and the boron molality is [B] = 0.43 mM.  For $P<P_2= 1.79\,\times\, 10^{-5}$ bar, more than half of the dissolved inorganic carbon, DIC, is carbonate ions, ${\rm CO}_3^{2-}$. For $P_2<P<P_1$ more than half is bicarbonate ions, ${\rm HCO}_3^-$. For $P_1<P$ more than half is uncharged species, CO$_2^*$.
The contemporary partial pressure of CO$_2$ is $P_{\rm c} = 430\ \mu$b. As discussed in (\ref{cb22}), for $P=0$ the hydroxyl molality is very nearly [OH$^-$]$_0$ = [A] - [B].}
\label{CP}
\end{centering}
\end{figure}

Fig. \ref{CP} shows a carbonation curve for seawater at a temperature of 25 C. The curves are found by solving the seawater polynomial equation (\ref{al34}) for $[{\rm H}^+]$ as a function of the CO$_2$ partial pressure $P$, with a total alkalinity, [A] = 2.4 mM, with a boron molality [B] = 0.43 mM, and with the equilibrium constants of Table \ref{K012} at a salinity $S=35$\textperthousand.  The values of $[{\rm H}^+]$ and $P$ are used to find the molalities [CO$_2^*$] of (\ref{in24}), $[{\rm HCO}_3^-]$ of (\ref{in38}), $[{\rm CO}_3^{2-}]$ of (\ref{in46}), and their sum, [DIC] of (\ref{if2}).
Fig. \ref{CP}  shows that for very low atmospheric partial pressures $P$ of CO$_2$, representative seawater is very basic because of the  alkalinity [A]. The boric acid is nearly completely ionized to borate ions, B(OH)$_4^{-}$, which together with hydroxyl ions, OH$^-$, constitute nearly all of the negative charge density of the water. The positive charge density is nearly all due to the total alkalinity [A] of (\ref{al6}). So  one can accurately approximate the hydroxyl molality in the limit $P\to 0$ as 
\begin{equation}
[{\rm OH}^-]_{{\rm s}0}= [{\rm A}]-[{\rm B}].
\label{cb22}
\end{equation}
Using (\ref{cb22}) in (\ref{if2}) we find that the molality of dissolved inorganic carbon approaches the low-pressure limit
\begin{eqnarray}
[\hbox{DIC}]_{{\rm s}0}&=&\left(1+\frac{K_1 [{\rm OH}^-]_{{\rm s}0}}{K_{\rm w}}+\frac{K_2K_1 [{\rm OH}^-]_{{\rm s}0}^2}{K_{\rm w}^2}\right )K_0 P\nonumber\\
&=&2.21\,\times\, 10^4 \hbox{ M bar}^{-1} P\quad \hbox{for saltwater at $T=25$ C and}\quad P\to 0.
\label{cb28}
\end{eqnarray}

From inspection of Fig. \ref{CPf} and Fig. \ref{CP} we see that  in the high pressure limit, the dissolved inorganic carbon is nearly all uncharged species,
\begin{eqnarray}
[\hbox{DIC}]_{\infty}&=& K_0 P\quad \hbox{for freshwater or saltwater and}\quad P\to \infty,
\label{cb10}
\end{eqnarray}
for both freshwater and seawater.

In making Fig. (\ref{CP}), we have assumed that even if the changing partial pressure $P$ of CO$_2$ causes the solution to become supersaturated with respect to calcite, CaCO$_3$(s), brucite Mg(OH)$_2$ or other crystals, no precipitates are formed, as discussed in connection with Fig. \ref{MP}.

Comparing (\ref{cb8}) and (\ref{cb28}) we see that in the limit of low CO$_2$ partial pressure, $P\to 0$,  for every CO$_2$ molecule dissolved in freshwater,  143 thousand CO$_2$ molecules  will dissolve in seawater,
\begin{equation}
\frac{[\hbox{DIC}]_{s0}}{[\hbox{DIC}]_{f0}}= 1.19\,\times\, 10^5\quad \hbox{for}\quad P\to 0.
\label{cb30}
\end{equation}
From (\ref{cb10}) and Table \ref{K012}  we see that in the limit of high CO$_2$ partial pressure, $P\to \infty$,  for every CO$_2$ molecule dissolved in freshwater, 0.835  CO$_2$ molecules  will dissolve in seawater,
\begin{equation}
\frac{[\hbox{DIC}]_{s0}}{[\hbox{DIC}]_{f0}}=\frac{K_{0s}}{K_{0f}}.
=0.835
\quad \hbox{for}\quad P\to \infty,
\label{cb32}
\end{equation}
At low CO$_2$ partial pressures, $P$, and a temperature of $T=25$ C,  CO$_2$ is orders-of-magnitude more soluble in seawater than freshwater.
At high CO$_2$ partial pressures, $P$, and a temperature of $T=25$ C,  CO$_2$ is slightly less soluble in seawater than freshwater.
\begin{figure}[t]
\begin{centering}
\includegraphics[height=90mm,width=1\columnwidth]{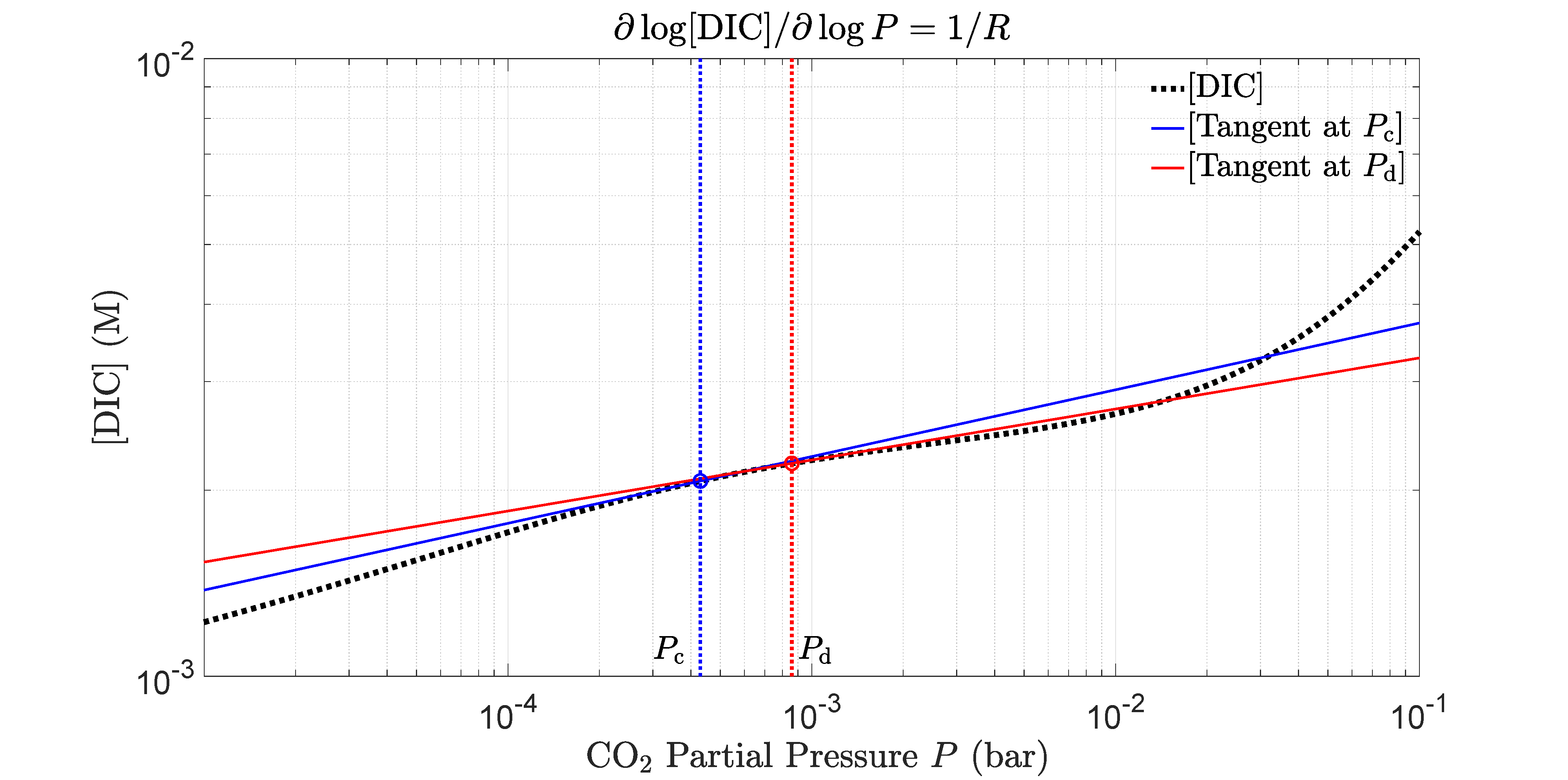}
\caption {The inverse Revelle factor $1/R$ of (\ref{rf2}) is the slope of the tangent line to  a log-log plot of the dissolved inorganic carbon [DIC] versus the CO$_2$ partial pressure $P$. For this example of seawater at a temperature of 25 C, the Revelle factors at the contemporary partial pressure $P_{\rm c} = 430\ \mu$b , and double that value, $P_{\rm d} = 860\ \mu$b,  are $R_{\rm c} =9.3$ and $R_{\rm d} = 12.1$.}
\label{Revelle2}
\end{centering}
\end{figure}
\begin{figure}[t]
\begin{centering}
\includegraphics[height=90mm,width=1\columnwidth]{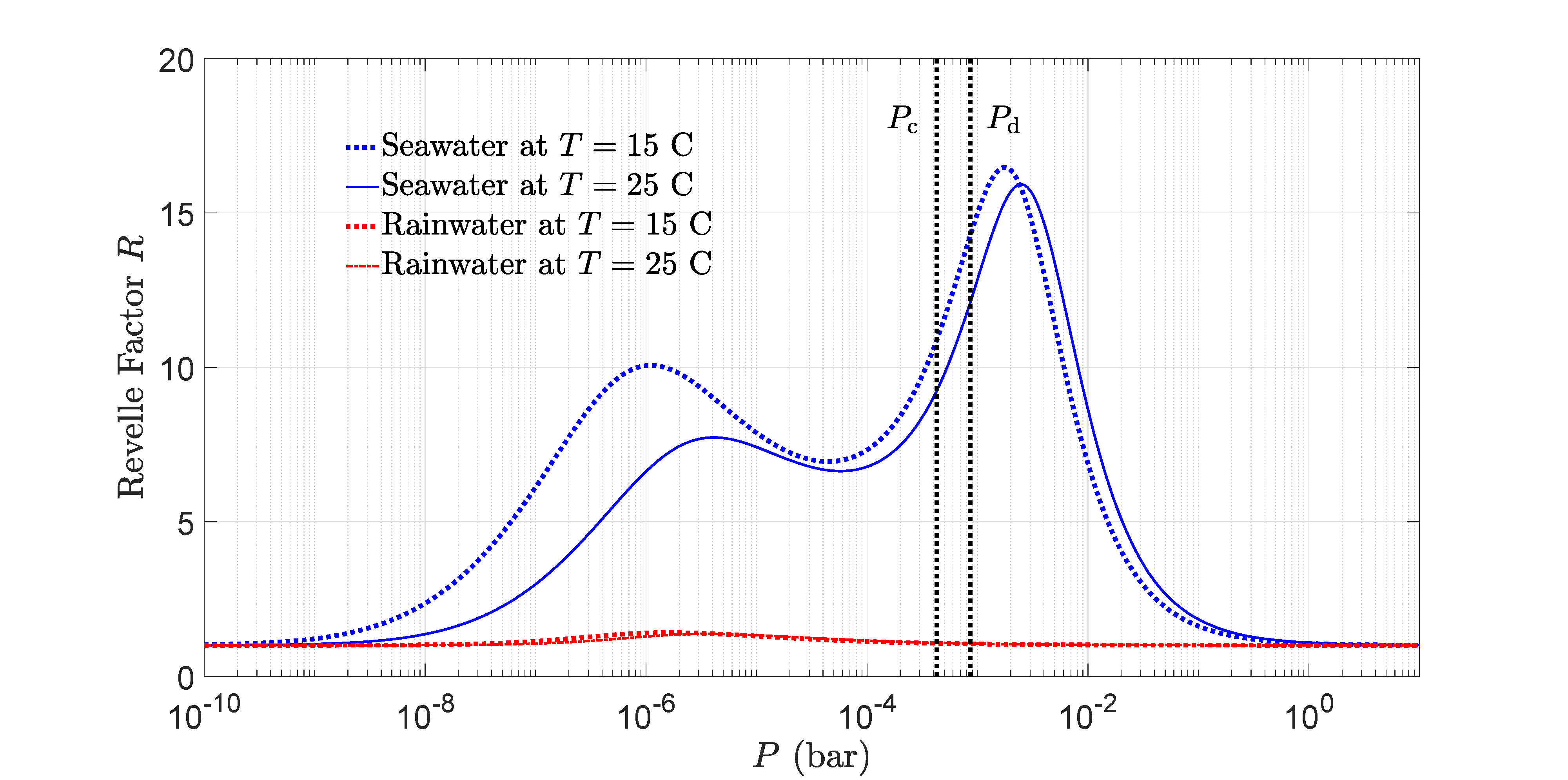}
\caption {The Revelle factor $R$ of (\ref{rf2}) measures how hard it is for atmospheric CO$_2$ to dissolve in water.  Alkaline water sucks in the weak acid CO$_2$. But more dissolved CO$_2$ makes the water less alkaline and less able retain CO$_2$.}
\label{Revelle1}
\end{centering}
\end{figure}
\subsection{The Revelle factor\label{rf}}
Fig. \ref{CP} shows that for intermediate CO$_2$  partial pressures, $10^{-7}\hbox{ bar}<P<10^{-1}\hbox{ bar}$, the rate of increase $\log [\hbox{DIC}]$ with increasing $\log P$  is around ten times slower than for higher or lower pressures. This slowing down is described quantitatively by  {\it Revelle factor}, $R$, defined by
\begin{equation}
\frac{1}{R} = \frac{\partial\ln[\hbox{DIC}]}{\partial \ln P}= \frac{\partial\log[\hbox{DIC}]}{\partial \log P}.
\label{rf2}
\end{equation}
In other words, the inverse Revelle factor $R$ defined by (\ref{rf2}) is the slope of the tangent line to  a log-log plot of the dissolved inorganic carbon [DIC] versus the CO$_2$ partial pressure $P$, as illustrated in Fig. \ref{Revelle2}.

The Revelle factor is the ratio of a small fractional change,  $\partial \ln(P)=\partial P/P$ of the CO$_2$ partial pressure $P$, to the corresponding small fractional change, $\partial \ln [{\rm DIC}]=\partial\, [{\rm DIC}]/[{\rm DIC}]$, of dissolved inorganic carbon, DIC. The partial derivative symbols $\partial$ indicate that the temperature, salinity, alkalinity and other properties of the water are held constant as $P$ and $[\hbox{DIC}]$ change. Revelle factors for the contemporary CO$_2$ partial pressure $P_{\rm  c} = 430\ \mu$b\, and for twice that value,  $P_{\rm  d} = 860\ \mu$b, are shown in Fig. \ref{Revelle2}.  We have plotted the Revelle factor (\ref{rf2}) versus a large range of partial pressure $P$ of CO$_2$ in Fig. \ref{Revelle1}.

\begin{figure}[t]
\begin{centering}
\includegraphics[height=90mm,width=1\columnwidth]{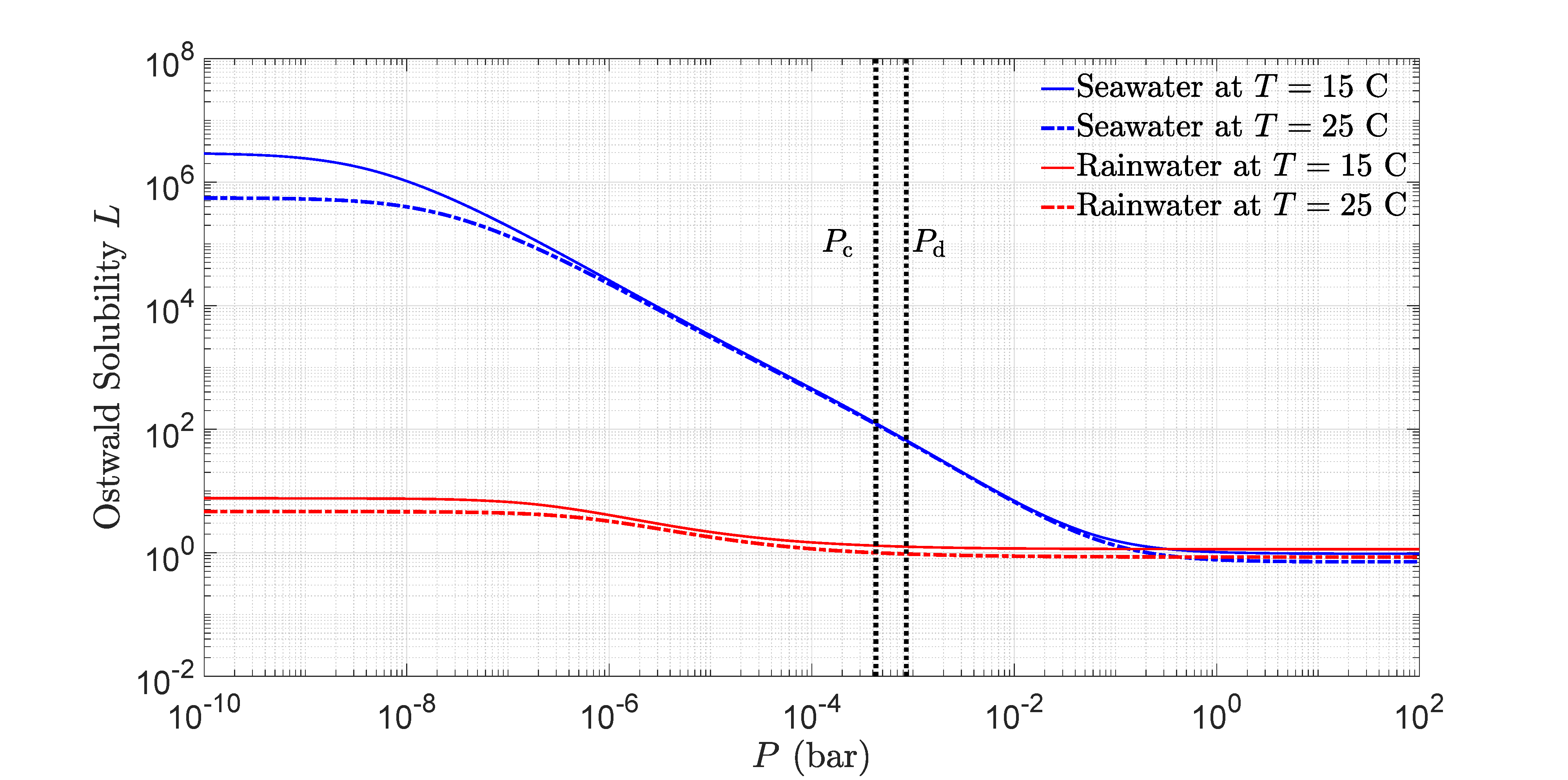}
\caption {The Ostwald Solubilities $L$ of (\ref{os2}) for seawater (blue) and freshwater (red) for a total pressure  $p=1$ bar and for temperatures $T=15$ C and $T = 25$ C. The solubilities are ratios of the equilibrium number density of inorganic carbon atoms in water $\{$DIC$\}$, to the number density $\{$CO$_2\}_g$ in the atmosphere above.  The partial pressure of CO$_2$ is $P$, the contemporary  value is $P_{\rm c} = 430\ \mu$b\, and the doubled value is $P_{\rm d} = 860\ \mu$b.  For seawater the total alkalinity is [A] = 2.4 mM, the boron molality is [B] = 0.43 mM, and the salinity is $S = 35$\textperthousand.  Rainwater has only dissolved CO$_2$.}
\label{Ostwald1}
\end{centering}
\end{figure}
\subsection{The Ostwald solubility\label{os}}
For basic values of the pH,  like those of the oceans or soda lakes where pH  $> 7$,  Fig. \ref{CP} shows that most of the dissolved CO$_2$  is in the form of bicarbonate  ions ${\rm HCO}_3^-$ or carbonate ions ${\rm CO}_3^{2-}$.  High-pH, basic water can absorb  orders of magnitude more CO$_2$ from the air than low-pH, acidic water. For example, at a very small CO$_2$ partial pressure $P = 10^{-10}$ bar and  at $T= 25$ C,   the seawater molality of Fig. \ref{CP} is [DIC] = $ 2.12\,\times\, 10^{-6}$ M, about five orders of magnitude greater than freshwater at $T=15$ C in Fig. \ref{CPf} with [DIC] = $1.85\,\times\, 10^{-11}$ M.   An informative measure of solubility of gaseous CO$_2$ in water is the dimensionless {\it Ostwald coefficient}\,\cite{Ostwald},
\begin{equation}
L = \frac{\{\hbox{DIC}\}}{\,\,\,\{{\rm CO_2}\}_g}.
\label{os2}
\end{equation}
In (\ref{os2}) we have denoted the molarity (moles per liter) of dissolved inorganic carbon by $\{{\rm DIC}\}$ and the molarity of CO$_2$ in the atmosphere  by $\{{\rm CO_2}\}_g$.
For our purposes, the numerical values of the molarities of dissolved species in moles per liter ($\ell$), and the  molalities in moles per kilogram will be nearly the same.  So  it will be sufficiently accurate to write
\begin{equation}
\{\hbox{DIC}\} =[\hbox{DIC}]\frac{{\rm kg}}{\ell},\quad \{{\rm H}^+\} =[{\rm H}^+] \frac{{\rm kg}}{\ell},\quad\hbox{etc.}
\label{os4}
\end{equation}
We can write laws of mass action in terms  of molarities instead of molalities. For example, the uncharged-bicarbonate and  the bicarbonate-carbonate equilibrium equations of (\ref{in34a}) and (\ref{in42}) become
\begin{equation}
 \{{\rm H}^+\}\{{\rm HCO}_3^-\}=\tilde K_{1}\{\hbox{CO$_2^*$}\}\quad\hbox{and}\quad  \{{\rm H}^+\}\{{\rm CO}_3^{2-}\}=\tilde K_{2}\{{\rm HCO}_3^-\},
\label{os6}
\end{equation}
where
\begin{equation}
\tilde K_1 =K_1\frac{{\rm kg}}{\ell}\quad\hbox{and}\quad \tilde K_2 =K_2\frac{{\rm kg}}{\ell}.
\label{os8}
\end{equation}

We can use the ideal gas law to write the molarity of CO$_2$ in the gas phase in terms of the partial pressure $P$ as
\begin{equation}
\{\hbox{CO}_2\}_g=100\frac{ P }{N_{\rm A}k T}.
\label{os10}
\end{equation}
In (\ref{os10}) the prefactor of $100$ includes a factor  $10^{5}$ to convert our pressure unit, bar, to the MKS unit Pa (Pascal) and a factor of $10^{-3}$ to convert  
mol/m$^{3}$ to mol/${\ell}$. Boltzmann's constant $k$ is
\begin{equation}
 k=1.38\,\times\, 10^{-23} \hbox{ J K}^{-1},
\label{ss19}
\end{equation}
and $T$ is the absolute temperature in K.  Avogadro's number,  $N_{\rm A}$, was given by (\ref{int4}).
\begin{table}
\begin{center}
\begin{tabular}{|l|l|l|l|l|}
\hline
&$S = 0$\textperthousand &$S = 0$\textperthousand &$S = 35$\textperthousand &$S = 35$\textperthousand\\
&$T=15$ C& $T=25$ C&$T=15$ C& $T=25$ C\\
\hline\hline
$L_0$&7.28&4.59&$2.83\,\times\, 10^6$&$5.48\,\times\, 10^5$\\
\hline
$L_{\rm c}$&1.29&0.990&124 &119\\
\hline
$L_{\rm d}$&1.24&0.947&65.6&63.7\\
\hline
$L_{\infty}$&1.13&0.843&0.922&0.740\\
\hline
\end{tabular}
\end{center}
\caption{Numerical values of the Ostwald solubility coefficients $L$ from (\ref{os12}) for rainwater and seawater at  temperatures of $T=15$ C and $T= 25$ C, and at a total pressure $p = 1$ bar.  $L_0$, $L_c$, $L_d$ and $L_{\infty}$ are values for the CO$_2$ partial pressures $P=0$, $P=P_{\rm c}$, $P=P_{\rm d}$ and $P=\infty$.  For seawater with salinity $S = 35$\textperthousand, alkalinity [A] = 2.4 mM and boron molality [B] = 0.43 mM,  the  values of $[{\rm H}^+]$ produced by the CO$_2$ partial pressure $P$  are the solutions of (\ref{al34}).  For freshwater  with $S=0$, and [A] = 0, the value of $[{\rm H}^+]$ came from the  solution of (\ref{al38}). 
\label{L}}
\end{table}

A general expression for the Ostwald solubility (\ref{os2}) follows from (\ref{if2}), (\ref{os4}),  (\ref{os10}),  (\ref{os8})
\begin{equation}
L=\bigg([{\rm H}^+]^2+ K_1[{\rm H}^+]+K_2 K_1\bigg) \frac{\tilde K_0 N_{\rm A} k T}{100 [{\rm H}^+]^2}.
\label{os12}
\end{equation}
The limiting value  of (\ref{os12}) for high CO$_2$ partial pressures, $P\to \infty$ and $[{\rm H}^+]\to \infty$ is
\begin{equation}
L_{\infty}=\frac{\tilde K_0 N_{\rm A} k T}{100}.
\label{os14}
\end{equation}
The limiting value  of (\ref{os12}) for low CO$_2$ partial pressures, $P\to 0$ and $[{\rm H}^+]\to [{\rm H}^+]_0$ is
\begin{equation}
L_{0}=\bigg(1+\frac{ K_1[{\rm OH}^-]_0}{K_{\rm w}}+\frac{K_2 K_1[{\rm OH}^-]_0^2}{K_{\rm w}^2}\bigg) \frac{\tilde K_0 N_{\rm A} k T}{100}.
\label{os14a}
\end{equation}
From (\ref{cb2}) and (\ref{cb22}) we see that the limiting hydroxyl-ion molalities for $P\to 0$ are
\begin{equation}
[{\rm OH}^-]_0=\left \{\begin{array}{cl}\sqrt{K_{\rm w}}&\hbox{for freshwater} ,\\
$[{\rm A}] - [{\rm B}]$&\hbox{for seawater}. \end{array}\right .
\label{os14b}
\end{equation}

The dependence of the Ostwald coefficient $L$ of (\ref{os2}) or (\ref{os12})  on CO$_2$ partial pressure $P$ is shown in Fig. \ref{Ostwald1}. Representative numerical values of $L$ are shown in Table \ref{L}. Note the huge decrease in the Ostwald coefficient for seawater at a temperature $T = 25$ C  from the limiting value  $L_0= 5.48\times 10^5$ for very low partial pressures  of CO$_2$, $P\ll 10^{-10}$ bar,  to the limiting value $L_{\infty}= 0.740$ for $P\gg 1$ bar. For rainwater at a temperature $T = 25$ C, the limiting values are  $L_0=4.59$ for low CO$_2$ partial pressure and $L_{\infty} = 0.843$ for high pressure, a very much smaller difference than for seawater.

Taking logarithms of both sides of (\ref{os2}) and using (\ref{os10})
\begin{equation}
\ln L = \ln \{\hbox{DIC} \}-\ln P -\ln 100+\ln(N_{\rm A}k T).
\label{rf4}
\end{equation}
Differentiating (\ref{rf4})  at constant temperature $T$ we find
\begin{equation}
\partial \ln L = \partial \ln \{\hbox{DIC} \}-\partial \ln P. 
\label{rf6}
\end{equation}
Dividing both sides of (\ref{rf6}) by $\partial \ln P$ we find that $L$, $P$ and $R$ are related by
\begin{equation}
\frac{\partial \ln L}{\partial \ln P} =\frac{\partial \log L}{\partial \log P} =  \frac{1}{R}-1. 
\label{rf8}
\end{equation}
\begin{figure}[t]
\begin{centering}
\includegraphics[height=90mm,width=1\columnwidth]{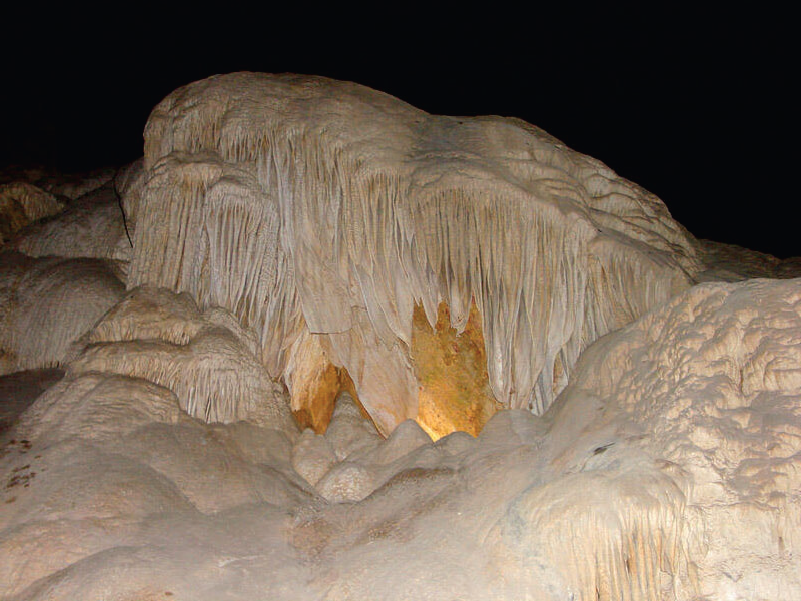}
\caption {Flowstone formations in Carlsbad Caverns, New Mexico\,\cite{Carlsbad}. Such deposits are formed by transport of CaCO$_3$, dissolved from limestone sediments by carbonated groundwater with a high partial pressure of CO$_2$, $P\gg P_{\rm c}=430\ \mu$b. This water originates as rain or snow, and it is carbonated by oxidation of organic matter in the soil. When the water comes back into contact with the air inside the cave, a little more than half of the dissolved inorganic carbon is degassed as CO$_2$ and a little less than half is precipitated as calcite crystals, CaCO$_3$(s). Details of  how this happens are  outlined in Fig. \ref{Limestone2} below. (carlsbad.eps)} 
\label{carlsbad}
\end{centering}
\end{figure}
\section{Groundwater \label{LL}}
Groundwaters can have much larger partial pressures of CO$_2$ than surface waters\,\cite{Macpherson}. The respiration of plant roots, or the oxidation of humus and other organic matter by soil microorganisms, can inject CO$_2$ into groundwater and into the air of soil pores.  For example, a representative partial pressure of CO$_2$ in air exhaled from human lungs can be taken to be about
\begin{equation}
P_{\rm b} = 100 P_{\rm c}= 43,000\, \mu{\rm b},
\label{gw0}
\end{equation}
one hundred  times larger than that of the well-mixed, free atmosphere. If acidic,  biologically-carbonated groundwater percolates through limestone, CaCO$_3$(s), or dolomite, CaMg(CO$_3$)$_2$(s), some of the rock can dissolve. This adds more dissolved inorganic carbon (DIC) to the solution, as well as Ca$^+$ and Mg$^+$ ions, which increases the alkalinity of the water. The carbonate chemistry of groundwater leads to remarkable terrestrial phenomena like Karst topography, travertine formations, caves, stalagmites,  flowstone, {\it etc.}, which we will discuss in more detail in Section {\bf \ref{tgw}}.  An example\,\cite{Carlsbad} of the transport of CaCO$_3$ by groundwater is shown in Fig. \ref{carlsbad}.

\begin{figure}[t]
\begin{centering}
\includegraphics[height=90mm,width=1\columnwidth]{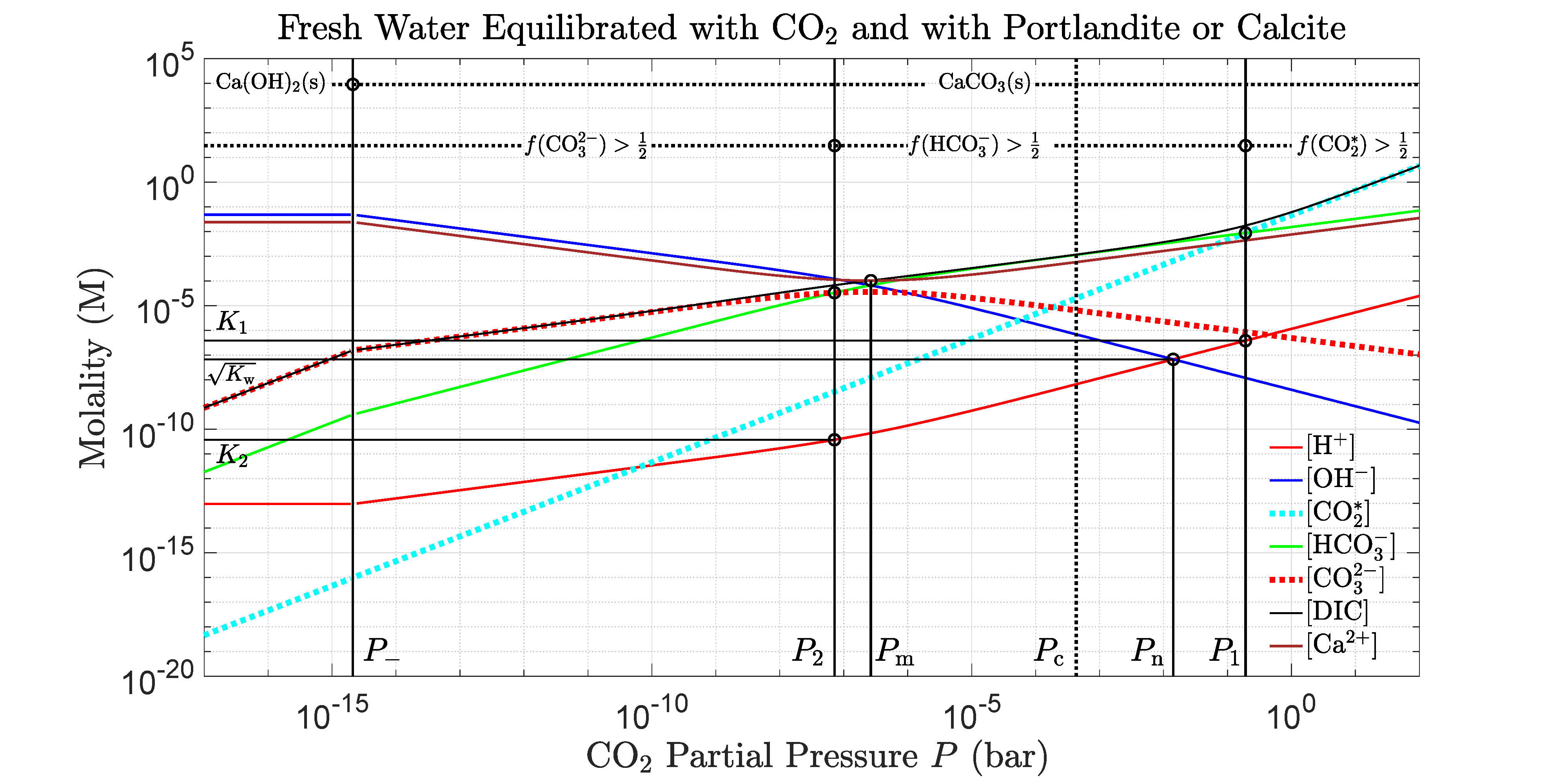}
\caption {Saturated solutions of groundwater, at the temperature $T= 15$ C in equilibrium with gaseous CO$_2$(g) at a partial pressure $P$. For extremely low partial pressures,  $P<P_{-}= 2.11\times 10^{-15} $ bar, the solution is in equilibrium with portlandite, Ca(OH)$_2$(s), cystals.  For higher pressures, $P>P_{-}$ the solution is in equilibrium with calcite, CaCO$_3$(s), crystals. For low pressures, $P<P_2 = 7.17\times 10^{-8} $ bar, most dissolved CO$_2$ is carbonate ions CO$_3^{2-}$.   For intermediate pressures, $P_2<P<P_1 =1.91\times 10^{-1} $ bar, most dissolved CO$_2$ is bicarbonate ions HCO$_3^{-}$. For  high pressures, $P_1<P$ most dissolved CO$_2$ is uncharged, dissolved molecules, CO$_2^{*}$.  At the partial pressure $P_{\rm m} = 2.60 \times 10^{-7}$ bar,  the  minimum calcium molality  is [Ca$^{2+}$]$_{\rm m} = 1.02 \times 10^{-4}$ M and the maximum carbonate molality is [CO$_3^{2+}$]$_{\rm m} = 3.62\times 10^{-5}$ M.  The CO$_2$ partial pressure of the sealevel atmosphere in the year 2025 was $P_c = 430\ \mu$b.}
\label{Limestone15}
\end{centering}
\end{figure}
\subsection{Saturated CaCO$_3$(s) solutions \label{LL}}
Consider groundwater in limey strata.  We will assume the water is a saturated solution of calcite cyrstals,  CaCO$_3$(s), with negligibly small salinity, $S = 0$\textperthousand, and with a temperature $T= 15$ C, close to that of many caves. Then the only pH-independent ions in groundwater are the Ca$^{2+}$ ions from the reaction (\ref{ss2}).  The alkalinity (\ref{al6}) becomes
\begin{eqnarray}
[{\rm A}]&=&\hbox{2[Ca$^{2+}$]}\nonumber\\
&=&\frac{2K_{\rm sp}}{\hbox{[CO$_3^{2-}$]}}\nonumber\\
&=&\frac{2K_{\rm sp}[{\rm H}^+]^2}{K_2K_1K_0 P }.
\label{LL6}
\end{eqnarray}
The second line of (\ref{LL6}) follows from (\ref{ss4}) and the third line follows from (\ref{in46}). Using (\ref{LL6}) in the basic equation (\ref{al32}), and assuming negligibly small boron molality, [B] = 0, we find the quartic polynomial equation in $[{\rm H}^+]$,
\begin{equation}
\frac{2K_{\rm sp}}{K_2K_1K_0 P} [{\rm H}^+]^4+
[{\rm H}^+]^3-\left(K_{\rm w}+ K_1K_0  P \right)[{\rm H}^+] - 2K_2K_1K_0 P=0.
\label{LL8}
\end{equation}
Alternatively, we can think of (\ref{LL8}) as a quadratic polynomial in $P$,
\begin{equation}
\bigg(2K_2K_1K_0+K_1K_0[{\rm H}^+]\bigg) P^2 + \bigg(K_w[{\rm H}^+]-[{\rm H}^+]^3\bigg)P-\frac{2K_{\rm sp}[{\rm H}^+]^4}{K_2K_1K_0}=0.
\label{LL10}
\end{equation}
In (\ref{if10}) we showed that for $[{\rm H}^+]=K_1$, the molalities [CO$_2^*$] of uncharged species and $[{\rm HCO}_3^-]$ of bicarbonate ions are equal. Setting $[{\rm H}^+]=K_1$ 
in (\ref{LL10}), we use the {\it quadratic equation} together with the equilibrium constants for $S = 0$\textperthousand\ and $T= 15$ C from Table \ref{K012} to find the physically acceptable (positive) solution
\begin{equation}
 P_1 = 1.91\times 10^{-1} \hbox{ bar}\quad\hbox{for}\quad [\hbox{CO$_2^*$}]=[\hbox{HCO$_3^-$}].
\label{LL14}
\end{equation}
In (\ref{if12}) we showed that for $[{\rm H}^+]=K_2$, the molalities $[{\rm HCO}_3^-]$ of bicarbonate ions and  $[{\rm CO}_3^{2-}]$ of carbonate ions are equal. 
Setting $[{\rm H}^+]=K_2$ 
in (\ref{LL10}), we use the {\it quadratic equation} to find the physically acceptable (positive) solution
\begin{equation}
 P_2 = 7.17\times 10^{-8} \hbox{ bar}\quad\hbox{for}\quad [\hbox{HCO$_3^-$}]= [\hbox{CO$_3^{2-}$}].
\label{LL16}
\end{equation}
Setting   $[{\rm H}^+]= \sqrt{K_{\rm w}}$, the hydrogen-ion molality for neutral pH in freshwater, in (\ref{LL10}), we use the {\it quadratic equation} to find the physically acceptable (positive) solution
\begin{equation}
P_{\rm n} = 1.42\times 10^{-2} \hbox{ bar}, \quad\hbox{for}\quad \hbox{pH}=-\log \sqrt{K_{\rm w}}.
\label{LL18}
\end{equation}
Setting the contemporary CO$_2$ partial pressure, $P_{\rm c}=430\ \mu$b, in (\ref{LL8}) we use modern mathematical software to find the physically acceptable (positive) solution to the quartic equation,
\begin{equation}
[{\rm H}^+]_{\rm c}=6.54 \times 10^{-9}\hbox{ M}\quad\hbox{or}\quad \hbox{pH}_{\rm c} = 8.18 \quad\hbox{for}\quad P_{\rm c}= 430\ \mu{\rm b}.
\label{LL22}
\end{equation}
At the contemporary partial pressure $P_{\rm c} = 430\ \mu$b\, of CO$_2$ and at the temperature $T=15$ C,  the pH (\ref{LL22}) of calcite-saturated freshwater is (accidently)  the same as the contemporary seawater pH, shown in Fig. \ref{Carbonated}, at the temperature $T=25$ C.

Fig. \ref{Limestone15} shows the hydrogen ion molalities [${\rm H}^+$] for a large range of the CO$_2$ partial pressure, $P$ for water saturated with calcite. The numbers come from solving  (\ref{LL8}) or (\ref{LL10}).  Also shown in Fig. \ref{Limestone15} are the molalities of other dissolved species that follow from the values of $[{\rm H}^+]$, $P$ and the equilibrium constants.  For example, [${\rm OH}^-$] follows from (\ref{in4}); [CO$_2^*$] follows from  (\ref{in24}); [${\rm HCO}_3^-$] and [${\rm CO}_3^{2-}]$ follow from (\ref{in38}) and (\ref{in46}); [DIC] follows from (\ref{if2});  [Ca$^{2+}$] follows from (\ref{ss4}). 

From inspection of Fig. \ref{Limestone15}, we see that the carbonate molality  [${\rm CO}_3^{2-}$] maximizes (and because of (\ref{ss4}) the calcium ion molality [Ca$^{2+}$] minimizes)  at a pressure 
\begin{equation}
P_{\rm m} = 2.65\times 10^{-7} \hbox{ bar}. 
\label{LL20}
\end{equation}
 For very large partial pressures, $P\gg P_{\rm m}$, carbonate ions are converted to bicarbonate ions, ${\rm HCO}_3^-$, or uncharged species, CO$_2^*$, and this drives down the carbonate molality [${\rm CO}_3^{2-}$] even though the dissolved inorganic carbon molality, [DIC], increases with increasing partial pressure.  For very small  partial pressures, $P\ll P_{\rm m}$, the solution degasses so much CO$_2$ that the carbonate molality [${\rm CO}_3^{2-}$] decreases, along with the other dissolved inorganic carbon molalities, [${\rm HCO}_3^-$] and [CO$_2^*$].
The maximum molality of carbonate ions, [${\rm CO}_3^{2-}$],  at the pressure $P_{\rm m}$ of (\ref{LL20}) is
\begin{equation}
[{\rm CO}_3^{2-}]_{\rm m}=  3.62\times 10^{-5} \hbox{ M}. 
\label{LL23}
\end{equation}
The minimum molality of calcium ions also occurs at the pressure $P_{\rm m}$ of (\ref{LL20}), 
\begin{equation}
[\hbox{Ca}^{2+}]_{\rm m}=  1.02\times 10^{-4} \hbox{ M}. 
\label{LL24}
\end{equation}
The curious minimum solubility of calcite in freshwater at a CO$_2$ partial pressure, $P_m\approx 10^{-6}$ bar, is shown by a table in reference\,\cite{Wiki}.

In Section {\bf \ref{bl}} we show algebraic methods to find the numerical values of $P_{\rm m}$ of (\ref{LL20}),  $[{\rm CO}_3^{2-}]_{\rm m}$ of (\ref{LL23}), and $[\hbox{Ca}^{2+}]_{\rm m}$ of (\ref{LL24}) from the values of the equilbrium constants of Table \ref{K012}.

\subsection{Saturated Ca(OH)$_2$(s) solutions\label{SL}}
Imagine that the partial pressure $P$  in equilibrium with a saturated solution of CaCO$_3$(s) is steadily reduced below the pressure $P_m$ of (\ref{LL20}) for maximum [${\rm CO}_3^{2-}$]. This will allow more CO$_2$ to outgas from the solution. This  diminishes the molalities of all dissolved species of CO$_2$, including  carbonate, [${\rm CO}_3^{2-}$]. The calcium ion molality  [Ca$^{2+}$] must increase to keep the solubility product (\ref{ss4}) constant.  Reducing the  molalities [${\rm CO}_3^{2-}$] and [${\rm HCO}_3^-$] reduces the negative-charge molality of the  solution at the same time that the positive-charge molality is increasing, due to the dissolution of more [Ca$^{2+}$]. To maintain charge neutrality, the molality of hydroxyl ions [${\rm OH}^-$] must increase with decreasing $P$.  The  increase of the molalities of both  [Ca$^{2+}$] and [${\rm OH}^-$] with decreasing $P$, for $P<P_{\rm m}$,  can be seen in Fig. \ref{Limestone15}.  Eventually the partial pressure will be reduced to a value $P= P_{-}$ where the solubility product of calcium hydroxide,  Ca(OH)$_2$(s) (portlandite or slaked lime), is reached. Then calcium hydroxide crystals, Ca(OH)$_2$(s), will begin to precipitate in the reaction (\ref{SL2}).
At the partial pressure $P= P_{-}$, attempts to further decrease $P$ cause calcium carbonate to convert to calcium hydroxide, and the solution will be saturated for both CaCO$_3$(s) and Ca(OH)$_2$(s) crystals. Then equations (\ref{SL4}) and (\ref{ss4}) must both be satisfied. We can equate the expressions that each of these equations gives for [Ca$^{2+}$] to find
\begin{equation}
\frac{K_{\rm sl}}{\hbox{[OH$^{-}$]$^2$}} =\frac{K_{\rm sp}}{[{\rm CO}_3^{2-}]}.
\label{SL6}
\end{equation}
Using (\ref{in4}) and (\ref{in46}) in (\ref{SL6}) we find that the partial pressure at which Ca(OH)$_2$(s) and CaCO$_3$(s) crystals can coexist in solution is
\begin{eqnarray}
P_{-}&=& \frac{K_{\rm sp}K_{\rm w}^2}{K_{\rm sl}K_2K_1K_0}\nonumber\\
&=&2.11\times 10^{-15}\hbox{ bar}.
\label{SL8}
\end{eqnarray}
The numerical value on the second line of (\ref{SL8}) follows from the equilibrium constants of from Table \ref{K012} with $T=15$ C and $S = 0$\textperthousand.  The very small partial pressure of (\ref{SL8}) corresponds to an ultra high vacuum. It quantifies the well-known fact that quicklime is an excellent sorbent for CO$_2$. As long as the solution contains both CaCO$_3$(s) and Ca(OH)$_2$(s) crystals, further removal  of CO$_2$ from the atmosphere above the solution will produce no further decrease of $P$. CO$_2$ molecules removed from the atmosphere are replaced by CO$_2$ molecules degassed from the solution. CO$_2$ molecules degassed from the solution are replaced by dissolving CaCO$_3$(s) crystals.  The excess  Ca$^{2+}$ from the dissolving CaCO$_3$ is  precipitated as Ca(OH)$_2$(s). Only after all of the CaCO$_3$(s) has been converted to Ca(OH)$_2$(s) will further removal of CO$_2$ from the atmosphere lead to further decreases of $P$. 

For partial pressures $P<P_{-}$, we assume that all CaCO$_3$(s) crystals have been converted to Ca(OH)$_2$(s) crystals.  Then the alkalinity of (\ref{al6}) becomes
\begin{eqnarray}
[{\rm A}]&=&\hbox{2[Ca$^{2+}$]}\nonumber\\
&=&\frac{2K_{\rm sl}}{[{\rm OH}^-]^2}\nonumber\\
&=&\frac{2K_{\rm sl}[{\rm H}^+]^2}{ K_{\rm w}^2}.
\label{SL10}
\end{eqnarray}
The second line of (\ref{SL10}) follows from (\ref{SL4}) and the third line follows from (\ref{in4}). Substituting the alkalinity $[{\rm A}]$ from (\ref{SL10}) into groundwater polynomial equation (\ref{al38}), we find the quartic polynomial equation
\begin{equation}
\frac{2K_{\rm sl}[{\rm H}^+]^4 }{K_{\rm w}^2}+
[{\rm H}^+]^3-\left(K_{\rm w}+ K_1K_0 P \right)[{\rm H}^+] - 2K_2K_1K_0 P=0.
\label{SL12}
\end{equation}
The physically acceptable solutions $[{\rm H}^+]$ of (\ref{SL12}), and the molalities of other dissolved species that follow, are shown for $P<P_{-}$ on  the left of Fig. \ref{Limestone15}. 

From inspection of Fig. \ref{Limestone15}, we see that for $P<P_{-}$ the equilibrium constants of Table \ref{K012} are such that the tiny amount of dissolved CO$_2$ has almost no effect on the saturated solution of limewater, for which the molalities of calcium and hydroxyl ions dominate the charge density of the solution, and therefore are given by (\ref{SL4}) as 
\begin{equation}
[\hbox{Ca}^{2+}]=\left(\frac{K_{\rm sl}}{4}\right)^{1/3}\quad\hbox{and}\quad[\hbox{OH}^{-}]=\left(2K_{\rm sl}\right)^{1/3}.
\label{SL14}
\end{equation}
The molalities of the three species of dissolved CO$_2$ follow from (\ref{SL14}), (\ref{in4}), (\ref{in24}), (\ref{in38}) and (\ref{in46}),   
\begin{equation}
\hbox{[CO$_2^*$]}=K_0P,\quad \hbox{[HCO$_3^-$]}=\frac{\left(2K_{\rm sl}\right)^{1/3}K_1K_0}{K_{\rm w}}P,\quad\hbox{and}\quad
\hbox{[CO$_3^{2-}$]}=\frac{\left(2K_{\rm sl}\right)^{2/3}K_2K_1K_0}{K_{\rm w}^2}P.
\label{SL16}
\end{equation}

\begin{figure}[t]
\begin{centering}
\includegraphics[height=90mm,width=1\columnwidth]{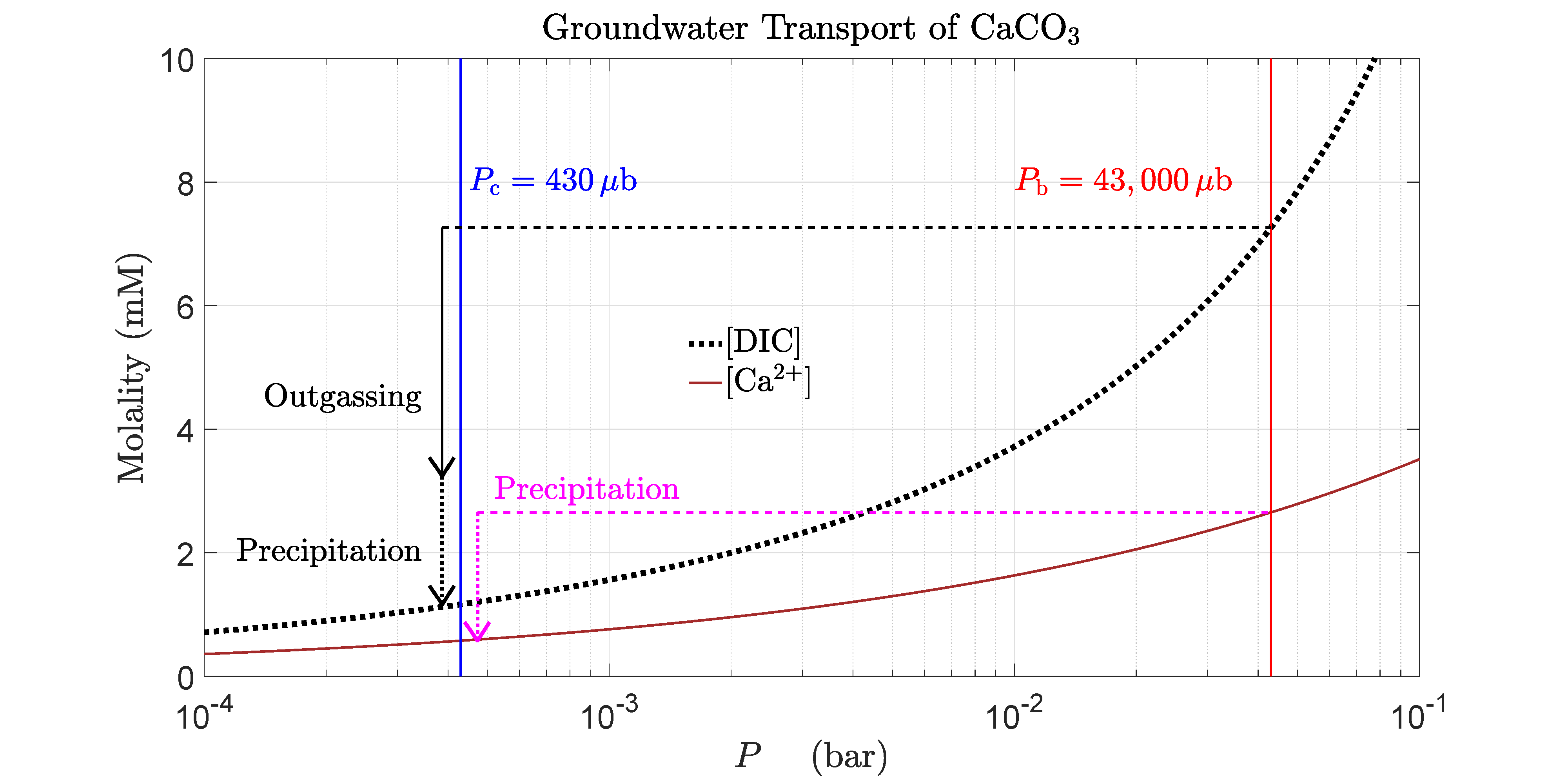}
\caption {A magnified part of Fig. \ref{Limestone15}, with a linear vertical scale. This shows how groundwater at a temperature $T=15$ C from limestone strata can form flowstone like that of Fig. \ref{carlsbad} in caves and spring outlets. CaCO$_3$(s) is more soluble at the higher partial pressures $P$ of CO$_2$ that often characterize groundwater, in this case $P_{\rm b} = 43,000\ \mu$b, than at contemorary atmospheric partial pressure, $P_{\rm c} = 430\ \mu$b. The emergent groundwater must precipitate part of the dissolved Ca$^{2+}$ and CO$_3^{2-}$ ions onto  CaCO$_3$(s) flowstone, and release part of the dissolved CO$_2$ to the atmosphere.  See the text for more discussion.}
\label{Limestone2}
\end{centering}
\end{figure}
\subsection{CaCO$_3$ transport by groundwater \label{tgw}}
A hypothetical example of CaCO$_3$ transport in groundwater is shown in Fig. \ref{Limestone2}, which is a magnified version of part of Fig. \ref{Limestone15}.  We assume that groundwater at a temperature $T= 15$ C is carbonated to $P_{\rm b}= 43,000\ \mu$b, one hundred times the pressure in the year 2025, and about the same partial pressure of CO$_2$ as  in exhaled human breath. The source of the CO$_2$ is respiration by plant roots and aerobic decomposition of organic matter by soil microorganisms. The water flows through limey strata and becomes  saturated with Ca$^{2+}$ and ${\rm CO}_3^{2-}$ with respect to calcite CaCO$_3$(s). The groundwater emerges into a cave, like that of Fig. \ref{carlsbad}, where it is exposed to atmospheric air of partial pressure $P_{\rm c} = 430\ \mu$b\, of CO$_2$. To reestablish equilibrium between the water, air, and flowstone,  a little less than half of the dissolved inorganic carbon must be precipitated as calcite flowstone, CaCO$_3$(s), as we will show below, 0.207 g per kg of groundwater. The remainder is released to the atmosphere as CO$_2$ gas, in this case 96 cm$^3$  kg$^{-1}$ of pure CO$_2$ at a pressure of $p = 1$ bar and temperature of $T= 15$ C. 

To get these numbers we use the molalilities of ions and molecules that can be calculated  as a function of $P$ with (\ref{LL8}), (\ref{in4}), (\ref{in24}), (\ref{in38}), (\ref{in46}), (\ref{if2}), and (\ref{ss4}). As can be seen from Fig. \ref{Limestone2}, we find that the molality of the dissolved inorganic carbon in the carbonated groundwater at the CO$_2$ partial pressure $P_{\rm b}$ is
\begin{equation}
[\hbox{DIC}]_{\rm b}= 7.26 \hbox{ mM}.
\label{gwt2}
\end{equation}
After equilibration with flowstone and with atmospheric CO$_2$, having a partial pressure $P_{\rm c} = 430\ \mu$b, the molality decreases to
\begin{equation}
[\hbox{DIC}]_{\rm c}= 1.16 \hbox{ mM}.
\label{gwt4}
\end{equation}
The incremental decrease is 
\begin{eqnarray}
\Delta[\hbox{DIC}]&=&[\hbox{DIC}]_{\rm b}-[\hbox{DIC}]_{\rm c}\nonumber\\
&=&6.10 \hbox{ mM}.
\label{gwt4}
\end{eqnarray}
Part of $\Delta[\hbox{DIC}]$, the decrease of dissolved inorganic carbon in the water, is outgassed as CO$_2$ into the cave air, and the remainder is deposited as additional flowstone CaCO$_3$(s).
The part removed as flowstone can be calculated by noting that the calcium-ion molality of the carbonated groundwater is
\begin{equation}
[\hbox{Ca}^{2+}]_{\rm b}= 2.65 \hbox{ mM},
\label{gwt6}
\end{equation}
The emergent water that has equilibrated with the air and flowstone in the cave has a calcium-ion molality
\begin{equation}
[\hbox{Ca}^{2+}]_{\rm c}= 0.576 \hbox{ mM}.
\label{gwt8}
\end{equation}
The decrease is
\begin{eqnarray}
\Delta[\hbox{Ca}^{2+}]&=&[\hbox{Ca}^{2+}]_{\rm b}-[\hbox{Ca}^{2+}]_{\rm c}\nonumber\\
&=&2.07\hbox{ mM}.
\label{gwt10}
\end{eqnarray}
Every mole of CaCO$_3$(s) flowstone deposited requires a mole of dissolved inorganic carbon.
Therefore, the fraction of dissolved inorganic carbon that is removed as flowstone is
\begin{eqnarray}
f\big(\hbox{CaCO$_3$(s)}\big)&=&\frac{\Delta[\hbox{Ca}^{2+}]}{\Delta[\hbox{DIC}]}\nonumber\\
&=&0.340.
\label{gwt12}
\end{eqnarray}
We used (\ref{gwt10}) and (\ref{gwt4}) to write the number on the second line of (\ref{gwt12}). The mass of additional flowstone released per kilogram of emergent groundwater is therefore
\begin{eqnarray}
m(\hbox{CaCO$_3$})\Delta[\hbox{Ca}^{2+}]=0.207 \hbox { g kg}^{-1},
\label{gwt14}
\end{eqnarray}
where the gram molecular weight, $m(\hbox{CaCO$_3$})= 100$ g mole$^{-1}$, of CaCO$_3$ was given by (\ref{hc6}).  The fraction of dissolved inorganic carbon that is released as gaseous CO$_2$ molecules to the cave air is
\begin{eqnarray}
f\big(\hbox{CO$_2$}\big)&=&1-f\big(\hbox{CaCO$_3$(s)}\big)\nonumber\\
&=&0.660.
\label{gwt16}
\end{eqnarray}
In summary, for this representative example, about 2/3 of the dissolved inorganic carbon from groundwater is released as CO$_2$ gas into the cave air, and about 1/3 is released as calcite flowstone.

A grand example of CaCO$_3$ transport by groundwater is the Mississippi River. 
According to reference\,\cite{Mississippi} the Mississippi River  
releases approximately $1\times 10^{13}$ grams of carbon  per year to the atmosphere, as gaseous CO$_2$. It carries an approximately equal mass of carbon incorporated into bicarbonate ions, ${\rm HCO}_3^-$, into the Gulf of Mexico, along with the Ca$^{2+}$ and Mg$^{2+}$ ions released when the limestone and dolomite in its watershed are dissolved. Although these are large numbers, they are only one thousandth of that released from burning fossil fuels, about $1\times 10^{16}$ grams of carbon in the year 2025\,\cite{emissions}
\begin{figure}[t]
\begin{centering}
\includegraphics[height=90mm,width=.6\columnwidth]{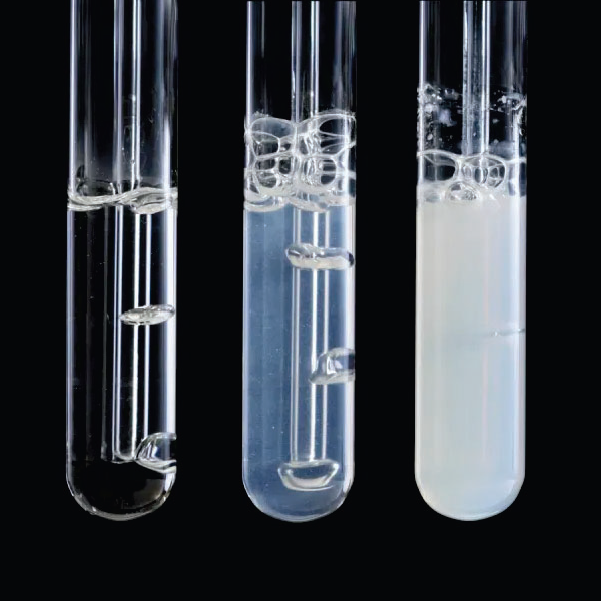}
\caption {A well known  experiment from reference\,\cite{breath} is to bubble human exhaled breath, which is about 4\% CO$_2$ by volume, through limewater.  The  test tube on the left shows initially clear limewater, a subsaturated solution of Ca(OH)$_2$. The middle test tube  is produced by the first few bubbles. It contains a sparse milky suspension calcite crystallites, CaCO$_3$(s).  Blowing still more bubbles makes the dense, milky suspension of the test tube on the right.}
\label{breath}
\end{centering}
\end{figure}
\begin{figure}[t]
\begin{centering}
\includegraphics[height=90mm,width=1\columnwidth]{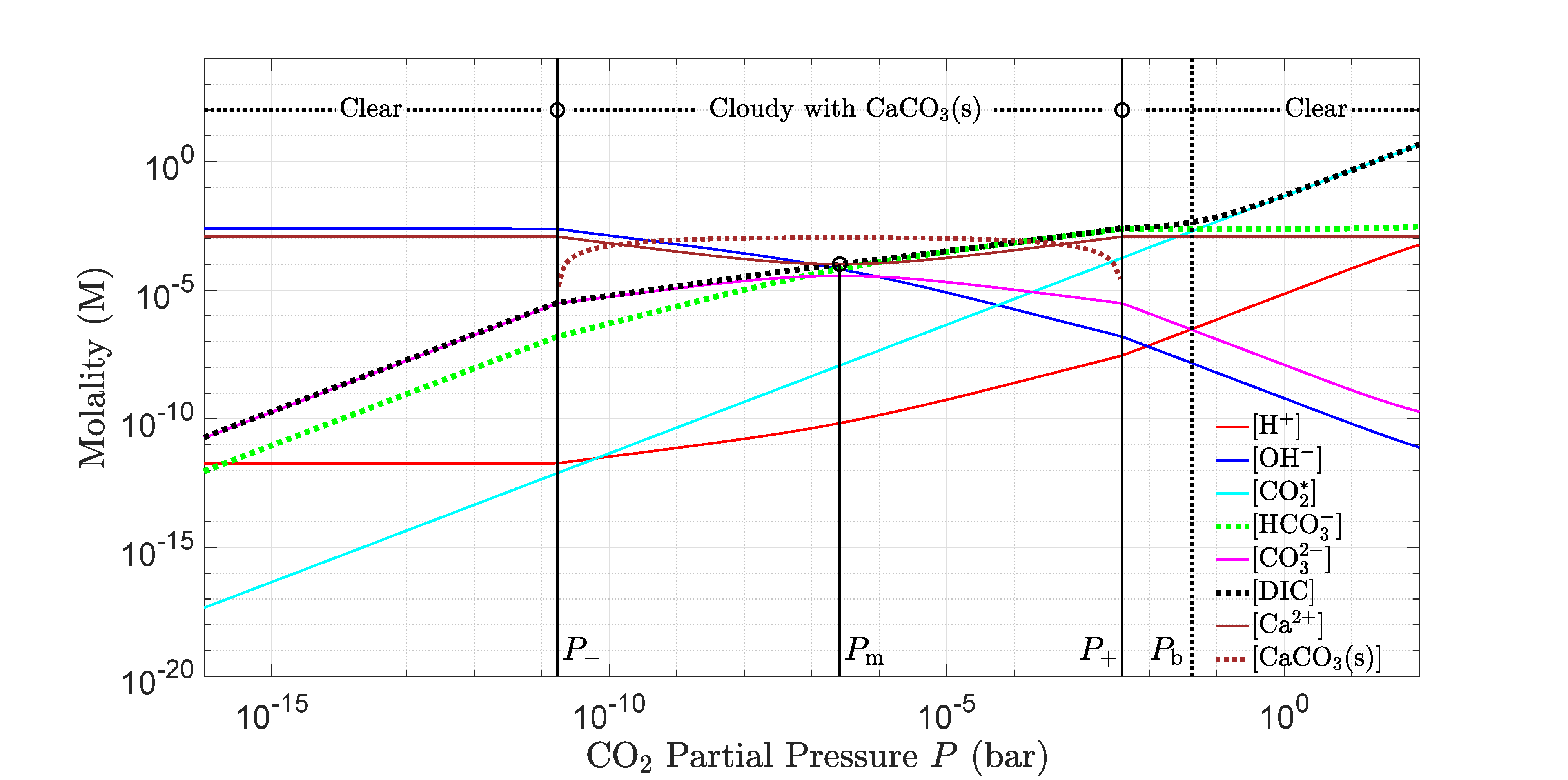}
\caption {A quantitative interpretation of Fig. \ref{breath}.  Limewater at a temperature, $T = 15$ C, and with  an initial alkalinity, [A]$_{\rm i}=2.4$ mM, due to calcium ions of initial molality [Ca$^{2+}$]$_{\rm i} = 1.2$ mM, is equilibrated with increasing partial pressures $P$ of gas-phase CO$_2$. For pressures $P>P_{-}=1.70\times 10^{-11}$ bar (but for $P<P_{+}$), the solution becomes supersaturated in calcite, CaCO$_3$(s), and a milky suspension of calcite crystals forms.  
At the pressure $P_{\rm m} = 2.60\, \times\, 10^{-7}$ bar,  the carbonate molality reaches a maximum value  [CO$_3^{2+}$]$_{\rm m} = 3.62\, \times\, 10^{-5}$ M. As $P$ increases beyond $P_{\rm m}$, the carbonate molality decreases because of the increasing acidity of the solution. For pressures $P>P_{+}=3.97\,\times\, 10^{-3}$ bar the carbonate molality becomes so small that the solution is subsaturated. Then all of the suspended CaCO$_3$(s) particulates dissolve and the solution becomes clear again. A representative   partial pressure of CO$_2$ in human breath  is  $P_{\rm b} = 4.3\,\times\, 10^{-2}$  bar. The text describes how the molalities are calculated as a function of $P$.}
\label{bubbles}
\end{centering}
\end{figure}

\subsection{Bubbles in limewater\label{bl}}
Closely related to the formation of flowstone by carbonated groundwater, which we discussed in the previous section, is the well known experiment where human breath bubbled through initially clear limewater causes clouds of milky white CaCO$_3$(s) crystallites   to form in the solution. Such an experiment is shown in Fig. \ref{breath}.

Limewater is a subsaturated solution of Ca(OH)$_2$.  Exhaled human breath is about 4\% CO$_2$. If one blows bubbles into lime water,  CO$_2$ from the bubbles dissolves in the highly alkaline lime water. Most of the CO$_2$ from the first few bubbles is converted to carbonate ions CO$_3^{2-}$. If the limewater has a sufficiently large molality of calcium ions, [Ca$^{2+}$],   the carbonate molality [CO$_3^{2-}$] can become large enough to exceed the solubility product  (\ref{ss4}).
Then a milky cloud of CaCO$_3$(s) crystallites forms in the solution. As one continues to blow bubbles, the molality of dissolved inorganic carbon increases, but the pH of the water steadily decreases. Bicarbonate ions and uncharged species become a larger fraction of the dissolved inorganic carbon, as shown in Fig. \ref{Fraction}. If the initial calcium ion molality, [Ca$^{2+}$], of the limewater was not too high, subsequent  bubbles can reduce the carbonate concentration enough that the solution is no longer saturated with respect to  CaCO$_3$(s). Then the milky suspension dissolves with further blowing, and the solution, now acidic because of all of the absorbed CO$_2$, becomes clear again.

A quantitative description of the bubble experiment is shown in Fig. \ref{bubbles}.  If the CO$_2$ partial pressure $P$ is small enough, $P<P_{-}$, or large enough, $P>P_{+}$, to prevent precipitation of CaCO$_3$(s),  the hydrogen-ion concentration,  $[{\rm H}^+]$, is given by (\ref{al38}), with a pH-independent-alkalinity
\begin{equation}
[{\rm A}]_{\rm i}=2[\hbox{Ca}^{2+}]_{\rm i}.
\label{bl2}
\end{equation}
For the intermediate range of  CO$_2$ partial pressures,
\begin{equation}
P_{-}<P<P_{+},\label{bl4}
\end{equation}
the solution ``clouds up" with calcite crystallites, since it is saturated with respect to calcite, CaCO$_3$(s). The alkalinity changes with changing partial pressure $P$ of CO$_2$ since  [Ca$^{2+}$]  decreases or increases as CaCO$_3$(s) crystals precipitate or dissolve. For this range of pressures, Section {\bf \ref{LL}} describes how to find the molalities of the various dissolved species.

At the upper and lower bounds $P=P_{\pm}$ of CO$_2$ partial pressures $P$, where the solution is saturated but no crystals of calcite remain undissolved in solution, the alkalinity will be [{\rm A}] = [{\rm A}]$_{\rm i}$ and (\ref{LL6}) implies that
\begin{equation}
P=\frac{2K_{\rm sp}[{\rm H}^+]^2}{K_2K_1K_0 [{\rm A}]_i}.
\label{bl4a}
\end{equation}
Substituting (\ref{bl4a}) into (\ref{LL8}) we find a quadratic equation for the hydrogen-ion concentration $[{\rm H}^+]$,
\begin{eqnarray}
0&=&\left(1-\frac{2K_{\rm sp}}{[{\rm A}]_{\rm i} K_2} \right)[{\rm H}^+]^2+
\left([{\rm A}]_{\rm i}-\frac{4K_{\rm sp}}{[{\rm A}]_{\rm i}} \right)[{\rm H}^+] - K_w\nonumber\\
&=&a[{\rm H}^+]^2+b[{\rm H}^+]+c.\label{bl10}
\end{eqnarray}
For typical limewater, the numerical values of the first two coefficients of the quadratic equation (\ref{bl10}) are both negative, $a<0$ and $b<0$. We can therefore write the largest solution, $[{\rm H}^+]_{+}$ and smallest solution   $[{\rm H}^+]_{-}$  of (\ref{bl10}) as
\begin{equation}
[{\rm H}^+]_{\pm}=\frac{-b\mp \sqrt{b^2-4ac}}{2a}.
\label{bl12}
\end{equation}
Substituting (\ref{bl12}) into (\ref{bl4a}) we find the upper and lower bounding pressures for cloudy solutions,
\begin{equation}
P_{\pm}=\frac{2K_{\rm sp}[{\rm H}^+]_{\pm}^2}{K_2K_1K_0[{\rm A}]_{\rm i}}.
\label{bl14}
\end{equation}
\subsection{Mininum dissolved Ca(OH)$_2$ to detect gaseous CO$_2$ with limewater}
If the calcium-ion molality [Ca$^{2+}$] of the initial limewater is too small, no CO$_2$ partial pressure $P$ will cause precipitation of calcite.  The critical concentration, [Ca$^{2+}$]$_{\rm m}$ = [A]$_{\rm m}/2$, below which no precipitate can be formed, is the value [A]$_{\rm i}$ = [A]$_{\rm m}$ that makes the discriminant, $\sqrt{b^2-4ac}$, of the quadratic formula (\ref{bl12}) equal to zero. This implies that [A]$_{\rm m}$ satisfies the quartic polynomial equation
\begin{equation}
0=[{\rm A}]_{\rm m}^4+\left(4K_{\rm w}-8K_{\rm sp}\right)[{\rm A}]_{\rm m}^2-\frac{8 K_{\rm w} K_{\rm sp}}{K_2}[{\rm A}]_{\rm m} +16 K_{\rm sp}^2.
\label{bl16}
\end{equation}
The corresponding carbonate molality will be
\begin{equation}
[\hbox{CO$_3^{2-}$}]_{\rm m}=\frac{2K_{\rm sp}}{[{\rm A}]_{\rm m}},
\label{bl18}
\end{equation}
and the corresponding value of $[{\rm H}^+]$ will be
\begin{equation}
[{\rm H}^+]_{\rm m} =-b/2a= \frac{4K_{\rm sp}-[{\rm A}]_{\rm m}^2}{2[{\rm A}]_{\rm m}-4K_{\rm sp}/K_2}.
\label{bl20}
\end{equation}
 Of the four  possible solutions [A]$_{\rm m}$ of (\ref{bl16}) only one is physically acceptable, the one for which  [A]$_{\rm m}$ and the corresponding $[{\rm H}^+]_{\rm m}$ of (\ref{bl20})  are both real, positive numbers,
\begin{equation}
\Re [{\rm A}]_{\rm m}>0, \quad\Re[{\rm H}^+]_{\rm m} >0,\quad\hbox{and}\quad\Im [{\rm A}]_{\rm m}= \Im [{\rm H}^+]_{\rm m} =0.
\label{bl22}
\end{equation}
For a temperature $T= 15$ C, the physically acceptable numerical solution  [A]$_{\rm m}$ of (\ref{bl22}) is 
\begin{equation}
[{\rm A}]_{\rm m}=0.205 \hbox{ mM}.
\label{bl24}
\end{equation}
The corresponding minimum concentration  of Ca(OH)$_2$ that must be dissolved in pure water  for the resulting limewater to detect CO$_2$ in air is
\begin{equation}
\hbox{[Ca(OH)$_2$]$_{\rm m}$}=\frac{1}{2}[{\rm A}]_{\rm m}=0.102 \hbox{ mM}\quad\hbox{or about}\quad 0.0075 \hbox{ g kg}^{-1},
\label{bl24}
\end{equation}
since the molecular weight of Ca(OH)$_2$ is 74 g mol$^{-1}$.
Fully saturated  limewater at a temperature of $T = 15$ C contains about 1.8 g kg$^{-1}$ of Ca(OH)$_2$\,\cite{Solubility_table}, some 240 times more than the minimum required value (\ref{bl24}).
\subsection{Hill Country Fare Texas Spring Water\label{hc}} 
Commercial, bottled water, {\it Hill Country Fare Texas Spring Water}, was used for instructive experiments in reference \cite{Buffering}.  Here we add some quantitative discussions of those experiments. The bottled water has been reported\,\cite{HEB} to have a pH of 6.9.  However, the addition of bromothymol blue generates a color consistent with a pH of 6.4 which we use in the analysis below.
\begin{equation}
 \hbox{pH}_{\rm bw}= 6.4\quad\to\quad [{\rm H}^+]_{\rm bw}=10^{-\rm pH_{\rm bw}}= 3.98 \times 10^{-7} \hbox{ M}.
\label{hc2}
\end{equation}
Here and later we will use the subscript ``bw" to refer to the bottled water.
The total alkalinity, expressed as an equivalent concentration of CaCO$_3$, is reported\,\cite{HEB} to be
\begin{equation}
(\hbox{CaCO$_3$})_{\rm bw}=53 \hbox{ mg L}^{-1}\approx 53 \hbox{ mg kg}^{-1} .
\label{hc4}
\end{equation}
The gram molecular weight of CaCO$_3$ is
\begin{equation}
m(\hbox{CaCO$_3$})=100  \hbox{ g mol$^{-1}$},
\label{hc6}
\end{equation}
so the molality of Ca$^{2+}$ ions in the spring water must be
\begin{eqnarray}
[\hbox{Ca}^{2+}]_{\rm bw}&=&\frac{(\hbox{CaCO$_3$})_{\rm bw}}{m(\hbox{CaCO$_3$})}=0.53  \hbox{ mM}.
\label{hc10}
\end{eqnarray}
If the only pH-independent ions in the water are doubly charged calcium ions, Ca$^{2+}$,
the total alkalinity (\ref{al6}) should have been
\begin{equation}
[\hbox{A}]_{\rm bw}=2 [\hbox{Ca$^{2+}$}]_{\rm bw}=1.06  \hbox{ mM}.
\label{hc8}
\end{equation}
This is less than half the value [A]= 2.4 mM  shown in Table \ref{table1} for seawater. 
Substituting the value of $[{\rm H}^+]$ from (\ref{hc2}) and the value of [A]$_{\rm bw}$ from (\ref{hc8}) into (\ref{al40}), with [B] = 0, and using equilibrium constants from Table\, \ref{table2} for a temperature $T=25$ C and vanishing salinity,  $S = 0$\textperthousand,  we see that the bottled water would have been in equilibrium with a CO$_2$ partial pressure of 
\begin{equation}
P_{\rm bw}=24355 \ \mu{\rm b},
\label{hc12}
\end{equation}
about 57 times greater than the CO$_2$ partial pressure $P_{\rm c}= 430\,\mu {\rm b}$\, in Earth's atmosphere in the year 2025.  According to Macpherson\,\cite{Macpherson}, ``Groundwater CO$_2$ partial pressures are typically $\sim 10-100$ times higher than atmospheric," so a factor of 57 is not surprising.  But the carbonation is probably adjusted at the bottling plant to produce bottled water with the ``fizz'' that customers expect.

The analysis\,\cite{HEB} also reports a  {\it Bicarb. \!Alkalinity} of $63 \hbox{ mg } l^{-1}$, so to good approximation
\begin{equation}
(\hbox{Bicarb})_{\rm bw}= 63 \hbox{ mg kg}^{-1} .
\label{hc14}
\end{equation}
We assume that this quantity is the mass of bicarbonate ions ${\rm HCO}_3^-$ per kg of solution. Using  numerical values of equilibrium constants from the first line of Table\,\ref{K012},  (\ref{hc2}) and (\ref{hc12}) in (\ref{in38}), we find that the bicarbonate molality is
\begin{eqnarray}
[{\rm HCO}_3^-]_{\rm bw}&=& \frac{K_1K_0P_{\rm bw}}{[{\rm H}^+]_{\rm bw}}=1.06 \hbox{ mM}.
\label{hc16}
\end{eqnarray}
Noting that the molecular weight of ${\rm HCO}_3^-$ ions is
\begin{equation}
m({\rm HCO}_3^-)=61  \hbox{ g mol$^{-1}$},
\label{hc18}
\end{equation}
we see that the bicarbonate alkalinity should be
\begin{eqnarray}
(\hbox{Bicarb})_{\rm bw}&=&m({\rm HCO}_3^-)[{\rm HCO}_3^-]_{\rm bw}=64 \hbox{ mg kg}^{-1}.
\label{hc20}
\end{eqnarray}
The value (\ref{hc14}) from the experimental water analysis and the value  (\ref{hc20}) that follows  from the reported  pH (\ref{hc2}) and implied CO$_2$ partial pressure (\ref{hc12}) are nearly the same.

\begin{figure}[t]
\begin{centering}
\includegraphics[height=70mm,width=.8\columnwidth]{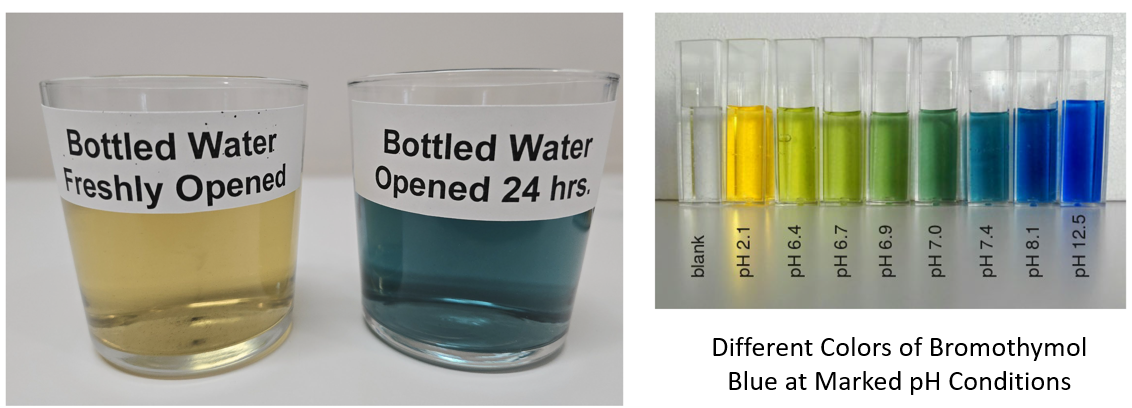}
\caption {Texas Spring Water\,\cite{HEB} is carbonated at the bottling plant and is slightly acidic, with pH$_{\rm bw}$ = 6.4. The dissolved CO$_2$ is in equilibrium with a CO$_2$ partial pressure of about $P_{\rm bw} = 24355\,\mu$b of (\ref{hc12}). The indicator dye, bromothymol blue \cite{bromo}, added to newly unsealed Texas Spring Water, gives the solution a pale shade of yellow, as one would expect from the indicator color chart on the right from reference\,\cite{bromo}. After the water has been exposed to the open air for 24 hours, much of the excess CO$_2$ escapes. The partial pressure with which the solution is in equilibrium approaches the open-air value $P_{\rm c} = 430\,\mu$b. Now the indicator dye turns the solution blue, indicating a more basic solution with less cancellation of the alkalinity by the weak acid CO$_2$. According to (\ref{hc34}), the pH should increase to pH$_{\rm c}$ = 8.15. From reference\,\cite{Buffering}.}
\label{Decarbonation}
\end{centering}
\end{figure}

Using (\ref{hc2}) and (\ref{hc12}) with freshwater equilibrium coefficients from Table \ref{K012} in (\ref{in46}), we find that the carbonate molality should have been 
\begin{eqnarray}
[{\rm CO}_3^{2-}]_{\rm bw}&=& \frac{K_2K_1K_0P_{\rm bw}}{[{\rm H}^+]_{\rm bw}^2}=9.89 \times 10^{-8} \hbox{ M}.
\label{hc22}
\end{eqnarray}
In like manner, (\ref{in24}) gives the molality of uncharged DIC as
\begin{eqnarray}
[\hbox{CO$_2^*$}]_{\rm bw}&=& K_0P_{\rm bw}=1.11\times 10^{-3} \ \hbox{ M}.
\label{hc24}
\end{eqnarray}
As shown in Fig. \ref{Decarbonation}, when the indicator dye, bromothymol blue\,\cite{bromo}, is added to Texas Spring Water from a newly opened bottle, the solution has the yellow color characteristic of the slightly acidic bottling value, pH = 6.4 and the bottling partial pressure $P_{\rm bw} = 24355\ \mu$b of (\ref{hc12}). 

After the water has been exposed to the air for 24 hours, much of the CO$_2$ from the water escapes. Since less of the alkalinity, [A]$_{\rm bw}$ = 1.06 mM of (\ref{hc8}), is cancelled by the weak acid, CO$_2$, the pH of the water increases to pH $>$ 7.4, as indicated by the color chart in Fig. \ref{Decarbonation}. Using the CO$_2$ partial pressure, $P_{\rm c} = 430\ \mu$b, with the equilibrium constants of Table \ref{K012} at $T = 15$ C and $S = 0$\textperthousand\, the solution of (\ref{al38}) gives
\begin{equation}
[{\rm H}^+]_{\rm c}=7.03\,\times\, 10^{-9} \hbox{ M}\quad\hbox{or}\quad \hbox{pH}_{\rm c}= 8.15.
\label{hc34}
\end{equation}
When Texas Spring Water goes ``flat" after long exposure to air, its hydrogen-ion molality decreases by a factor $[{\rm H}^+]_{\rm bw}/[{\rm H}^+]_{\rm c}\approx 18$. 

Using (\ref{in46}) with the contemporary CO$_2$ partial pressure $P_{\rm c}$ of (\ref{int6}) and the atmosphere-equilibrated hydrogen-ion molality $[{\rm H}^+]_{\rm c}$ of (\ref{hc34}) with the freshwater equilibrium coefficients from Table \ref{K012} for $T =15$ C, we find the carbonate molality is
\begin{eqnarray}
[{\rm CO}_3^{2-}]_{\rm c}&=& \frac{K_2K_1K_0P_{\rm c}}{[{\rm H}^+]_{\rm c}^2}=5.59\, \times\, 10^{-6} \hbox{ M}.
\label{hc36}
\end{eqnarray}

Substituting values of of [Ca$^{2+}$] from (\ref{hc10}), [CO$_3^{2-}$]$_{\rm bw}$ from (\ref{hc22}), [CO$_3^{2-}$]$_{\rm c}$ from (\ref{hc36}), and the freshwater solubility constant $K_{\rm sp}$ of Table \ref{K012} for $T = 15$ C into (\ref{ss8}), we see that the saturation states of newly unsealed Texas Spring Water and water equilibrated with atmospheric CO$_2$ are 
\begin{equation}
\Omega_{\rm bw}(P_{\rm bw}) =0.0141\quad\hbox{and}\quad\Omega_{\rm bw}(P_{\rm c}) =0.798.
\label{hc30}
\end{equation}
Calcite crystals CaCO$_3$(s) crystals would tend to dissolve in water with $\Omega_{\rm bw}(P_{\rm c}) =0.798$.
Yet freshwater bivalves and snails with calcium carbonate shells can thrive in streams with subsaturated water, similar to Texas Spring Water. Even though their shells are mostly calcium carbonate crystals, they do not dissolve because they are coated with small amounts of organic polymers like {\it conchiolin},\,\cite{conchiolin} that keep the crystals intact in water where physical chemistry alone would permit them to dissolve. 

The calcium-ion molality, $[\hbox{Ca}^{2+}]_{\rm bw}=0.53 \hbox{ mM}$, of (\ref{hc10}) exceeds the molality $[\hbox{Ca}^{2+}]_{\rm m}=0.102 \hbox{ mM}$ of minimum solubility of calcite in fresh water of temperature $T = 15$ C. Therefore, further reductions of the CO$_2$ partial pressure below $P_c$ would lead to precipitation of calcite in the Texas Spring Water.
At the very small CO$_2$ partial presssure pressure $P_{\rm m} = 0.265\,\mu$b of (\ref{LL20}), and the corresponding carbonate molality $[{\rm CO}_3^{2-}]_{\rm c} =3.62\,\times\, 10^{-5} \hbox{ M}$ of (\ref{LL23}) for minimum calcite solubility, the Texas Spring Water would be supersaturated,
\begin{equation}
\Omega_{\rm bw}(P_{\rm m}) =5.17,
\label{hc32}
\end{equation}
and calcite crystals would tend precipitate. Sealevel CO$_2$ partial pressures as small as $P_{\rm m} = 0.265\,\mu$b are not naturally encountered on Earth.

\section{Chemical thermodynamics}
In this final section, we review those parts of chemical thermodynamics which give useful insight into ocean acidifiation.
\subsection{Chemical potentials\label{g}}
Reactions like (\ref{ss2}) can be understood quantitatively with aid of chemical potentials\,\cite{mu} of the {\it reactant} molecules, CaCO$_3$(s) on the left side of the equation and {\it product} molecules Ca$^{2+}$ and CO$_3^{2-}$ on the right. The chemical potential is the Gibbs energy per molecule, often denoted with the symbol $\mu$. For a molecule of type X at the absolute temperature $T$, the chemical potential can be written as
\begin{eqnarray}
\mu&=&h-Ts\nonumber\\
&=&\mu^{\circ}+kT\ln a.
\label{g2}
\end{eqnarray}
On the first line of (\ref{g2}), 
the enthalpy per molecule is $h$, the entropy per molecule is $s$.
On the second line of (\ref{g2})  $\mu^{\circ}$  the chemical potential at the temperature $T$ and at unit {\it activity}, $a = 1$
when the concentration [X] of molecule X has its standard value [X]$^{\circ}$, and when other independent thermodynamic values also have their standard values, denoted by the superscipt $^{\circ}$. The value of Boltzmann's constant $k$ was given by (\ref{ss19}).

The ``effective concentration" of the molecule X is the activity $a$. For molecules in solution, we will measure molality [X] in moles per kg (M), and we will define the activity in terms of the molality and the dimensionless activity coefficient $\gamma$ as
\begin{equation}
 a=\gamma\frac{[{\rm X}]}{{\rm M}}\quad\hbox{and}\quad [{\rm X}]^{\circ}= \gamma {\rm M}\quad\hbox{for a dissolved molecule.}
\label{ss19a}
\end{equation}
where $\gamma \approx 1$ for ideal solutions.
For gaseous molecules $X$ of partial pressure $P$ the activity can  be defined as
\begin{equation}
 a=\gamma\frac{P}{{\rm bar}}.
\label{ss19b}
\end{equation}
For molecules in a solid, the activity  is usually defined to be unity,
\begin{equation}
 a=1.
\label{ss19c}
\end{equation}

The conversion of a single set of reactant molecules to a single set of product molecules, as in (\ref{ss2}), will be called a {\it forward} reaction. The inverse, the conversion of a set of product molecules to reactant molecules will be called a {\it reverse} reaction.
We will use the symbol $\Delta\mu$ to denote the difference between the sum of the chemical potentials of the product molecules and the chemical potentials of the reactant molecules. For example, in the reaction (\ref{ss2})
\begin{eqnarray}
\Delta \mu &=&\mu({\rm Ca}^{2+}) + \mu({\rm CO}_3^{2-}) - \mu({\rm CaCO}_3({\rm s}))\nonumber\\
%\Delta \mu &=&\mu(\hbox {Ca$^{2+}$})+\mu( \hbox{{\rm CO}_3^{2-}})-\mu(\hbox {CaCO$_3$(s)})\nonumber\\
&=&\Delta h-T\Delta s
\label{g4}
\end{eqnarray}
In accordance with  (\ref{g2}), the enthalpy of the set of product molecules exceeds that of the reactant molecules by  
\begin{equation}
\Delta h = h({\rm Ca}^{2+}) + h({\rm CO}_3^{2-}) - h({\rm CaCO}_3({\rm s})).
%\Delta h =h(\hbox {Ca$^{2+}$})+h( \hbox{{\rm CO}_3^{2-}})-h(\hbox {CaCO$_3$(s)}).
\label{g6}
\end{equation}
 The entropy of the set of product molecules exceeds that of the reactant molecules by  
\begin{equation}
\Delta s = s({\rm Ca}^{2+}) + s({\rm CO}_3^{2-}) - s({\rm CaCO}_3({\rm s})).
%\Delta s =s(\hbox {Ca$^{2+}$})+s( \hbox{{\rm CO}_3^{2-}})-s(\hbox {CaCO$_3$(s)}).
\label{g8}
\end{equation}
The heat $\Delta h$ that must be absorbed or released for each forward reaction comes from other molecules, with which the reactant and product molecules are in contact.  The heat $\Delta h$ can be positive, $\Delta h >0$, for {\it endothermic} reactions or negative, $\Delta h<0$, for {\it exothermic } reactions. As discussed  in Section {\bf \ref{K}}, the solubility reaction (\ref{ss2})  is exothermic for freshwater, but for seawater it is slightly endothermic for temperatures less than about  12 C, and slightly exothermic for higher temperatures, as shown by the bottom left panel of Fig. \ref{KTpS} and by the Van't Hoff equation (\ref{K6}).

We assume that all molecules involved in reactions like (\ref{ss2}) are characterized with the same temperature $T$ and total pressure $p$.  The reacting molecules are in contact with surrounding molecules, which we will call the {\it environment} and label with the subscript $e$.  From the first law of thermodynamics, conservation of energy, the enthalpy change $\Delta h$ for one forward reaction must be equal and opposite to the enthalpy change, $\Delta h_e$, of the environment
\begin{equation}
\Delta h_e =-\Delta h.
\label{g10}
\end{equation}
We assume that heat is transferred at constant pressure and temperature between the environment and reacting molecules. Using enthalpy change per molecule,
 $\Delta h$,  rather than internal energy change per molecule, $\Delta u = \Delta h-p\Delta v$, avoids the  need to account for the ``$p\, \Delta v$" work done by changes $\Delta v$ in the volumes of the reacting molecules. 

Each forward reaction (\ref{ss2}) changes the entropy of the environment by
\begin{equation}
\Delta s_e =\frac{\Delta h_e}{T} =-\frac{\Delta h}{T}.
\label{g12}
\end{equation}
Using (\ref{g12}) in (\ref{g4}) we find
\begin{equation}
\Delta \mu=-T\Delta s_t.
\label{g14}
\end{equation}
where the change in the total entropy, that of the environment and that of the reacting molecules, is 
\begin{equation}
\Delta s_t =\Delta s_e+\Delta s.
\label{g16}
\end{equation}
The second law of thermodynamics,  states that the total entropy of spontaneously reacting molecules and the environment must increase for irreversible reactions
 $\Delta s_t> 0$, or remain constant for reversible reactions,  $\Delta s_t=0$. Therefore (\ref{g14}) implies that 

\begin{eqnarray}
\Delta \mu &<& 0\quad\hbox{for spontaneous  or exergonic forward reactions,}\nonumber\\
\Delta \mu &=& 0\quad\hbox{for chemical equilibrium,}\nonumber\\
\Delta \mu &>& 0\quad\hbox{for spontaneous reverse or endergonic forward reactions.}
\label{g18}
\end{eqnarray}
It is customary, especially in biochemistry, to use the term {\it exergonic} to describe a forward reaction with $\Delta \mu <0$, or {\it endergonic} to describe a forward reaction with $\Delta\mu>0$\,\cite{ergonic}. A spontaneous forward reaction must be exergonic, with $\Delta \mu<0$, and it is often exothermic as well, with $\Delta h<0$. But a spontaneous reaction can be endothermic with $\Delta h >0$ if the entropy increase is sufficiently large, $\Delta s>\Delta h/T$.  An example is the dissolution of portlandite,  Ca(OH)$_2$,  in subsaturated water\,\cite{solubility}. This cools the water and leads to a retrograde temperature dependence of the solubility product, as discussed in Section {\bf\ref{K}}. 

We can write the chemical potential change per forward reaction,  (\ref{ss2}), as
\begin{eqnarray}
\Delta \mu
&=&\Delta \mu^{\circ}+kT \ln\left(a(\hbox {Ca$^{2+}$})a({\rm CO}_3^{2-})\right)\nonumber\\
&=&\Delta \mu^{\circ}+kT \ln\left(\gamma(\hbox {Ca$^{2+}$})\gamma({\rm CO}_3^{2-})[\hbox {Ca$^{2+}$}][{\rm CO}_3^{2-}] \hbox{ M}^{-2}\right)\nonumber\\
&=&\Delta \mu^{\circ}+kT \ln\left(\gamma(\hbox {Ca$^{2+}$})\gamma({\rm CO}_3^{2-})\Omega K_{\rm sp} \hbox{ M}^{-2}\right).
\label{ss26}
\end{eqnarray}
To write the first line of (\ref{ss26}), we used the second line of (\ref{g2}) and we noted from (\ref{ss19c}) that $a$\big( CaCO$_3$(s) \big) = 1. To write the second line of (\ref{ss26}) we used the expression (\ref{ss19a}) for the activities.
To write the third line of (\ref{ss26}) we used the definition (\ref{ss8}) of the saturation state $\Omega$.
The standard chemical potential change of (\ref{ss26}) is
\begin{eqnarray}
\Delta \mu^{\circ}
&=&\mu^{\circ}(\hbox {Ca$^{2+}$})+\mu^{\circ}(\hbox {CO$_3^{2-}$})-\mu^{\circ}(\hbox {CaCO$_3$(s)}).
\label{ss28}
\end{eqnarray}
For a saturated solution with $\Omega =1$, there will be no tendency for crystals to dissolve or grow, and therefore $\Delta \mu =0$. Then the third line of
 (\ref{ss26}) gives
\begin{equation}
\Delta \mu^{\circ}=-kT\ln\left (\gamma(\hbox {Ca$^{2+}$})\gamma({\rm CO}_3^{2-})K_{\rm sp} \hbox{ M}^{-2}\right).
\label{ss30}
\end{equation}
Dividing both sides of (\ref{ss30}) by $kT$ and exponentiating, we find that the solubility product can be writtten as
\begin{equation}
K_{\rm sp}=\frac{\mathcal{K}_{\rm sp}}{\gamma(\hbox {Ca$^{2+}$})\gamma({\rm CO}_3^{2-})} \hbox{ M}^{2}.
\label{ss30a}
\end{equation}
Here the dimensionless, thermodynamic solubility product is  
\begin{equation}
\mathcal{K}_{\rm sp}=e^{-\Delta \mu^{\circ}/kT}.
\label{ss30a}
\end{equation}
For a saturated solution of CaCO$_3$, the analog of (\ref{ss4}) is the dimensionless solubility product equation,
\begin{equation}
a(\hbox {Ca$^{2+}$}) a({\rm CO}_3^{2-})=\mathcal{K}_{\rm sp}.
%\label{ss30b}
\end{equation}

Substituting (\ref{ss30}) into (\ref{ss26}) we find that the chemical potential change for each forward reaction (\ref{ss2}) is
\begin{eqnarray}
\Delta \mu= k T\ln\Omega.
\label{ss32}
\end{eqnarray}
Substituting the numerical value  (\ref{ss11c}) of the saturation state of calcite in seawater  into (\ref{ss32}) we find 
\begin{equation}
\Delta \mu=1.65\,\times\, k T\hbox{ reaction}^{-1}\label{ss33} 
\end{equation}
The positive chemical-potential increments of (\ref{ss33})  mean that for the oceans of the year 2025, the dissolution reaction (\ref{ss2}) is {\it uphill}. For spontaneous reactions, there will be more reverse, precipitation reactions than forward, dissolution reactions. For the Texas Spring Water discussed in Section {\bf \ref{hc}} we can substitute the saturation state (\ref{hc30}) into (\ref{ss32}) to find 
\begin{equation}
\Delta \mu=- 0.226\,\times\, kT \hbox{ reaction}^{-1}\label{ss36}
\end{equation}
The negative value of $\Delta\mu$ in (\ref{ss36}) mean that the dissolution reaction (\ref{ss2}) is {\it downhill} and  calcite  crystals would dissolve in Texas Spring Water.

To get some feeling for how big the driving energies $\Delta\mu$ of (\ref{ss33})  or (\ref{ss36}) are, with respect to representative biological energy exchanges, we consider the hydrolysis  of adenosine triphosphate  (ATP) to adenosine diphosphate (ADP)\,\cite{ATP}, which provides the energy to run cellular metabolism for all of life.
The hydrolysis of ATP  is described with the chemical equation
\begin{equation}
\hbox {ATP}+\hbox{H$_2$O} \rightleftharpoons\hbox{ADP}\, +P_i.
\label{ss12}
\end{equation}
Unlike (\ref{ss2}), Eq. (\ref{ss12}) does not balance, because the inorganic phosphate, denoted by the symbol $P_i$, stands for several different phosphate ions, the most abundant of which are hydrogen phosphate,  HPO$_4^{2-}$, and dihydrogen phosphate H$_2$PO$_4^{-}$, 
\begin{equation}
P_i = \hbox{HPO$_4^{2-}$} \quad \hbox{and}\quad \hbox{H$_2$PO$_4^{-}$}.
\label{ss12a}
\end{equation}
According to reference\,\cite{ATP}, for physiological conditions in living cells,  hydrolysing ATP molecules, as described by (\ref{ss12}) releases a Gibbs free energy per mole  of about
\begin{equation}
\Delta G= -64 \hbox{ kJ mol}^{-1}.
\label{ss40}
\end{equation}
The corresponding increment of chemical potential for the hydrolysis of a single ATP molecule is 
\begin{eqnarray}
\Delta \mu &=& \frac{\Delta G}{N_{\rm A}}\nonumber\\
&=&-1.06\,\times\, 10^{-19} \hbox { J reaction}^{-1}\nonumber\\
&=&-26\, \times\, k T\hbox{ reaction}^{-1} \qquad\qquad \hbox{for ATP$\to$ ADP}.
\label{ss42}
\end{eqnarray}
The number on the last line is for the standard temperature $T = 25 \hbox{ C} = 298.15$ K. 

According to (\ref{ss42}) and (\ref{ss36}), the hydrolysis of 1 ATP molecule provides  26/0.226 = 115 times more free energy than the dissolution of one pair of Ca$^{2+}$ and CO$_3^{2-}$ ions from calcite into subsaturated Texas Spring Water. The oxidative metabolism of a single glucose sugar molecule produces about 30 ATP molecules\,\cite{Citric}, and would provide enough free energy to precipitate 30$\,\times\,$115 = 3450 pairs of  Ca$^{2+}$ and CO$_3^{2-}$ ions from subsaturated Texas Spring Water onto biomineralized calcite crystals.  This is one of the reasons that freshwater bivalves or snails have little trouble growing calcite or aragonite shells in regions with limestone or dolomite bedrock.

\subsection{Dependence of $K_{\rm sp}$ on $T$, $p$ and $S$ \label{K}}
For the ranges of temperatures $T$, pressures $p$ and salinity $S$ encountered in natural waters, 
the solubility product constant $K_{\rm sp}$ of (\ref{ss4}) for calcite or aragonite depends weakly on temperature and strongly on pressure and salinity.  The rate of change of $K_{\rm sp}$ with temperature $T$ can be used to determine how much heat $\Delta h^{\circ}$ is absorbed or released by a single forward reaction (\ref{ss2})  if  the products and reactants have their standard concentrations. The rate of change of $K_{\rm sp}$ with total pressure $p$ can be used to determine the volume change $\Delta v^{\circ}$.

Van't Hoff's equation, which comes from basic chemical thermodynamics\,\cite{VH}, can be written as
\begin{equation}
 \Delta h^{\circ}=kT^2\frac{\partial \ln K_{\rm sp}}{\partial T}. 
\label{K6}
\end{equation}
For  (\ref{K6}) to be exact, one should use the thermodynamic solubility product $\mathcal{K}_{\rm sp}$ of (\ref{ss30a}),  rather than than the concentration solubility product $K_{\rm sp}$.  But if the activity constants $\gamma$ of (\ref{ss19b}) have negligible dependence on temperature $T$, either solubility constant can be used.
The {\it standard enthalpy change} per forward reaction (\ref{ss2})
\begin{equation}
\Delta h^{\circ}=h^{\circ}(\hbox{Ca$^{2+}$})+h^{\circ}(\hbox{CO$_3^{2-}$})-h^{\circ}(\hbox{CaCO$_3$(s)}).
\label{K4}
\end{equation}
In (\ref{K4}),  $h^{\circ}(\hbox{Ca$^{2+}$})$ or $h^{\circ}(\hbox{CO$_3^{2-}$})$ are the enthalpy increases caused by adding single $\hbox{Ca$^{2+}$}$ or $\hbox{CO$_3^{2-}$}$ ions to the solution, and $h^{\circ}(\hbox{CaCO$_3$(s)})$ is the enthalpy increase caused by precipitating a pair of $\hbox{Ca$^{2+}$}$ and $\hbox{CO$_3^{2-}$}$ ions onto a calcite crystal, where the ion molalities in solution have their standard values,  [Ca$^{2+}$]$^{\circ}$ = 1 M and  [CO$_3^{2-}$]$^{\circ}$ = 1 M.

Mucci\,\cite{Mucci} has shown that experimental measurements of the solubility product $K_{\rm sp}$ of calcite at a pressure, $p=1$ bar  for various salinities $S$ (in units of \textperthousand) and  temperatures $T$ (in units of K), are described accurately by the  empirical equation,
\begin{eqnarray}
\log K_{\rm sp}&=&-171.9065 - 0.077993 \,T +2839.319/T +71.595 \log T\nonumber\\
&&+\big(-0.77712+0.0028426\,T+178.34/T\big)S^{1/2}\nonumber\\
&&-0.07711\,S+0.0041249\,S^{3/2}.
\label{K8}
\end{eqnarray}
We have used (\ref{K8}) to make panels (a), (b) and (d) of Fig. \ref{KTpS}.  Van't Hoff's equation (\ref{K6}), together with the steady decrease of $K_{\rm sp}$ for the temperature range, 0 C $<T<$ 40 C  of  Fig. \ref{KTpS} (a), show that dissolving calcite in freshwater is exothermic for natural conditions on Earth. From Fig. \ref{KTpS} (b),  we see that dissolving calcite is slightly endothermic for $T<12 $ C and  slightly exothermic for $T> 12 $ C in seawater.  For a temperature $T = 25$ C, we can use 
(\ref{K8}) with (\ref{K6}) to find that the enthalpy change of reacting molecules, per forward reaction (\ref{ss2}) is 
\begin{equation}
 \Delta h^{\circ}=kT\,\times\, \left \{\begin{array}{rl}-3.88, &\hbox{ for $S = 0{^o\!\!/_{\!\!oo}}$ or freshwater},\\
-0.48,&\hbox{ for $S = 35 {^o\!\!/_{\!\!oo}}$ or seawater}. \end{array}\right .
\label{K10}
\end{equation}
The negative values of the enthalpy changes of (\ref{K10}) mean that the thermal energy needed to detach Ca$^{2+}$ and CO$_3^{2-}$ ions from a calcite crystal in solution, with unit molalities,  [Ca$^{2+}$] =  [CO$_3^{2-}$] = 1 M, and at constant temperature and pressure, is  more than compensated by the heat released when the ions form hydration shells\,\cite{hydration}.

\begin{figure}[t]
\begin{centering}
\includegraphics[height=90mm,width=1\columnwidth]{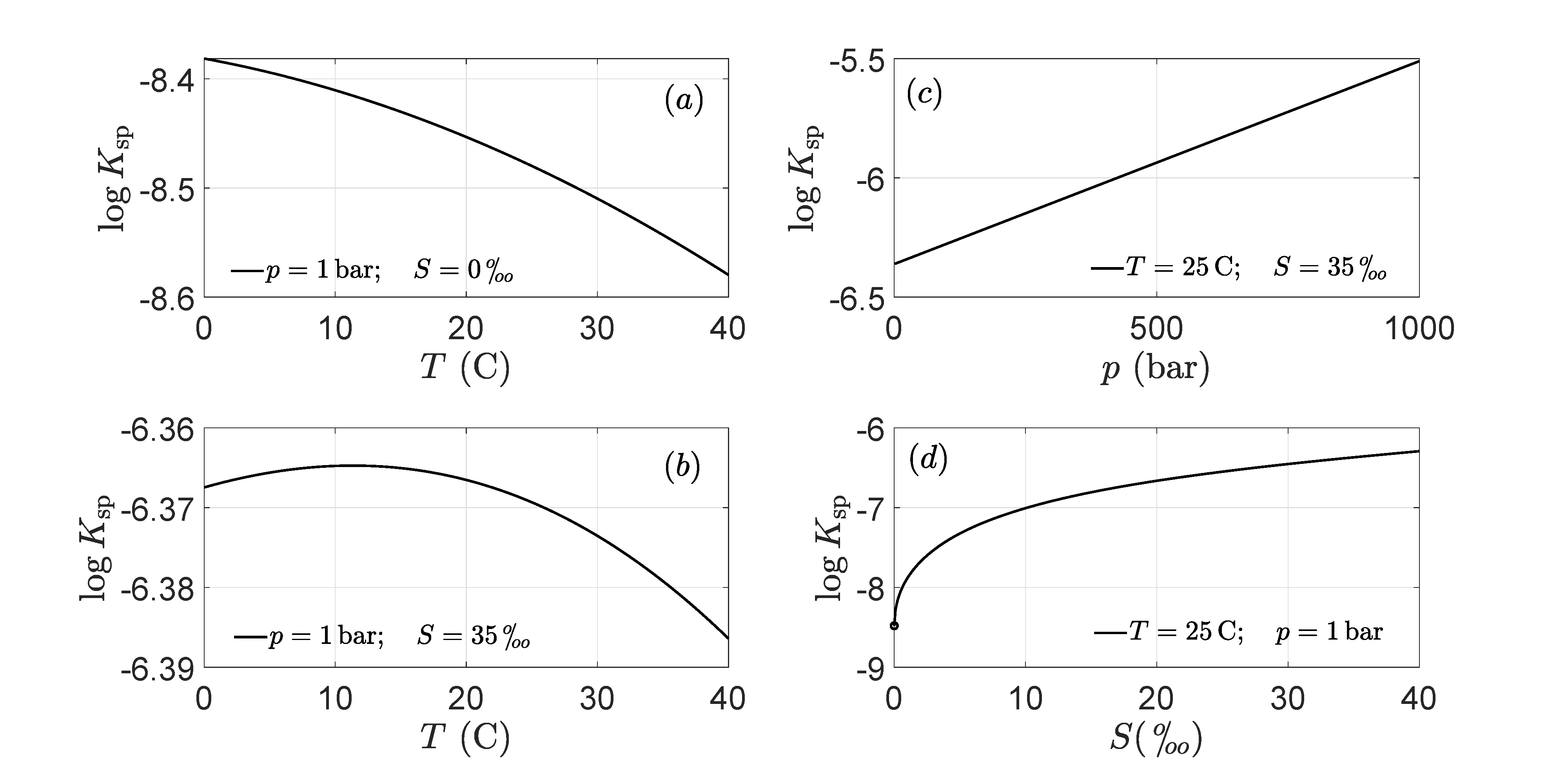}
\caption {For the ranges of temperature $T$, pressure $p$ and salinity $S$ encountered in natural waters, the solubility product constant $K_{\rm sp}$ for calcite depends very  weakly on increasing temperature $T$ in seawater and weakly on increasing temperature in freshwater. $K_{\rm sp}$ increases strongly with increasing pressure $p$ and very strongly with increasing salinity $S$.}
\label{KTpS}
\end{centering}
\end{figure}

An analog of van't Hoff's equation (\ref{K6})  is\,\cite{VH}
\begin{equation}
\Delta v^{\circ}=-kT \frac{\partial \ln K_{\rm sp}}{\partial p}.
\label{K20}
\end{equation}
In (\ref{K20}) the {\it standard volume change} per forward reaction (\ref{ss2}),
\begin{equation}
\Delta v^{\circ}=v^{\circ}(\hbox{Ca$^{2+}$})+v^{\circ}(\hbox{CO$_3^{2-}$})-v^{\circ}(\hbox{CaCO$_3$}),
\label{K22}
\end{equation}
is the amount that the volume of the solution must increase (if  $\Delta v^{\circ}>0$), or decrease (if  $\Delta v^{\circ}<0$) for each forward reaction (\ref{ss2}). 

Boudreau {\it et al.} \,\cite{Boudreau} have shown that experimental measurements of the solubility product $K_{\rm sp}$ of calcite in seawater versus pressure, adjusted to an idealized profile of temperature versus depth, is described accurately by the  empirical equation,
\begin{eqnarray}
K_{\rm sp}=4.3513\,\times\, 10^{-7}e^{c\, p} \hbox{ M}^2,\quad\hbox{where}\quad c=0.0019585\,\hbox{ bar}^{-1}.
\label{K26}
\end{eqnarray}
We have used the empirical formula (\ref{K26}) to make panel (c) of Fig. \ref{KTpS}. At a pressure $p=1$ bar, (\ref{K26}) gives $K_{\rm sp}=4.36\,\times\, 10^{-7}$. This is  about 2\% larger than the value  $K_{\rm sp}=4.27\,\times\, 10^{-7}$ from Table \ref{K012},  which comes from the empirical formula (\ref{K8}). For the discussions of this paper, the difference is too small to matter.

We can use (\ref{K20}) and (\ref{K26})  to write the volume change, per forward reaction (\ref{K22}), at temperature $T = 25$ C as
\begin{eqnarray}
 \Delta v^{\circ}&=&-kT c=-8.06\,\times\, 10^{-23} \hbox{ cm}^3\nonumber\\
&=&-2.70\,v(\hbox{H$_2$O}).
\label{K28}
\end{eqnarray}
Here $v({\rm H_2O})$ is a representative volume  of a liquid H$_2$O molecule. Taking the density of liquid water to be to $\rho = 1$ g cm$^{-3}$, and the gram molecular weight of H$_2$O molecules to be  $M({\rm H_2O})=18.015$ g, we find
\begin{equation}
v({\rm H_2O})=\frac{M({\rm H_2O})}{\rho N_{\rm A}}=3.0\, \times\, 10^{-23}\hbox{ cm}^3.
\label{K30}
\end{equation}
According to (\ref{K28}), at constant temperature and pressure, each pair of ions, Ca$^{2+}$ and CO$_3^{2-}$, dissolved from calcite into solution, decreases the overall volume of the solution and submerged calcite crystal by 2.7 times the representative volume (\ref{K30}) of a liquid water molecule. This is mostly due to the formation of dense hydration shells around the newly released ions\,\cite{hydration}.

The increase in solubility of calcite with pressure resulting from the decrease in volume (\ref{K28}) is consistent with Le Chatelier's Principle\,\cite{LeC}. Dissolving more calcite, at constant temperature and pressure, decreases the overall volume of calcite-seawater mixture and counteracts the pressure increase.

\section{Conclusion}
We have reviewed the inorganic carbonate chemistry of natural waters, seawater, groundwater and rainwater, in equilibium with the partial pressures $P$ of CO$_2$ in the atmosphere.  We have shown that even very large changes of $P$, like doubling from $P_c = 430\ \mu$b \,to $P_d = 860\ \mu$b, change the  pH of natural waters less than the day-to-night changes due photosynthesis of aquatic organisms, or due to local variations of seawater alkalinity.  The pH changes are so small that any credible future increases of atmospheric CO$_2$ will cause no harm to aquatic life. In a subsequent paper we will discuss details of how increasing concentrations of atmospheric CO$_2$ affect biomineralization and other biochemistry of aquatic organisms. Carbon from CO$_2$ is the main building block of life, and it can be a limiting nutrient when sunlit water has adequate amounts of phosphorous, nitrogen, iron and other essential elements. It is highly likely that more atmospheric CO$_2$ has already been and will continue to be a net benefit to aquatic life.

\section*{Acknowledgements}
We are grateful to Ferdinand Engelbeen for carefully reviewing drafts of this paper and for useful criticisms.  W. Happer and W. van Wijngaarden thank their respective universities, Princeton and York, for facilitating their work on this paper.

\end{document}